\documentclass[%
 reprint,
 nofootinbib,
 superscriptaddress,
 amsmath,amssymb,
 aps,
 nobalancelastpage,
]{revtex4-2}

\usepackage{amssymb}
\usepackage{graphicx}
\usepackage{dcolumn}
\usepackage{bm}
\usepackage[colorlinks=true,citecolor=blue,urlcolor=blue,linkcolor=red]{hyperref}
\usepackage{fontawesome5}
\usepackage{siunitx}
\usepackage{pgf}
\usepackage[normalem]{ulem}
\usepackage{xspace}
\usepackage{makecell}
\usepackage{physics}
\usepackage{enumitem}
\usepackage{multirow}
\usepackage{orcidlink}

\newcommand\numberthis{\addtocounter{equation}{1}\tag{\theequation}}

\usepackage{listings}
\usepackage{xcolor}

\definecolor{listingbg}{RGB}{248,248,245}
\definecolor{listingrule}{RGB}{225,225,220}

\lstdefinestyle{settings}{
  basicstyle=\ttfamily\footnotesize,     
  backgroundcolor=\color{listingbg},
  numberstyle=\ttfamily\footnotesize\color{gray},
  numbersep=10pt,
  frame=single,                   
  rulecolor=\color{listingrule},
  framexleftmargin=6pt,           
  xleftmargin=6pt,
  xrightmargin=6pt,
  aboveskip=1em, belowskip=1em,
  breaklines=true,                
  columns=fullflexible,           
  keepspaces=true,                
  showstringspaces=false,
}

\newcommand{\planck}{{\it Planck}\xspace}
\newcommand{\lcdm}{\mbox{$\Lambda$CDM}}
\newcommand{\sptthreeg}{SPT-3G}
\newcommand{\ito}{It\^{o}\xspace}

\newcommand{\mnras}{Mon. Not. Roy. Astron. Soc.}

\newcommand{\code}[1]{{\texttt{\detokenize{#1}}}}

\newcommand{\eavg}[1]{\langle{#1}\rangle}

\makeatletter
\newcommand\thefontsize[1]{{#1 The current font size is: \f@size pt\par}}
\makeatother

\begin{document}

\title{Limits on the size of electron density fluctuations in the \\ early universe and implications for the Hubble tension}

\author{Gabriel P. Lynch\,\orcidlink{0009-0004-3143-1708}}
\email[]{gplynch@ucdavis.edu}
\affiliation{Department of Physics and Astronomy, University of California, Davis, California, USA 95616}

\author{Jens Chluba\,\orcidlink{0000-0003-3725-6096}}
\affiliation{Jodrell Bank Centre for Astrophysics, University of Manchester, Manchester M13 9PL, UK}

\author{Richard Battye\,\orcidlink{0000-0002-4269-515X}}
\affiliation{Jodrell Bank Centre for Astrophysics, University of Manchester, Manchester M13 9PL, UK}

\author{Geoffrey Vasil\,\orcidlink{0000-0002-8902-5030}}
\affiliation{School of Mathematics and the Maxwell Institute for Mathematical Sciences, University of Edinburgh, EH9 3FD, UK}

\date{\today}

\begin{abstract}
Little is known about the early universe on scales smaller than those resolved by the cosmic microwave background (CMB), where electron density fluctuations on kiloparsec to megaparsec scales are expected in many extensions of $\Lambda$CDM. These fluctuations nonetheless leave two imprints on the CMB. They accelerate recombination, the effect behind previous constraints and proposals to raise the inferred Hubble constant, $H_0$. Through correlations along the line of sight, they also reduce the effective Thomson scattering rate, enhancing diffusion damping by an amount that grows with their coherence length. We implement both effects in a modified Boltzmann code, \textsc{classito}, and use CMB data from \planck, ACT DR6, and SPT-3G D1 to place the first limits on such fluctuations that include this modified photon transport. We find no evidence for them: the variance of the electron density contrast is below $0.21$ ($95\%$ CL), falling to $0.033$ for a coherence length of $2\,{\rm Mpc}$, and we present these limits in terms of the electron power spectrum at $1\,{\rm Mpc}^{-1}\lesssim k\lesssim10^2\,{\rm Mpc}^{-1}$. We also reassess small-scale electron density fluctuations as a resolution of the Hubble tension. In \planck\ data the transport effect is partially degenerate with accelerated recombination, widening the range of $H_0$ allowed: combined with a distance-ladder prior, \planck\ gives $H_0 = 71.1\pm0.7\,{\rm km\,s^{-1}\,Mpc^{-1}}$ and an improved fit relative to $\Lambda$CDM. ACT and SPT-3G break this degeneracy, chiefly through the temperature--$E$-mode correlation in the damping tail, and with the same prior they give only $H_0 = 69.3\pm0.5\,{\rm km\,s^{-1}\,Mpc^{-1}}$ and no preference over $\Lambda$CDM. With our treatment and publicly available code, models that generate small-scale electron density fluctuations can be directly confronted with CMB data through the variance and coherence length that they predict.
\end{abstract}

\maketitle

\section{Introduction}\label{sec:intro}

We know very little about the small-scale universe at early times. Anisotropies in the cosmic microwave background (CMB), which on larger scales provide a wealth of cosmological data, are erased by Silk damping below $\sim10$ comoving Mpc. Our knowledge of smaller scales --- in particular, of the existence and character of perturbations at these scales --- must therefore rely on less direct probes, which are often sensitive to only one component of the plasma at one epoch. For example, CMB spectral distortions are sensitive to photon perturbations that dissipated well before last scattering \citep{Chluba:2012we,Chluba:2013dna,Nakama:2014vla,Chluba:2019kpb}, the primordial deuterium abundance is sensitive to baryon inhomogeneities at nucleosynthesis \citep{Barrow:2018yyg,Inomata:2018htm,Bagherian:2025puf}, and the abundance of high-redshift galaxies is sensitive to small-scale matter power long after recombination \citep{Co:2026klr}. The focus of this paper is fluctuations in the free-electron density at and before recombination, on comoving scales of $\sim10\,{\rm kpc}$ to $\sim1\,{\rm Mpc}$, which we probe through their averaged effect on the ionization history and evolution of photon perturbations \citep{Jedamzik:2013gua,Lee:2021bmn,Chluba:2025jqt}.

What could generate such fluctuations? Writing $\delta_{\rm e} \simeq \delta_{\rm b}+\delta X_{\rm e}/X_{\rm e}$ identifies two general routes: either directly sourcing a baryon perturbation $\delta_{\rm b}$, or modulating the free-electron fraction at fixed baryon density. In the first case, $\delta_{\rm b}$ could arise from a primordial perturbation in the baryon-to-photon ratio, such as baryon or compensated isocurvature, which Compton drag holds fixed until kinematic decoupling \citep{Lee:2021bmn}. One example is that of baryogenesis modulated by a light field during inflation \citep{Affleck:1984fy,Enqvist:1998pf, Kasuya:2008xp,Inomata:2018htm}, provided the resulting spectrum is blue enough to evade large-scale isocurvature limits \citep{Planck:2018jri}. Alternatively, instead of a primordial perturbation, a baryon overdensity could be driven after horizon entry by a force able to overcome the Compton drag. The Lorentz force supplied by primordial magnetic fields (PMFs) is one such force and is perhaps the best-studied example \citep{Jedamzik:2013gua,Jedamzik:2023rfd,Ralegankar:2023pyx} in this context. Another possibility is ``gravitational seeding'' by collapsed, non-linear structures, such as primordial black holes and the haloes they nucleate \citep{Ali-Haimoud:2016mbv,Ricotti:2009bs,Inman:2019wvr}, cosmic string loops \citep{Cyr:2025tkt}, or the first haloes of a cosmology with strongly enhanced small-scale power \citep{Hirano:2015wla,Qin:2025ymc}. Finally, the ionization fraction can spatially vary due to locally deposited energy injected by compact sources \citep{Dvorkin:2013cga, Ali-Haimoud:2016mbv, Jensen:2021mik}, or due to spatial variations in the fundamental constants at the scales in question \citep{Smith:2018rnu}.

The CMB is sensitive to small-scale electron fluctuations in two ways. First, because the recombination rate is quadratic in the electron density while the photoionization rate is linear, clumping accelerates the mean recombination history \citep{Jedamzik:2013gua}. This effect has been used to constrain baryon clumping from PMFs \citep{Jedamzik:2018itu} and small-scale isocurvature \citep{Lee:2021bmn}, but it depends only on the variance of the fluctuations and not on their size. Second, as shown recently by \citet[hereafter CVB]{Chluba:2025jqt}, fluctuations correlated along the line of sight also modify photon transport: the effective Thomson scattering rates entering the Boltzmann hierarchy are reduced in a multipole-dependent way for temperature and polarization, and diffusion damping is enhanced. This effect grows with the coherence length of the fluctuations, and no modification of the mean ionization history reproduces it. Because the free-electron fraction responds non-linearly to the local density and to any heating, the electron fluctuations need not directly inherit the amplitude or coherence length of the perturbation that seeded them \citep{Chluba:2025jqt}; we therefore parameterize the electron field directly, by its variance and coherence length as functions of redshift. 

Small-scale electron fluctuations are also of interest for the Hubble tension, the $>6\sigma$ disagreement between the value of $H_0$ measured with the distance ladder and that inferred from the CMB within \lcdm\ \citep[see][for a review]{CosmoVerseNetwork:2025alb}. Among proposed resolutions, modifying cosmological recombination, which is commonly treated as a fixed theoretical ingredient obtained from detailed recombination codes \citep{Chluba:2010ca, PhysRevD.83.043513}, is one of the more conservative avenues \citep{Chiang:2018xpn,Hart:2019dxi,Sekiguchi:2020teg}. The CMB fixes the angular size of the sound horizon, $\theta_s^\star = r_s^\star/D_A^\star$, at the $0.02\%$ level, so a smaller $r_s^\star$ admits a larger $H_0$ \citep{Knox:2019rjx}, and an earlier recombination is one way to obtain it. Model-independent reconstructions confirm that recombination histories degenerate with a high $H_0$ exist \citep{Lee:2022gzh,Lynch:2024gmp,Mirpoorian:2024fka,Lee:2026hlo}; the question is whether a physical mechanism produces them. Because inhomogeneity accelerates recombination, small-scale clumping is one candidate, first proposed with PMFs as the source \citep{Jedamzik:2020krr} and later studied with more flexible clumping models and with small-scale CMB data \citep{Rashkovetskyi:2021rwg,Thiele:2021okz,Galli:2021mxk,Lee:2021bmn,Schiff:2025vjf}.\footnote{An alternative candidate is varying fundamental constants \citep{Hart:2019dxi,Sekiguchi:2020teg}, which have scored well in model comparisons even with the inclusion of small-scale CMB data \citep{Schoneberg:2021qvd, AtacamaCosmologyTelescope:2025nti,SPT-3G:2025bzu, Schoneberg:2026eys}.} These analyses find that clumping accommodates $H_0 \sim 70\,{\rm km}\,{\rm s}^{-1}\,{\rm Mpc}^{-1}$ with \planck\ data, but that small-scale data significantly restrict the high-$H_0$ tail of the posteriors. Indeed, the recent standardized comparison of \citet{Schoneberg:2026eys} 
ranks a clumping-inspired model \citep{Mirpoorian:2024fka} below other proposed solutions.

The limitation faced by these models has a general origin: the CMB measures the damping scale in addition to the sound horizon, and a model that lowers $r_s^\star$ must lower $r_{\rm d}^\star$ in proportion \citep{Knox:2019rjx,Sekiguchi:2020teg,Smith:2025zsg}, which clumping does not do \citep{Rashkovetskyi:2021rwg}. All of these analyses, however, included only the recombination channel. The transport channel acts directly on photon diffusion and therefore on $r_{\rm d}^\star$, and any physical clumping mechanism produces both channels. A reassessment that includes both, confronted with the latest small-scale measurements from ACT DR6 and \sptthreeg, is therefore warranted.

In this paper we present the first constraints on small-scale electron density fluctuations that include the transport channel. We implement the CVB formalism in \textsc{classito}\footnote{Available at \url{https://github.com/gplynch619/classito}}, a fork of the Einstein--Boltzmann code \textsc{class} \citep{Blas:2011rf}, and confront it with CMB data from \planck, ACT DR6, and \sptthreeg\ D1, together with BAO data from DESI DR2 and a distance-ladder prior on $H_0$. Our main results are as follows. First, we obtain joint limits on the variance and coherence length of electron density fluctuations at recombination, which translate into limits on the fluctuation power spectrum over $1\,{\rm Mpc}^{-1}\lesssim k\lesssim 10^2\,{\rm Mpc}^{-1}$. We compare our limits with those from the deuterium abundance and from previous recombination-channel analyses, and forecast constraints for future CMB experiments. We show that the two channels are partially degenerate in \planck\ data, so that an analysis including only the recombination channel overstates its constraining power. Marginalizing over the transport channel is necessary for a conservative bound. The small-scale data from ACT and \sptthreeg\ break this degeneracy. Third, we reassess clumping solutions to the Hubble tension in light of both the transport channel and the latest small-scale data, and find that the high-$H_0$ solutions allowed by \planck\ are excluded by ACT and \sptthreeg.

Given the number of mechanisms that could produce such fluctuations, and the difficulty of following any one of them from its generation to the free electrons at last scattering, we do not tie our analysis to a particular source. \textsc{classito} is instead parameterized by the variance and coherence length of the electron density fluctuations as functions of redshift. We release it publicly, together with an interface to the \textsc{cobaya} sampling framework through which a user supplies these two functions for a model of their choice (Appendix~\ref{app:code}), so that any mechanism predicting them can be confronted with CMB data directly.

The paper is organized as follows. Section~\ref{sec:model} describes the physical model and its two effects on the CMB power spectra; Section~\ref{sec:data_analysis} our datasets and methods; Section~\ref{sec:constraints} our constraints on small-scale fluctuations and a comparison with previous limits; and Section~\ref{sec:hubble} the implications for the Hubble tension. We conclude in Section~\ref{sec:conclusion}. Appendix~\ref{app:validity} assesses the range of validity of our model and estimates its systematic theoretical uncertainty, and Appendix~\ref{app:code} describes \textsc{classito} and its interface.

\section{Physical effects of small-scale electron fluctuations}\label{sec:model}

In the standard treatment of CMB anisotropies, the photons scatter off electrons that are coupled via the Coulomb interaction to the baryons. The baryon density perturbations are linear perturbations of the same primordial, adiabatic origin as the photon anisotropies, erased by Silk damping below $\sim10$ comoving Mpc. The fluctuations we consider here are of a different kind. They are coherent on comoving scales $L_{\rm e} \sim \mathcal{O}(10\,{\rm kpc}) - \mathcal{O}(1\,{\rm Mpc})$, far below the damping scale, may be non-linear, and need not be primordial or adiabatic in origin. Projected from last scattering, these scales correspond to multipoles $\ell \approx 10^4 - 10^{6}$, beyond the resolution of current CMB surveys. Future experiments like the proposed CMB-HD \citep{Sehgal2019CMB-HD} could reach the lower end of this range, but there the signal will be dominated by secondary anisotropies \citep[see][for recent forecasts]{MacInnis2026CMB-HD}. However, as we shall see, the data at observable scales are sensitive to fluctuations in this regime in a statistical way.

In our analyses we adopt the formalism of CVB. We begin in Section~\ref{sec:motivation} with a description of the physical model for the small-scale fluctuations, and motivate the statistical treatment presented in CVB. This will also serve to introduce our notation and new model parameters. After motivating our approach, we then discuss two distinct effects of small-scale fluctuations: changes to the ionization fraction $X_{\rm e}(z)$ (Section~\ref{sec:modified_rec}), and modifications to the photon Boltzmann hierarchy in the presence of correlated fluctuations (Section~\ref{sec:modified_BH}). The latter is a qualitatively new effect derived in CVB, and its inclusion is new to our analysis.

\subsection{Motivation and overview}\label{sec:motivation}

The main features of the problem can be seen by first writing the real-space photon Boltzmann equation:
\begin{equation}\label{eqn:real_space_boltzmann}
    \left(\partial_\eta + \vu*{\gamma} \vdot \nabla \right) \Theta = S_{\rm grav} - \bar{\Gamma}\left(1 + \delta_{\rm e}(\eta, \vb{x}) \right) \mathcal{C}[\Theta]\,,
\end{equation}
where $\Theta = \Theta(\eta, \vb{x}, \vu*{\gamma})$ is the photon temperature perturbation, $S_{\rm grav} = -\partial_\eta \Phi - \vu*{\gamma}\vdot\nabla \Psi$ is the gravitational source term, $\bar{\Gamma} = a \sigma_{T}\bar{N}_{\rm e}(\eta)$ is the mean photon scattering rate and  $\delta_{\rm e}(\eta,\vb{x})$ is the electron density fluctuation. The collision term is given by 
\begin{equation}\label{eqn:collision_kernel}
    \mathcal{C}[\Theta] = \Theta - \Theta_0 - \vu*{\gamma} \vdot \vb{v}_{\rm b} + \frac{1}{2}P_2(\vu*{\gamma}\vdot\vu{k})\,\Pi\,,
\end{equation}
with $\Theta_0$ being the monopole of $\Theta$, $\vb{v}_{\rm b}$ the baryon bulk velocity, and $\Pi$ the combination of the temperature quadrupole and the polarization monopole and quadrupole that Thomson scattering re-emits (Section~\ref{sec:modified_BH}). The last term is written for a single Fourier mode $\vb{k}$, with multipoles defined by the $(-i)^\ell(2\ell+1)$ expansion of \citet{Ma:1995ey}, which fixes its sign. The collision operator annihilates any photon distribution in kinetic equilibrium with respect to the Thomson scatterers, so $\mathcal{C}[\Theta] \neq 0$ represents the departure of $\Theta$ from local equilibrium.

\subsubsection{Motivation for a statistical approach}

\begin{figure}[t]
    \centering
    \includegraphics{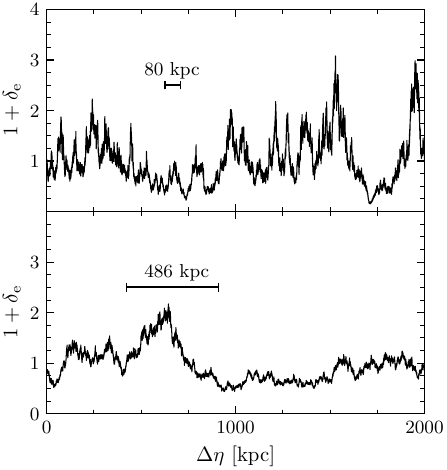}
    \caption{Two snapshots of an OU process (Equation~\ref{eqn:ou_process}) for a log-normal electron density fluctuation $\delta_{\rm e}(\eta) = e^{\xi_{\rm e}} -1$. We illustrate a process with constant $\sigma_{\rm e}^2 = 0.3$ but redshift-dependent $\Delta_c(z) = \Delta_c^{\rm ref} (1+z)^{-3} / (1+z_{\rm ref})^{-3}$, with $z_{\rm ref} = 1100$ and $\Delta_c^{\rm ref} = 525\, {\rm kpc}$. {\it Top:} A snapshot of the process at $z = 2000$ ($\eta_{\rm start} = 175\, {\rm Mpc}, \Delta_c(\eta_{\rm start})=87\, {\rm kpc}$), showing $1 + \delta_{\rm e}(\eta_{\rm start} + \Delta \eta)$ for a window of $2$ Mpc. We have labeled the physical correlation scale $\eta_c^{\rm phys}$, which is slightly smaller than the OU correlation time for a log-normal field. Since $\Delta_c$ is slowly varying over the interval shown, we simulate the process with $\Delta_c$ fixed to its value at the beginning of the interval, $\Delta_c(\eta_{\rm start})$. {\it Bottom:} The same as the top panel, but for $z = 1100$ ($\eta_{\rm start} = 278\, {\rm Mpc}, \Delta_c(\eta_{\rm start})= 525\,{\rm kpc}$).}\label{fig:realizations}
\end{figure}

We begin by noting that the free-streaming operator on the left-hand side of Equation~\ref{eqn:real_space_boltzmann}, $\qty(\partial_\eta + \vu*{\gamma} \vdot \nabla)$, is diagonalized if we impose 
\begin{equation}
    \vb{x}(\eta) = \vb{x}_0 + \vu*{\gamma}(\eta - \eta_0).
\end{equation}
Equation~\ref{eqn:real_space_boltzmann} can, therefore, be decomposed into a family of ODEs along different lines of sight, labeled by $(\vb{x}_0, \vu*{\gamma})$:

\begin{equation}\label{eqn:characteristic_decomposition}
    \eval{\dv{\Theta}{\eta}}_{\vb{x}(\eta)} = \eval{S_{\rm grav}}_{\vb{x}(\eta)} - \bar{\Gamma}(\eta)\qty[1 + \delta_{\rm e}(\eta, \vb{x}(\eta))]\eval{\mathcal{C}[\Theta]}_{\vb{x}(\eta)}\,,
\end{equation}
where we emphasize that the various quantities are evaluated along a definite spatial trajectory. This perspective highlights that the electron fluctuations result in a modulation of the scattering rate that, along a given line of sight, depends on conformal time alone:
\begin{equation}
    \delta_{\rm e}^{(\vb{x}_0, \vu*{\gamma})}(\eta) \equiv \delta_{\rm e}(\eta, \vb{x}(\eta))\,.
\end{equation}
The spatial structure of $\delta_{\rm e}(\eta, \vb{x})$ is therefore encoded in the temporal structure of $\delta_{\rm e}^{(\vb{x}_0, \vu*{\gamma})}(\eta)$, that is, two times along a line of sight correspond to a spatial separation $\abs{\eta_2 - \eta_1}$, so the correlation time can be identified with the spatial coherence length, $\Delta_c \sim L_{\rm e}$. 

Each $\delta_{\rm e}^{(\vb{x}_0, \vu*{\gamma})}(\eta)$ is naturally different, as each line of sight will skewer a different sequence of over- and underdense regions. This renders Equation~\ref{eqn:characteristic_decomposition} intractable, as solving it would require detailed knowledge of $\delta_{\rm e}(\eta, \vb{x})$ at the scales where the fluctuations are present. However, we can make the following observation, which follows from the large {\it separation of scales} we have assumed: a single Fourier mode $\Theta_k = \int \dd[3]x \Theta(\eta, \vb{x}, \vu*{\gamma}) \exp(-i \vb{k} \vdot \vb{x})$ averages $(k_{\rm e}/k)^3 \sim 10^9$ independent patches of the electron field, where $k_{\rm e} \sim 1/L_{\rm e}$ is the scale of a typical fluctuation. This motivates us to adopt a statistical prescription for $\delta_{\rm e}^{(\vb{x}_0, \vu*{\gamma})}(\eta)$, where we anticipate that the evolution of $\Theta_k$ will be described well by averaging over realizations of the small-scale electron density field.

\subsubsection{Gaussian and log-normal processes for $\delta_{\rm e}$}

The basic workhorse in our approach is the Ornstein--Uhlenbeck (OU) stochastic process \citep{Uhlenbeck:1930zz} for some stochastic variable $\xi_{\rm e}\equiv\xi_{\rm e}(\delta_{\rm e})$
\begin{equation}\label{eqn:ou_process}
    \dd{\xi_{\rm e}} = \alpha (\mu - \xi_{\rm e}) \dd \eta + \sigma \dd W\,.
\end{equation}
This describes a deterministic, mean-reverting drift along with stochastic (Brownian) fluctuations, which are characterized by the term proportional to the Gaussian process $\dd W$. The stationary variance and correlation time of this process are given, respectively, by
\begin{equation}\label{eqn:new_params}
    \sigma_{\rm e}^2 \equiv \frac{\sigma^2}{2\alpha} \qand \Delta_c \equiv \frac{1}{\alpha}\,.
\end{equation}
These expressions hold strictly for OU processes with constant coefficients $\alpha$, $\mu$ and $\sigma$. We will also allow for time-varying $\alpha(\eta)$ and $\sigma(\eta)$, with the assumption that they are slowly varying in the sense that $\abs{\dot{\alpha}/\alpha}\Delta_c \ll 1$ and $\abs{\dot{\sigma}/\sigma}\Delta_c \ll 1$. The OU process reaches stationarity much faster than $\alpha(\eta)$ and $\sigma(\eta)$ change, and Equation~\ref{eqn:new_params} holds locally at each $\eta$. 

We must now relate this process to the electron density fluctuations $\delta_{\rm e}$. The simplest assumption is $\xi_{\rm e}\equiv \delta_{\rm e}$, which models the field with Gaussian statistics. In this case, solving Equation~\ref{eqn:ou_process} yields
\begin{equation}
    \delta_{\rm e}(\eta) = \int_{0}^{\eta} e^{(\gamma(\eta^{\prime}) - \gamma(\eta))}\,\sigma(\eta')\,\dd W(\eta'),\qq{(Gaussian)}
\end{equation}
where $\gamma(\eta)= \int_0^\eta \alpha(\eta') \dd\eta'$ \citep{Chluba:2025jqt}. Because the stochastic integral is a sum of independent Gaussian increments, $\delta_{\rm e}$ is itself Gaussian. We can compute the variance, time-time correlation $C_{\delta_{\rm e}}(\eta, \eta') \equiv \eavg{\delta_{\rm e}(\eta) \delta_{\rm e}(\eta')}$, and physical correlation time as:
\begin{subequations}
\begin{align}
    \eavg{\delta_{\rm e}^2(\eta)} &= \sigma_{\rm e}^2(\eta)\,,\qq{(Gaussian)}\\
    C_{\delta_{\rm e}}(\eta, \eta') &= \sigma_{\rm e}^2 e^{-\abs{\gamma - \gamma'}}\,,\qq{(Gaussian)}\label{eqn:gaussian_correlator}\\
    \eta_c^{\rm phys}(\eta) &= \Delta_c(\eta)\,.\qq{(Gaussian)}
\end{align}
\end{subequations}
Here, we have defined the physical correlation time $\eta_c^{\rm phys}$ as the integral of the normalized correlation function, which for an exponential correlation is the time required for $C_{\delta_{\rm e}}(\eta, \eta')$ to drop by an e-fold: 
\begin{equation}\label{eqn:phys_corr}
    \eta_c^{\rm phys} \equiv \frac{\int_0^{\infty} C_{\delta_{\rm e}}(\eta, \eta+\Delta \eta) \dd \Delta \eta}{\eavg{\delta_{\rm e}^2}}\,.
\end{equation}
In the Gaussian case, the physical correlation time and the OU process correlation time coincide (as is the case for the physical variance $\eavg{\delta_{\rm e}^2}$ and OU variance $\sigma_{\rm e}^2$).

While the Gaussian case is conceptually straightforward, it is not the optimal choice for describing physical fluctuations, as in principle the unphysical $1 + \delta_{\rm e}<0$ is possible. To avoid this, we can instead say $\delta_{\rm e} \equiv e^{\xi_{\rm e}} - 1$, so that $\ln(1 + \delta_{\rm e})$ is the normally distributed quantity described by Equation~\ref{eqn:ou_process}. In this case Equation~\ref{eqn:new_params} still holds, but the OU variance and correlation time have a more complicated relationship with the variance and correlation of the physical $\delta_{\rm e}$ field, due to the non-linear relationship between the two. For a log-normal (LN) field,

\begin{subequations}
\begin{align}
    \eavg{\delta_{\rm e}^2} &= \exp(\sigma_{\rm e}^2) -1\,,\qq{(LN)}\\
    C_{\delta_{\rm e}}(\eta, \eta')&= \exp(\sigma_{\rm e}^2 e^{-\abs{\gamma - \gamma '}}) - 1\,,\qq{(LN)} \label{eqn:LN_corr}\\
    \eta_{c}^{\rm phys} &= \Delta_c \frac{ {\rm Li}(e^{\sigma_{\rm e}^2}) - \ln\left(\sigma_{\rm e}^2\right) - \gamma_{\rm E}}{\exp(\sigma_{\rm e}^2)-1}\,,\qq{(LN)}
\end{align}
\end{subequations}
where ${\rm Li}(x)$ is the logarithmic integral and $\gamma_{\rm E} = 0.577\dots$ is the Euler constant \citep{Chluba:2025jqt}. In these expressions, the (slow) time dependence of $\sigma_{\rm e}^2$ and $\Delta_c$ is understood.

In either case, we assume that $\delta_{\rm e}$ is an isotropic random field so that all $\delta_{\rm e}^{(\vb{x}_0, \vu*{\gamma})}(\eta)$ have the same statistics, which are described by Equation~\ref{eqn:ou_process}. Still assuming a large separation of scales, we can then drop the label $(\vb{x}_0, \vu*{\gamma})$ and think of $\delta_{\rm e}(\eta)$ as the fluctuation along a typical line of sight. We show two realizations of $\delta_{\rm e}(\eta)$ as a log-normal field in Figure~\ref{fig:realizations}. To produce these realizations, we simulated Equation~\ref{eqn:ou_process} using the Euler--Maruyama method for $2 \times 10^{5}$ steps, with a step size of $10\, {\rm pc}$. In each case, we used $\sigma_{\rm e}^2 = 0.3$, with $\Delta_c = 87\, {\rm kpc}$ and $\Delta_c = 525\, {\rm kpc}$ for the top and bottom panels respectively. We note that a strong separation of scales is crucial to this treatment and is violated at long coherence lengths, as we discuss in Appendix~\ref{app:validity}.

\subsubsection{Corrections from correlations}

Equation~\ref{eqn:characteristic_decomposition} implies that $\Theta$ is itself a stochastic process, driven by the fluctuating $\delta_{\rm e}$. Taking the ensemble average directly gives
\begin{equation}\label{eqn:eavg_boltzmann}
    \dv{\eavg{\Theta}}{\eta} = S_{\rm grav} - \bar{\Gamma}\mathcal{C}[\eavg{\Theta}] - \bar{\Gamma}\eavg{\delta_{\rm e} \mathcal{C}[\Theta]}\,.
\end{equation}
This equation is not closed: the evolution of $\eavg{\Theta}$ depends on the correlation $\eavg{\delta_{\rm e}\Theta}$ through the final term, whose evolution in turn depends on $\eavg{\delta_{\rm e}^2\Theta}$, and so on. Defining the moments
\begin{equation}
    \kappa_p(\eta, \vb{x}, \vu*{\gamma}) \equiv \eavg{\delta_{\rm e}^p \Theta(\eta, \vb{x}, \vu*{\gamma})}\,,
\end{equation}
we obtain an infinite hierarchy, and we are ultimately interested in the observable $\kappa_0 = \eavg{\Theta}$. CVB close this hierarchy by Fourier and Legendre transforming Equation~\ref{eqn:eavg_boltzmann} and expanding in the small parameter $\bar{\Gamma}\Delta_c$; the result is the modified photon Boltzmann hierarchy of Section~\ref{sec:modified_BH}, with an $\ell$-dependent correction to the scattering term. We do not reproduce that calculation, but instead give a heuristic derivation of the leading correction that exposes its physical origin and its dependence on the fluctuation statistics. The full derivation can be found in \citet{Chluba:2025jqt}.

To begin, write $\Theta = \bar{\Theta} + \var{\Theta}$, where $\bar{\Theta}$ solves the unperturbed ($\delta_{\rm e}=0$) equation,
\begin{equation}
    \dv{\bar{\Theta}}{\eta} = S_{\rm grav} - \bar{\Gamma}\mathcal{C}[\bar{\Theta}]\,.
\end{equation}
Linearizing in $\var{\Theta}$, the perturbation obeys
\begin{equation}
    \dv{\var{\Theta}}{\eta} + \bar{\Gamma}\mathcal{C}[\var{\Theta}]=  - \bar{\Gamma}\,\delta_{\rm e}\,\mathcal{C}[\bar{\Theta}]\,,
\end{equation}
which we integrate to write the total perturbation at $\eta$ as
\begin{equation}
    \var{\Theta}(\eta) = -\bar{\Gamma}\int_{\eta_{\rm ini}}^{\eta} G(\eta, \eta^\prime)\,\delta_{\rm e}(\eta^\prime)\,\mathcal{C}[\bar{\Theta}(\eta^\prime)]\dd \eta^\prime\,,\label{eqn:dtheta_solution}
\end{equation}
where the Green's function solves\footnote{Due to the angular couplings entering $\mathcal{C}[\var{\Theta}]$ via the monopole $\var{\Theta}_0$ and quadrupole $\var{\Pi}$, the evolution equation is in fact an integro-differential equation and $G$ is not truly a scalar. This necessitates the Fourier and Legendre decomposition carried out in CVB, in which $G$ becomes a matrix propagator ($\vu{X}(\eta', \eta)$ in the notation of \citealp{Chluba:2025jqt}) that acts differently on each $\Theta_\ell$.}
\begin{equation}\label{eqn:greens_perturbation}
    \dv{G}{\eta} = -\bar{\Gamma}\,\mathcal{C}[G]\,, \qquad G(\eta^\prime, \eta^\prime) = 1\,.
\end{equation}

The electron fluctuation sources photon perturbations by modulating the scattering rate: where $\delta_{\rm e}>0$ ($\delta_{\rm e}<0$), an existing anisotropy $\mathcal{C}[\bar{\Theta}]\neq0$ relaxes toward the local scattering equilibrium faster (slower) than in the unperturbed case. This creates a perturbation $\var{\Theta}$ in the occupancy of photons in the $\vu*{\gamma}$ direction compared to the $\delta_{\rm e} = 0$ case, which subsequently evolves via scattering with the mean field. The photon distribution along $\vu*{\gamma}$ therefore carries an imprint of $\delta_{\rm e}(\eta^\prime)$ for all $\eta^\prime \leq \eta$.

We now evaluate the last term of Equation~\ref{eqn:eavg_boltzmann}. Using $\eavg{\delta_{\rm e}}=0$,
\begin{align}\label{eqn:correction_term_heuristic}
    -\bar{\Gamma}\eavg{\delta_{\rm e} \mathcal{C}[\Theta]} &= -\bar{\Gamma}\eavg{\delta_{\rm e}(\eta)\, \mathcal{C}[\var{\Theta}(\eta)]}\notag \\
    \begin{split}
    &= \bar{\Gamma}^2 \int_{\eta_{\rm ini}}^{\eta} \eavg{\delta_{\rm e}(\eta)\delta_{\rm e}(\eta^\prime)} \\
    &\qquad \times \mathcal{C}\big[G(\eta, \eta^\prime)\,\mathcal{C}[\bar{\Theta}(\eta')]\big]\, \dd \eta^\prime\,.
    \end{split}
\end{align}
The correlator $\eavg{\delta_{\rm e}(\eta)\delta_{\rm e}(\eta')}$ falls off over the coherence time $\Delta_c$, whereas $G$ and $\bar{\Gamma}$ vary on the scattering time $1/\bar{\Gamma}$. In the optically thin limit $\tau_c \equiv \bar{\Gamma}\Delta_c \ll 1$ these are well separated, so over the support $\abs{\eta-\eta'}\lesssim\Delta_c$ of the correlator we may take $G(\eta,\eta')\simeq1$ and treat $\bar{\Gamma}$ and $\mathcal{C}[\bar{\Theta}]$ as constant. The correction then reduces to
\begin{align}\label{eqn:heuristic_correction}
    -\bar{\Gamma}\eavg{\delta_{\rm e} \mathcal{C}[\Theta]} &\approx \bar{\Gamma}^2\,\mathcal{C}^2[\bar{\Theta}(\eta)]\int_{\eta_{\rm ini}}^{\eta}  \eavg{\delta_{\rm e}(\eta)\delta_{\rm e}(\eta')}\, \dd \eta' \notag\\
    &= \bar{\Gamma}^2\,\mathcal{C}^2[\bar{\Theta}(\eta)] \left(\eavg{\delta_{\rm e}^2}\, \eta_{c}^{\rm phys} \right)\,,
\end{align}
where the second line uses Equation~\ref{eqn:phys_corr}. Here $\mathcal{C}^2[\bar{\Theta}]\equiv\mathcal{C}\big[\mathcal{C}[\bar{\Theta}]\big]$ is the collision operator applied twice, so the correction is linear in $\bar{\Theta}$. For $\ell\geq3$, where $\mathcal{C}$ acts as the identity, the scattering term becomes $-\bar{\Gamma}\big(1-\bar{\Gamma}\eavg{\delta_{\rm e}^2}\eta_c^{\rm phys}\big)\bar{\Theta}$: the fluctuations reduce the scattering rate by a fraction of order $\tau_c\eavg{\delta_{\rm e}^2}$, which is small but acts on every scattering before last scattering.

The correction is second order in $\delta_{\rm e}$: a fluctuation at $\eta'$ perturbs the photon distribution, and that perturbation persists when the photons encounter the correlated fluctuation at $\eta$, so $\eavg{\delta_{\rm e}\,\mathcal{C}[\var{\Theta}]}\neq0$ even though $\eavg{\var{\Theta}}=0$. At this order the correction depends on the fluctuation statistics only through the combination $\eavg{\delta_{\rm e}^2}\,\eta_c^{\rm phys}$, and not on the detailed shape of the correlation function --- a fact we use in Section~\ref{sec:constrained_quants} to project our constraints onto different fluctuation power spectra. The calculation of CVB extends this result beyond leading order in $\tau_c$, summing the series in closed form to obtain the shape functions of Section~\ref{sec:modified_BH}, and retains the dependence on multipole, which no rescaling of the mean scattering rate can capture.

The same result can be understood from the transmission of photons through the fluctuating medium. For the higher multipoles of the local photon distribution, the photons scattered into a given line of sight (the $\Theta_0$, $\Pi$, and $\vb{v}_{\rm b}$ terms of Equation~\ref{eqn:collision_kernel}) can be neglected, so that $\mathcal{C}[\Theta]\simeq\Theta$; this is exact for $\ell\geq3$. Neglecting also the gravitational source, Equation~\ref{eqn:characteristic_decomposition} describes pure attenuation along the line of sight, $\Theta\propto e^{-\tau}$, where $\tau=\bar{\Gamma}\int_0^T(1+\delta_{\rm e})\,\dd\eta$ is the optical depth accumulated over an interval $T$. Scattering isotropizes the radiation, so the surviving anisotropy is carried by the fraction $e^{-\tau}$ of photons that have not scattered. The mean optical depth is $\bar{\Gamma}T$ whether or not fluctuations are present, but each line of sight samples the electron density only along its own path, so $\tau$ scatters about this mean from one line of sight to the next. The quantity of interest is therefore the transmission averaged over lines of sight, $\eavg{e^{-\tau}}$.

Each line of sight provides a noisy sample of the optical depth, built from a finite number of patches. A path of length $T\gg\Delta_c$ crosses $N\sim T/\Delta_c$ independent patches. A patch at the mean density has optical depth $\tau_c=\bar{\Gamma}\Delta_c$, while one with density contrast $\delta_{\rm e}$ has $\tau_c(1+\delta_{\rm e})$. The accumulated optical depth therefore has mean $N\tau_c=\bar{\Gamma}T$ and, between lines of sight, variance $\mathrm{Var}(\tau)\sim N\tau_c^2\eavg{\delta_{\rm e}^2}=\bar{\Gamma}^2\eavg{\delta_{\rm e}^2}\Delta_c T$. A direct calculation fixes the coefficient,\footnote{For $T\gg\Delta_c$ and $\bar{\Gamma}$ approximately constant over $T$, $\mathrm{Var}(\tau)=\bar{\Gamma}^2\int_0^T\!\int_0^T\eavg{\delta_{\rm e}(\eta)\delta_{\rm e}(\eta')}\,\dd\eta\,\dd\eta'\simeq2\bar{\Gamma}^2\eavg{\delta_{\rm e}^2}\,\eta_c^{\rm phys}\,T$, using Equation~\ref{eqn:phys_corr}.} $\mathrm{Var}(\tau)\simeq2\bar{\Gamma}^2\eavg{\delta_{\rm e}^2}\,\eta_c^{\rm phys}\,T$. For Gaussian fluctuations $\tau$ is itself Gaussian, and
\begin{align}
    \eavg{e^{-\tau}} &= e^{-\bar{\Gamma}T+\mathrm{Var}(\tau)/2} \notag\\
    &= \exp\left[-\bar{\Gamma}\left(1-\bar{\Gamma}\eavg{\delta_{\rm e}^2}\eta_c^{\rm phys}\right)T\right]\,,
\end{align}
reproducing the reduced scattering rate of Equation~\ref{eqn:heuristic_correction}. Because $e^{-\tau}$ is a non-linear function of $\tau$, $\eavg{e^{-\tau}}>e^{-\eavg{\tau}}$ and the transmission is enhanced. Two limits bracket this behavior. As $\Delta_c\to0$ at fixed fluctuation amplitude, the number of samples along each path diverges, every line of sight accumulates the same optical depth, and the transmission of a uniform medium is recovered. In the opposite extreme, where the electrons are gathered into a few dense clumps, the scatter is as large as it can be: lines of sight that miss the clumps suffer no scattering at all, and the medium is far more transparent than a uniform one containing the same number of electrons. This is the transport counterpart of the effect that accelerates recombination (Section~\ref{sec:modified_rec}): there the rate scales as $N_{\rm e}^2$, and its average over an inhomogeneous medium, $\propto\eavg{N_{\rm e}^2}$, exceeds its value at the mean density, $\propto\eavg{N_{\rm e}}^2$.

\subsection{New model parameters}\label{sec:new_params}

The fluctuations are specified by the OU variance and correlation time of Equation~\ref{eqn:new_params}. The transport corrections, however, do not depend on $\Delta_c$ alone: as in Equation~\ref{eqn:heuristic_correction}, the coherence length always enters multiplied by the mean scattering rate, as the optical depth across one coherence length,
\begin{equation}\label{eqn:tauc_def}
    \tau_c(\eta) \equiv \bar{\Gamma}(\eta)\, \Delta_c (\eta)\,.
\end{equation}

Following CVB, we parametrize the model using $\sigma_{\rm e}^2$ and $\tau_c$. At leading order in $\tau_c$, Equation~\ref{eqn:heuristic_correction} shows that the fluctuations reduce the scattering rate for $\ell\geq3$ by the fraction $\bar{\Gamma}\eavg{\delta_{\rm e}^2}\eta_c^{\rm phys}$. For Gaussian fluctuations $\eavg{\delta_{\rm e}^2} = \sigma_{\rm e}^2$ and $\eta_c^{\rm phys} = \Delta_c$, which motivates defining
\begin{equation}\label{eqn:zeta_def}
    \zeta_{\rm e} \equiv \tau_c\, \sigma_{\rm e}^2\,.
\end{equation}
We use the same definition for log-normal fluctuations. In that case the fractional reduction is $\bar{\Gamma}\eavg{\delta_{\rm e}^2}\eta_c^{\rm phys} = \zeta_{\rm e}\,(1 + \sigma_{\rm e}^2/4 + \dots)$, so $\zeta_{\rm e}$ retains this interpretation only at leading order in $\sigma_{\rm e}^2$; at $\sigma_{\rm e}^2 = 0.2$ the reduction is $5\%$ larger than $\zeta_{\rm e}$. Since our main analysis below assumes log-normal fluctuations, $\zeta_{\rm e}$ should not be directly interpreted as the fractional reduction in scattering rate. Because the modified recombination history depends on $\sigma_{\rm e}^2$ alone while the transport corrections are controlled by $\zeta_{\rm e}$, we sample $\sigma_{\rm e}^2$ and $\zeta_{\rm e}$, so that each sampled parameter is tied to one of the two channels.

\begin{figure}[t]
    \centering
    \includegraphics{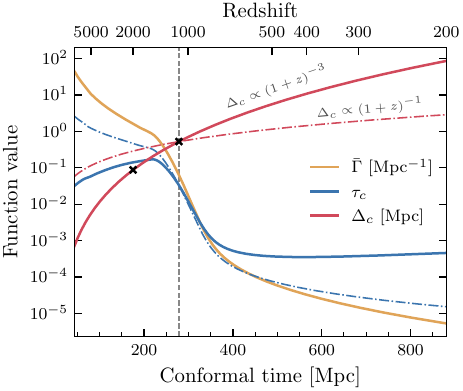}
    \caption{Time dependence of the quantities entering the transport corrections, shown against conformal time (bottom axis) and redshift (top axis): the mean scattering rate $\bar{\Gamma}$ (gold), the comoving coherence length $\Delta_c$ (red), and the optical depth across one coherence length, $\tau_c = \bar{\Gamma}\Delta_c$ (blue). We show two illustrative models, $\Delta_c \propto (1+z)^{-3}$ (solid) and $\Delta_c \propto (1+z)^{-1}$ (dash-dotted), both normalized to $\Delta_c = 525\,{\rm kpc}$ at $z_{\rm ref} = 1100$ (vertical dashed line). The black crosses mark the two epochs, $z = 2000$ and $z = 1100$, at which realizations of the $(1+z)^{-3}$ model are shown in Figure~\ref{fig:realizations}.}\label{fig:example_evolutions}
\end{figure}

While both $\sigma_{\rm e}^2$ and $\tau_c$ are in principle functions of time, in our baseline analysis we take both to be constant. Since $\bar{\Gamma}\propto X_{\rm e}(1+z)^2$, constant $\tau_c$ requires the comoving coherence length to track the ionization history, $\Delta_c\propto[X_{\rm e}(z)(1+z)^{2}]^{-1}$, which is unlikely to result from a physical mechanism for generating the fluctuations. We therefore also consider power laws for the coherence length,
\begin{equation}\label{eqn:time_dep_corr}
    \Delta_c (z) = \Delta_c^{\rm ref} \left(\frac{1+z}{1+z_{\rm ref}}\right)^{n}\,, \qquad \Delta_c^{\rm ref} \equiv \frac{\tau_c^{\rm ref}}{\bar{\Gamma}(z_{\rm ref})}\,,
\end{equation}
for which the sampled transport parameter is $\zeta_{\rm e}^{\rm ref} \equiv \tau_c^{\rm ref}\sigma_{\rm e}^2$. When presenting results we convert $\tau_c$ to a length. All quoted values of $\Delta_c$ are comoving and evaluated at $z_{\rm ref} = 1100$, near the peak of the visibility function: $\Delta_c \equiv \tau_c(z_{\rm ref})/\bar{\Gamma}(z_{\rm ref})$. This applies to both the constant and time-varying $\tau_c$ models. For reference, the standard recombination history (the Planck 2018 baseline cosmology \citep{Planck:2018vyg}) has $\bar{\Gamma}(z_{\rm ref}) \approx 0.065\, {\rm Mpc}^{-1}$, so $\Delta_c \approx 15\, {\rm Mpc} \times \tau_c$ provides a good rule-of-thumb conversion. 

Figure~\ref{fig:example_evolutions} shows $\bar{\Gamma}$, $\Delta_c$, and $\tau_c$ for two such models. Because $\bar{\Gamma}$ falls by several orders of magnitude as the electrons recombine, while $\Delta_c$ varies only as a power law, $\tau_c$ inherits the steep decline of $\bar{\Gamma}$. In each case, $\tau_c$ becomes negligible soon after $z_{\rm ref}$, and the models differ mainly in how large $\tau_c$ is before last scattering.

\subsection{Modified recombination history}\label{sec:modified_rec}

\begin{figure}[t]
    \centering
    \includegraphics{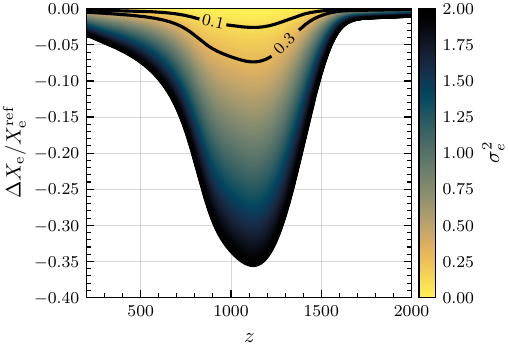}
    \caption{Fractional deviation of the mean recombination history from the standard, $(\bar{X}_{\rm e} - X_{\rm e}^{\rm std}) / X_{\rm e}^{\rm std}$, computed using our separate-universe prescription detailed in Section~\ref{sec:modified_rec}. We show the full range of the precomputed table used in \textsc{classito}, $\sigma_{\rm e}^2 \in [0,2]$, and for illustrative purposes we also highlight the deviations corresponding to $\sigma_{\rm e}^2 = 0.1$ and $\sigma_{\rm e}^2 = 0.3$, which are more representative of the values allowed by the data (see Section~\ref{sec:results}). }\label{fig:delta_xe}
\end{figure}

Baryon inhomogeneities will generically lead to an accelerated recombination process that terminates at an earlier redshift \citep{Jedamzik:2013gua}. For illustrative purposes, one can consider the three-level atom \citep{Peebles:1968ja,Zeldovich:1969qmq}, where the physical electron density evolves as 
\begin{multline}\label{eqn:peebles_model}
    \pdv{N_{\rm e}}{t} + 3 H N_{\rm e} =C \bigl[ - \alpha(T_{\rm m}) N_{\rm e}^2 \\ + \beta(T_\gamma) N_{\rm H} \exp (-E_{{\rm Ly}\alpha}/k_{\rm B}T_\gamma) \bigr]\,,
\end{multline}
where $N_{\rm H}$ is the number density of neutral hydrogen, $T_{\rm m}$ and $T_\gamma$ are the matter and radiation temperatures,\footnote{It is important to distinguish the two, as setting $T_\gamma=T_{\rm m}$ can have dramatic effects \citep{Chluba2015PMF}.} and $\alpha(T_{\rm m})$ and $\beta(T_\gamma)$ are the case-B recombination and photoionization rates, respectively. The leading prefactor is the Peebles $C$-factor, which encodes the suppression of the recombination rate due to resonant absorption of Lyman-$\alpha$ photons. It has the form 
\begin{equation}\label{eqn:C_factor}
    C = \left(1 + \frac{\beta(T_\gamma)}{\Gamma_{2s} + 3 P_{\rm esc} \Gamma_{2p}} \right)^{-1}\,,
\end{equation}
where the $\Gamma_{i}$ terms are transition rates from the labeled states to the $1s$ state, and $P_{\rm esc}$ is the probability that a Lyman-$\alpha$ photon will redshift out of resonance before re-exciting another hydrogen atom.

The key observation is that the recombination term (proportional to $\alpha(T_{\rm m})$) is quadratic in the density, while at fixed ionization fraction the photoionization term is only linear. Averaged over an inhomogeneous medium the former is enhanced, $\alpha \eavg{N_{\rm e}^2} \geq \alpha \eavg{N_{\rm e}}^2$, while the latter is not, so the net recombination rate is enhanced and recombination is accelerated \citep{Jedamzik:2013gua}. The density dependence of the $C$-factor through $P_{\rm esc}$, introduced below, partly offsets this, and is included in the calculation that follows.

To account for this, we modify the recombination history using a ``separate-universe'' approach. We consider recombination occurring independently in many different inhomogeneous regions, with the baryon density varying from region to region but the properties of the radiation bath (for example, the photon density) and the background expansion $H(z)$ being held fixed and common between them. Furthermore, we do not modify the local expansion of the gas in response to the perturbation, and we treat the baryon perturbation as time-independent. 

Concretely, we create a family of histories $N_{\rm e}(z; F_{\rm b})$, labeled by $F_{\rm b} \equiv 1 + \delta_{\rm b}$, by running a recombination code (in our case, \textsc{cosmorec} \citep{Chluba:2010ca, Chluba:2015gta}) for a grid of values $F_{\rm b} \in [10^{-3}, 10^{3}]$. We hold all other parameters fixed, except for the dark matter density, which we adjust so as to keep $H(z)$ unchanged from the $F_{\rm b} = 1$ case. Then, given a probability distribution for the baryon fluctuations, $P(F_{\rm b}, \sigma_{\rm b})$, we can compute the average electron density as
\begin{equation}\label{eqn:mean_electron}
    \bar{N}_{\rm e}(z) = \int N_{\rm e}(z; F_{\rm b}) P(F_{\rm b}, \sigma_{\rm b}) \dd F_{\rm b}\,.
\end{equation}
In our analyses, we adopt a log-normal distribution for $F_{\rm b}$, with the form 
\begin{equation}\label{eqn:lognormal_baryon}
    P(F_{\rm b}, \sigma_{\rm b}) = \frac{1}{\sqrt{2\pi}\sigma_{\rm b}F_{\rm b}} \exp\left(- \frac{\left(\ln{F_{\rm b}} + \sigma_{\rm b}^2 /2 \right)^2}{2 \sigma_{\rm b}^2} \right)\,.
\end{equation}
By construction, this distribution is normalized and has $\eavg{F_{\rm b}} = 1$ and ${\rm Var}(F_{\rm b}) = e^{\sigma_{\rm b}^2}-1$, in parallel with the log-normal electron field of Section~\ref{sec:motivation}.

There are a number of effects not captured by our simple prescription.
\begin{itemize}
    \item The calculation neglects any extra localized heating that could be caused by the clumping \citep[e.g.,][in the context of PMFs]{Sethi2005, Chluba2015PMF}. This could significantly alter the mapping $\delta_{\rm b}\rightarrow \delta_{\rm e}$.

    \item The escape probability $P_{\rm esc}$ for Lyman-$\alpha$ photons depends on both the local baryon density and local expansion rate of the gas. While the former is correctly incorporated in the separate-universe approach we employ, in the presence of baryon fluctuations the latter receives a contribution from the baryon velocity divergence, that is, $H \to H + \theta_b / 3a$ \citep{Senatore:2008vi,Lee:2021bmn}, which is not captured in our approach. Since $P_{\rm esc} \propto H / n_{\rm b}$, this modifies the escape probability.

    \item Adiabatic compression or rarefaction of the gas in the presence of a dynamical perturbation $\dot{\delta}_{\rm b}$ will accordingly modify the matter temperature $T_{\rm m}$. This is primarily relevant to the evolution of the free-electron fraction through $\alpha$ in Equation~\ref{eqn:peebles_model}, which depends on $T_{\rm m}$ \citep{hummerTotalRecombinationEnergyloss1994,Seager:1999km}. However, since heat transfer via Compton scattering is efficient for $z \gtrsim 100$, any deviation of $T_{\rm m}$ from the radiation temperature is quickly washed out and has little impact at the redshifts of interest \citep{Lee:2021bmn}.
    
    \item For perturbations at sufficiently small scales, that is, below the Lyman-$\alpha$ mean free path of $\sim {\rm kpc}$, the problem becomes non-local due to mixing between the different regions, and radiative transfer effects need to be taken into account \citep[e.g.,][]{Schiff:2025vjf}. 
\end{itemize}

While the third point can be ignored for the present analyses, in principle the other points could have a significant impact on the deduced mean recombination history. Properly accounting for these effects will require detailed work, tailored to specific mechanisms for generating the perturbations, which we deem out of scope for the present phenomenological exploration. However, existing studies taking into account the second \citep[e.g.,][]{Lee:2021bmn} and fourth \citep[e.g.,][]{Jedamzik:2023rfd,Schiff:2025vjf} effects indicate that they lead to an enhancement of the effects seen in our model; that is, an even more accelerated recombination process. Therefore, our model for modifying $X_{\rm e}$ can be viewed as conservative in this respect. 

In addition, none of the previous treatments account for the inevitable stochastic averaging process over the underlying electron density field. This should again be treated using \ito calculus rather than with a simplified multi-zone model. We defer a more detailed study to future work.

We use this prescription to precompute a table of $\Delta X_{\rm e}(z) / X_{\rm e}(z)$ as a function of the baryon variance $\sigma_{\rm b}^2 \in [0, 2]$, illustrated in Figure~\ref{fig:delta_xe}. In \textsc{classito} we identify $\sigma_{\rm b}^2 = \sigma_{\rm e}^2$, so the table is interpolated in $\sigma_{\rm e}^2$ and the deviation is added to the standard $X_{\rm e}(z)$ computed in the thermodynamics module using \textsc{hyrec-2} \citep{PhysRevD.83.043513,Lee:2021bmn}. Within the separate-universe model, however, the electron variance falls below the baryon variance through recombination, since denser regions recombine earlier and all regions freeze out at a similar electron density \citep[see Figure~6 of][]{Chluba:2025jqt}. More realistic treatments of the local recombination, heating, and ionization physics may change this mapping substantially. We therefore regard the common variance as a phenomenological parameter, and return to the consequences of this choice in Section~\ref{sec:conclusion}.

\subsection{Modified Boltzmann hierarchy}\label{sec:modified_BH}

As discussed in Section~\ref{sec:motivation}, the presence of line-of-sight correlations leads to additional effects beyond a change to the mean recombination history \citep{Chluba:2025jqt}. The net result is a modification of the scattering term that appears in the evolution for each $\Theta_\ell$ in the photon Boltzmann hierarchy. Furthermore, these modifications are $\ell$-dependent, with the implication that the total effect of the small-scale fluctuations {\it cannot} be captured via rescalings of the visibility function $g(\eta)$. These $\ell$-dependent corrections can be encoded with a shape function $f$ that depends on whether $\delta_{\rm e}$ is Gaussian or log-normal distributed. The shape functions are given by
\begin{equation}\label{eqn:f_gaussian}
    f_S(\tau_{\ell}, \sigma_{\rm e}^2) \equiv \tau_{\ell} \sigma_{\rm e}^2 \exp\left(-\tau_\ell^2 \sigma_{\rm e}^2 \right)\,,\qq{(Gaussian)}  
\end{equation}
and 
\begin{equation}\label{eqn:f_lognormal}
    \begin{split}
    f_S(\tau_{\ell}, \sigma_{\rm e}^2) &\equiv \tau_\ell \left(e^{\sigma_{\rm e}^2} -1 \right) e^{-\sigma_{\rm e}^2 / 4} \\&\qquad \times \exp\left(-(2 + \tau_\ell)\tau_\ell \sigma_{\rm e}^2 \right)\,.
    \end{split}\qq{(LN)}
\end{equation}
As $\tau_c\to0$ both shape functions reduce to the leading-order form of Equation~\ref{eqn:heuristic_correction}, $f_S\to\tau_\ell\eavg{\delta_{\rm e}^2}\eta_c^{\rm phys}/\Delta_c$: exactly in the Gaussian case, and to better than $0.2\%$ for $\sigma_{\rm e}^2\leq0.3$ in the log-normal case. We adopt the log-normal form throughout, as the recombination channel requires (Section~\ref{sec:modified_rec}). For the transport channel the two forms then differ at small $\tau_c$ only through the mapping from $(\sigma_{\rm e}^2,\Delta_c)$ to $(\eavg{\delta_{\rm e}^2},\eta_c^{\rm phys})$.

The argument $\tau_\ell$ encodes the angular dependence of the corrections: it is the optical depth across one coherence length at the rate at which scattering damps the mode in question. Multipoles $\ell\geq3$ are damped at the rate $\bar{\Gamma}$, so $\tau_{\ell\geq3} = \tau_c$. The photon--baryon relative velocity is damped at the rate $(1+R)\bar{\Gamma}$, where $R \equiv 4\bar{\rho}_\gamma/(3\bar{\rho}_{\rm b})$, so $\tau_1 = (1+R)\tau_c$. For clarity we write $f_\ell \equiv f_S(\tau_\ell,\sigma_{\rm e}^2)$. Scattering couples the temperature quadrupole to the polarization monopole and quadrupole through the combination $\Pi$ defined below, which is damped at the rate $(3/10)\bar{\Gamma}$, while the orthogonal combinations are damped at $\bar{\Gamma}$. The corrections in this sector therefore involve $f_3$ together with
\begin{equation}\label{eqn:f_pol}
    f_2^{\rm P}(\tau_c, \sigma_{\rm e}^2) = \frac{10}{7} f_S(\tau_c, \sigma_{\rm e}^2) - \frac{3}{7}f_S((3/10)\tau_c, \sigma_{\rm e}^2)\,.
\end{equation}
In Equations~\ref{eqn:f_gaussian}, \ref{eqn:f_lognormal}, and \ref{eqn:f_pol}, the time dependence of $\tau_c$, $\sigma_{\rm e}^2$, and $R$ is understood. 

We present the corrected equations in conformal Newtonian gauge, and we adopt the notation of \citet{Ma:1995ey}. The temperature hierarchy is:
\begin{subequations}
\begin{align}
    \begin{split}
    \partial_\eta \theta_{\rm b} &= 
    k^2 \Psi 
    - \mathcal{H} \theta_{\rm b} 
    + c_s^2 k^2 \delta_{\rm b} \\
    &\qquad + R \bar{\Gamma}(1 - f_1)(\theta_{\gamma} - \theta_{\rm b})\,,
    \end{split} \\
    \partial_\eta \theta_{\gamma} &= 
    k^2 \left(\frac{\delta_{\gamma}}{4} - \sigma_{\gamma}\right) 
    + k^2 \Psi -\bar{\Gamma}(1 - f_1)\left(\theta_{\gamma} 
    - \theta_{\rm b}\right)\,, \\
    \begin{split}
    2 \partial_\eta \sigma_{\gamma} &= 
    \frac{8 \theta_{\gamma}}{15} 
    - \frac{3kF_{\gamma 3}}{5} 
    - 2\bar{\Gamma}\sigma_\gamma 
    +\frac{\bar{\Gamma}}{10}\Pi \\ 
    &\qquad + 2\bar{\Gamma} f_3 \sigma_\gamma 
    - \frac{f_{2}^{\rm P} \bar{\Gamma}}{10} \Pi \,,
    \end{split} \\
    \begin{split}
    \partial_\eta F_{\gamma \ell} &= \frac{k}{2 \ell + 1}\left(\ell F_{\gamma(\ell -1)} - (\ell + 1)F_{\gamma(\ell + 1)}\right) \\&\qquad - \bar{\Gamma}(1-f_3) F_{\gamma \ell} \qq{($\ell \geq 3$)}\,.
    \end{split}
\end{align}\label{eqn:temperature_hierarchy}
\end{subequations}
In the baryon Euler equation, $c_{s}$ is the baryon sound speed and we have also defined the polarization source $\Pi \equiv (2 \sigma_\gamma + G_{\gamma 2} + G_{\gamma 0})$. The equations for $\delta_{\gamma}$ and $\delta_{\rm b}$ are unmodified. Likewise, the polarization hierarchy is:
\begin{subequations}
\begin{align}
    \begin{split}
        \partial_\eta G_{\gamma 0} &= -kG_{\gamma 1} - \bar{\Gamma}(G_{\gamma 0} - \frac{\Pi}{2}) \\ 
        &\qquad + \bar{\Gamma}\left(f_3 G_{\gamma 0} - \frac{f_2^{\rm P} \Pi}{2} \right)\,, 
    \end{split}\\
    \partial_\eta G_{\gamma 1} &= \frac{k}{3}G_{\gamma 0} - \frac{2k}{3}G_{\gamma 2} - \bar{\Gamma} G_{\gamma 1} + f_{3}\bar{\Gamma}G_{\gamma 1}\,,\\
    \begin{split}
    \partial_\eta G_{\gamma 2} &= \frac{2k}{5}G_{\gamma 1} - \frac{3k}{5}G_{\gamma 3} - \bar{\Gamma} \left( G_{\gamma 2} - \frac{\Pi}{10}\right)\\
    &\qquad+ \bar{\Gamma} \left(f_3 G_{\gamma 2} - \frac{f_2^{\rm P} \Pi}{10}  \right)\,,
    \end{split} \\
    \begin{split}
    \partial_\eta G_{\gamma \ell} &= \frac{k}{2\ell + 1}\left(\ell G_{\gamma (\ell -1)} - (\ell + 1)G_{\gamma(\ell + 1)}\right) \\
    &\qquad - \bar{\Gamma}(1 -f_3)G_{\gamma \ell} \qq{($\ell \geq 3$)}\,.
    \end{split}
\end{align}\label{eqn:polarization_hierarchy}
\end{subequations}
The small-scale fluctuations also contribute new source terms in the line-of-sight solutions \citep{Seljak:1996is} for the temperature and polarization multipoles, which can be modified as explained in CVB to give
\begin{align}\label{eqn:t_los}
    \begin{split}
        \Theta_{\ell} &= \int_0^{\eta_0} \dd \eta \left[ \left( \tilde{g} \left( \Theta_0 + \Psi \right) + \tilde{f} \left(\partial_\eta \Psi + \partial_\eta \Phi \right)\right) j_{\ell} \right. \\
        &\qquad + \left(\tilde{g}v_{\rm b} + \tilde{f}\bar{\Gamma}\left(3 \Theta_1 - v_{\rm b}\right)\left(f_1 - f_3 \right) \right) j_{\ell}^{(1,0)} \\
        &\qquad \left. + \left(\frac{\tilde{g}\Pi}{8} - \frac{\tilde{f}\bar{\Gamma}\Pi}{8}\left(f_2^{\rm P} - f_3 \right) \right) j_{\ell}^{(2,0)} \right]\,.
    \end{split}
\end{align}
Here $\Theta_0 = \delta_\gamma/4$, $\Theta_1 = \theta_\gamma/(3k)$, $v_{\rm b} = \theta_{\rm b}/k$, and the radial functions are $j_\ell^{(1,0)} = \partial_x j_\ell$, $j_\ell^{(2,0)} = \frac{1}{2}\left[3\partial_x^2 j_\ell + j_\ell\right]$, and $j_\ell^{(2,2)} = \sqrt{\frac{3}{8}\frac{(\ell+2)!}{(\ell-2)!}}\, j_\ell/x^2$, evaluated at $x = k(\eta_0-\eta)$. Similarly for polarization, we have that
\begin{equation}\label{eqn:e_los}
    \Theta_\ell^{\rm E} = \int_0^{\eta_0} \dd \eta \left[\tilde{g} - \tilde{f}\bar{\Gamma}\left(f_2^{\rm P} - f_3\right) \right] \frac{\sqrt{6} \Pi}{8}j_\ell^{(2,2)}\,.
\end{equation}
In Equations~\ref{eqn:t_los} and \ref{eqn:e_los}, $\tilde{f} \equiv \exp(-\tilde{\tau})$ and $\tilde{g} \equiv \tilde{\Gamma}\exp(-\tilde{\tau})$ are the transmission and visibility functions for the effective scattering rate $\tilde{\Gamma} \equiv \bar{\Gamma}(1-f_3)$, with $\tilde{\tau}(\eta) \equiv \int_\eta^{\eta_0}\tilde{\Gamma}\,\dd\eta'$.

\subsection{Impact on the CMB power spectra}\label{sec:cmb_impact}

\begin{figure*}[t]
    \centering
    \includegraphics{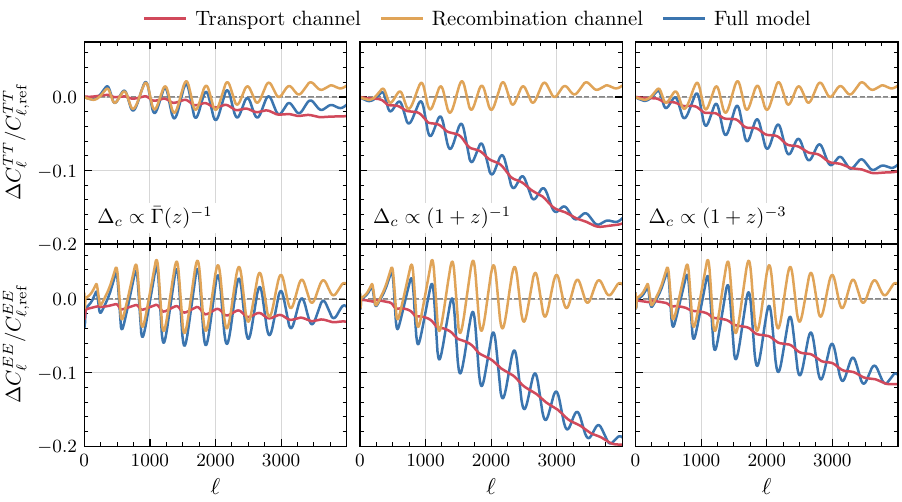}
    \caption{Fractional residuals in $C_{\ell}^{TT}$ (top row) and $C_{\ell}^{EE}$ (bottom row), relative to the $\lcdm$ best fit to \planck PR3, for three choices of the redshift dependence of the coherence length. From left to right: $\Delta_c \propto \bar{\Gamma}^{-1}$ (constant $\tau_c$), $\Delta_c \propto (1+z)^{-1}$, and $\Delta_c \propto (1+z)^{-3}$. All three are normalized to $\tau_c = 0.034$ at $z_{\rm ref} = 1100$, corresponding to $\Delta_c(z_{\rm ref}) \approx 525\,{\rm kpc}$ as in Figure~\ref{fig:example_evolutions}, with $\sigma_{\rm e}^2 = 0.3$. We assume log-normal statistics for the fluctuations. We show the residuals from the Boltzmann hierarchy modifications alone (red; Section~\ref{sec:modified_BH}), from the modified recombination history alone (gold; Section~\ref{sec:modified_rec}), and from the full model (blue). The cosmological parameters are held fixed.}\label{fig:power_spectra_illustration}
\end{figure*}

Figure~\ref{fig:power_spectra_illustration} illustrates how the two channels affect the CMB power spectra. In this plot, we hold the \lcdm\ parameters $\{\omega_{\rm b}, \omega_{\rm cdm}, n_s, \ln 10^{10}A_s, \tau_{\rm reio}, H_0\}$ fixed. The modified recombination history produces an oscillatory residual in both temperature and polarization. Accelerated recombination increases $z_\star$, which reduces $r_s^\star$ and hence $\theta_s^\star$; the acoustic peaks move to higher $\ell$, and the residual oscillates as they move out of phase with the fiducial. The oscillation sits on a small net enhancement of power at high $\ell$, at the percent level for $\sigma_{\rm e}^2 = 0.3$, consistent with a damping scale that shrinks along with the sound horizon (see Section~\ref{sec:hubble}).

The Boltzmann hierarchy modifications, on the other hand, mainly add damping. Each correction in Equations~\ref{eqn:temperature_hierarchy} and \ref{eqn:polarization_hierarchy} reduces the effective scattering rate of a multipole, $\bar{\Gamma}\to\bar{\Gamma}(1-f)$. Photons therefore diffuse further, and anisotropies are erased up to larger scales than in the fiducial model. Previous work has found that freedom to adjust the damping scale is important for maintaining the fit to CMB data at high $H_0$, and that this is difficult to achieve through changes to $X_{\rm e}$ during hydrogen recombination alone \citep{Sekiguchi:2020teg,Rashkovetskyi:2021rwg,Lynch:2024gmp}; we return to this point in Section~\ref{sec:hubble}.

We also compare different functional forms for $\Delta_c(z)$, all normalized to the same $\tau_c(z_{\rm ref}) = 0.034$ (see Figure~\ref{fig:example_evolutions}). The modifications to $X_{\rm e}$ do not depend on the coherence length and are the same in all three cases. The transport corrections, by contrast, differ substantially: at $\ell\sim3500$ the suppression of $C_\ell^{TT}$ is roughly $3\%$ for constant $\tau_c$, $11\%$ for $\Delta_c\propto(1+z)^{-3}$, and $18\%$ for $\Delta_c\propto(1+z)^{-1}$. In the power-law models $\tau_c$ rises above its value at $z_{\rm ref}$ in the epochs before last scattering, which increases the $f_\ell$ there and leads to more damping. The constraints on the transport channel therefore depend on the assumed evolution, which we quantify in Section~\ref{sec:timedep}. Note that $n=-1$ is shown for illustration only; the analysis uses $n=-3,-4,-5$.

\subsection{Constrained quantities and projection onto power spectra}\label{sec:constrained_quants}

The two channels constrain different properties of the fluctuations. The recombination channel is local and depends only on the one-point distribution of the density, which in our log-normal model is fixed by $\eavg{\delta_{\rm e}^2}$. The transport channel depends in addition on correlations along the line of sight, but in the optically thin limit only through the integral of the correlation function, $\int_0^\infty C_{\delta_{\rm e}}\,\dd r = \eavg{\delta_{\rm e}^2}\eta_c^{\rm phys}$ (Equation~\ref{eqn:heuristic_correction}): neither the form of the correlation function nor the non-Gaussianity of the field enters at leading order in $\tau_c$. A limit on $\eavg{\delta_{\rm e}^2}$ at fixed coherence length should therefore show two regimes. At short coherence lengths the transport corrections are negligible and the recombination channel sets a limit independent of $\Delta_c$, a ``recombination plateau''. At long coherence lengths the transport channel dominates, and because it constrains the product $\eavg{\delta_{\rm e}^2}\eta_c^{\rm phys}$, the limit on the variance falls as $1/\eta_c^{\rm phys} \simeq 1/\Delta_c$. Where the crossover lies depends on the relative sensitivity of the data to the two channels.

The quantities constrained in both regimes are integrals over the dimensionless power spectrum of the fluctuations, $\Delta_{\rm e}^2(k) \equiv k^3 P_{\rm e}(k) / 2 \pi^2$. For a statistically homogeneous and isotropic field (Appendix~\ref{app:power_spectrum}),
\begin{align}
    \langle \delta_{\rm e}^2 \rangle &= \int \Delta_{\rm e}^2(k) \dd\ln k \label{eqn:proj_var}\,,\\
    \langle \delta_{\rm e}^2 \rangle \eta_c^{\rm phys} &= \frac{\pi}{2} \int \frac{\Delta_{\rm e}^2(k)}{k} \dd\ln k\,.\label{eqn:proj_corr}
\end{align}
In the optically thin limit the data are thus sensitive to the power spectrum of the fluctuations only through these two integrals, and all spectra that share them, at fixed one-point distribution, form an equivalence class: they produce the same corrections to the CMB anisotropies. We compute the corrections using one representative of this class, the OU process, chosen because it is amenable to the \ito framework of CVB. Each spectrum in the class has its own shape functions at finite $\tau_c$, those of Section~\ref{sec:modified_BH} being the ones for the OU process, but all share the limit $f_\ell\to(\tau_\ell/\tau_c)\,\bar{\Gamma}\eavg{\delta_{\rm e}^2}\eta_c^{\rm phys}$ as $\tau_c\to0$. Our constraints therefore apply to the whole class when $\tau_c\ll1$.

To restate them for a given spectral shape, we map each point $(\sigma_{\rm e}^2,\Delta_c)$ to $(\eavg{\delta_{\rm e}^2},\eta_c^{\rm phys})$ using the log-normal expressions of Section~\ref{sec:motivation}, and then to the parameters of the spectrum using Equations~\ref{eqn:proj_var} and \ref{eqn:proj_corr}. For a $\delta$-function spike at wavenumber $k_s$ this gives (for example) an amplitude $\eavg{\delta_{\rm e}^2}$ and $k_s = \pi/(2\eta_c^{\rm phys})$; the relations for a spectrum that is log-normal in $k$ are given in Appendix~\ref{app:projection}, and the resulting constraints in Section~\ref{sec:projected_constraints}. The restatement is exact in the limit $\tau_c \to 0$ at fixed $\sigma_{\rm e}^2$; at finite $\tau_c$ the next-order correction to the transport channel depends on one further property of the spectrum, its mean separation $\bar{r}$, which enters the uncertainty band of Figure~\ref{fig:power_spectrum_constraints} as described in Appendix~\ref{app:error_budget}.

\section{Data analysis}\label{sec:data_analysis}

In Section~\ref{sec:model}, we introduced our model for small-scale electron density fluctuations and their impact on the CMB. We now confront this model with data. We begin in Section~\ref{sec:data} by describing the datasets used in our analysis. We then describe our analysis methods, models considered and priors adopted in Section~\ref{sec:methods}, and in Section~\ref{sec:fisher} the linear-response formalism we use to interpret the constraints and to forecast future ones.

\subsection{Data}\label{sec:data}

As described in Section~\ref{sec:model}, our primary sensitivity to electron density fluctuations comes via their impact on the CMB power spectra. We therefore consider two CMB datasets.
\begin{itemize}
    \item {\bf PR3:} \planck low-$\ell$ TT \texttt{Commander}, \planck low-$\ell$ EE \texttt{SimAll}, and \planck high-$\ell$ TTTEEE \texttt{plik\_lite} likelihoods \citep{Planck:2019nip,Planck:2018vyg}.
    \item {\bf SPA:} \planck PR3 data, in combination with high-$\ell$ data from ACT DR6 \citep{AtacamaCosmologyTelescope:2025vnj,AtacamaCosmologyTelescope:2025blo} and SPT-3G D1 \citep{SPT-3G:2025bzu}. Following \citet{AtacamaCosmologyTelescope:2025blo}, we restrict \planck TT data to $\ell \leq 1000$ and \planck TE/EE data to $\ell \leq 600$ (the ``P-ACT'' combination) when combining \planck and ACT DR6 data, in order to reduce correlations between the two datasets. We use the \texttt{ACT-lite} and \texttt{SPT-lite} implementations of the corresponding likelihoods.
\end{itemize}
We consider two additional non-CMB datasets:
\begin{itemize}
    \item {\bf DR2:} Baryon acoustic oscillation (BAO) measurements from DESI DR2 \citep{DESI:2025zgx}.
    \item {\bf SH0ES:} A Gaussian prior on the Hubble constant of $H_0 = (73.17 \pm 0.86)\,{\rm km}\,{\rm s}^{-1}\,{\rm Mpc}^{-1}$, from a recent SH0ES collaboration measurement of the expansion rate using the local distance ladder \citep{Breuval:2024lsv}.
\end{itemize}

We do not include CMB lensing reconstruction data, for compatibility with the emulator used for our SPA chains (Section~\ref{sec:methods}).

\subsection{Methods}\label{sec:methods}

We perform a Bayesian Markov chain Monte Carlo (MCMC) analysis using the Metropolis sampler implemented in \textsc{cobaya} \citep{Neal_dragging,Lewis:2002ah,Lewis:2013hha,Torrado:2020dgo}, and consider chains converged when the Gelman--Rubin statistic satisfies $R - 1 < 0.02$. The settings \textsc{classito} requires (Appendix~\ref{app:code_inputs}), combined with the high-precision settings needed for ACT and SPT-3G, make each evaluation slow. For chains including SPA data we therefore use the \textsc{ole} emulator \citep{Gunther:2025xrq}, which trains Gaussian-process emulators of the requested spectra on \textsc{classito} outputs computed at points accepted during sampling, and falls back to \textsc{classito} wherever its internal accuracy estimate is insufficient. Chains using PR3 data are run with \textsc{classito} directly. As a validation check, we also ran PR3 chains using \textsc{ole}, with the same settings used for the SPA chains, and found good agreement with the chains run using \textsc{classito}.  

\subsubsection{Parameters and priors}\label{sec:params_priors}

We vary the six parameters $\vec{\theta}_{\Lambda{\rm CDM}} = \{\omega_{\rm cdm}, \omega_{\rm b}, H_0, \ln(10^{10} A_s), n_s, \tau_{\rm reio}\}$, with the CMB temperature $T_0$ held fixed, and consider two extensions: the ``recombination-only'' model, \lcdm$+\sigma_{\rm e}^2$, which includes only the modified recombination history (Section~\ref{sec:modified_rec}); and the ``full'' model, \lcdm$+\sigma_{\rm e}^2+\zeta_{\rm e}$, which adds the modified transport (Section~\ref{sec:modified_BH}) and provides our main results. We assume log-normal statistics for the fluctuations, and our baseline results assume constant $\sigma_{\rm e}^2$ and $\tau_c$. We also consider the power-law coherence lengths of Equation~\ref{eqn:time_dep_corr} with $n = -3$, $-4$, and $-5$, for which $\zeta_{\rm e}^{\rm ref}$ is sampled in place of $\zeta_{\rm e}$ (Section~\ref{sec:new_params}). Table~\ref{tab:priors} lists the priors.

\begin{table}[t]
    \caption{\label{tab:priors} Uniform priors on the sampled parameters ($H_0$ in units of ${\rm km}\,{\rm s}^{-1}\,{\rm Mpc}^{-1}$). For the power-law models the last row applies to $\ln\zeta_{\rm e}^{\rm ref}$. Two further prior terms are described in Section~\ref{sec:params_priors}.}
    \centering
    \begin{tabular}{lc}
        \hline\hline
        Parameter & Prior range \\
        \hline
        $\omega_{\rm cdm}$ & $[0.01,\, 0.99]$ \\
        $\omega_{\rm b}$ & $[0.005,\, 0.1]$ \\
        $H_0$ & $[50,\, 80]$ \\
        $\ln(10^{10}A_s)$ & $[1.61,\, 3.91]$ \\
        $n_s$ & $[0.8,\, 1.2]$ \\
        $\tau_{\rm reio}$ & $[0.01,\, 0.8]$ \\
        \hline
        $\ln \sigma_{\rm e}^2$ & $[-8,\, \ln 2]$ \\
        $\ln \zeta_{\rm e}$ & $[-11,\, -3]$ \\
        \hline\hline
    \end{tabular}
\end{table}

Two further terms enter the prior. First, because we sample $\ln\sigma_{\rm e}^2$ and $\ln\zeta_{\rm e}$, we add $\ln\sigma_{\rm e}^2 + \ln\zeta_{\rm e}$ to the log-prior, so that it is flat in $\sigma_{\rm e}^2$ and $\zeta_{\rm e}$.

Second, we suppress a region in which the transport corrections vanish. For small $\sigma_{\rm e}^2$ the log-normal shape function (Equation~\ref{eqn:f_lognormal}) is
\begin{equation}
    f_S(\tau_c, \sigma_{\rm e}^2) \approx \zeta_{\rm e}\, e^{-2\zeta_{\rm e}}\, e^{-s}\,, \qquad s \equiv \frac{\zeta_{\rm e}^2}{\sigma_{\rm e}^2} = \tau_c^2 \sigma_{\rm e}^2\,,
\end{equation}
so at fixed $\zeta_{\rm e}$ the corrections are exponentially suppressed as $\sigma_{\rm e}^2 \to 0$. This is the regime $\tau_c \gtrsim \sigma_{\rm e}^{-1} \gg 1$, in which photons scatter many times within a single patch and lose sensitivity to the fluctuations \citep{Chluba:2025jqt}. Since the change to $X_{\rm e}$ is also small there, the region is a plateau of \lcdm-like likelihood spanning a large prior volume. It is of no interest here but degrades sampling, particularly with the emulator, whose training set fills with points from the plateau. We therefore apply a half-normal prior on $s$ with scale $0.05$, which removes the plateau while leaving the limit $\sigma_{\rm e}^2, \zeta_{\rm e} \to 0$ open. It becomes appreciable only for $\tau_c \sigma_{\rm e} \gtrsim 0.2$ ($\Delta_c \gtrsim 7$~Mpc at $\sigma_{\rm e}^2 = 0.24$), beyond the coherence lengths for which we report limits, and PR3 posteriors with and without it agree. We apply it in all analyses for consistency.

\subsubsection{Tension metrics}\label{sec:tension_metrics}

To assess the Hubble tension in Section~\ref{sec:hubble} we use three of the many available metrics \citep[see, e.g.,][]{Raveri:2021wfz,Leizerovich:2023qqt}, each answering a different question and similar to those of recent model comparisons \citep{Schoneberg:2026eys}.

The first, the difference of the maximum a posteriori \citep[called $Q_{\rm DMAP}$ in][]{Khalife:2023qbu,SPT-3G:2025bzu}, asks whether a single point in the model's parameter space fits a CMB dataset $\mathcal{D}$ (PR3 or SPA) and the distance ladder simultaneously:
\begin{equation}
    \Delta_{\rm DMAP} \equiv \sqrt{\chi^2_{\rm min, \mathcal{D}+SH0ES} - \chi^2_{\rm min, \mathcal{D}}}\,,
\end{equation}
quoted in units of $\sigma$. $\Delta_{\rm DMAP} = 0$ means adding the SH0ES prior costs nothing in fit. Despite the name, we evaluate $\chi^2$ at the maximum-likelihood point rather than the MAP, which differ because of our Jacobian prior. This keeps the metric independent of priors and posterior volume.

The second metric we use interrogates whether the $H_0$ posteriors from $\mathcal{D}$ and from SH0ES are mutually consistent. Drawing values of $H_0$ from each, the difference $\delta$ is distributed as \citep{Raveri:2021wfz,Khalife:2023qbu}
\begin{equation}
    \mathcal{P}(\delta) = \int \dd H_0\, \mathcal{P}_{\mathcal{D}}(H_0)\, \mathcal{P}_{\rm SH0ES}(H_0 - \delta)\,,
\end{equation}
and we convert the probability of the region $\{\delta: \mathcal{P}(\delta) \geq \mathcal{P}(0)\}$ to an equivalent Gaussian significance, $\Delta_{\rm shift}$. Unlike $\Delta_{\rm DMAP}$, this metric is sensitive to priors and posterior volume. A large value of $\Delta_{\rm shift}$ indicates that $\delta=0$ is not typical given the two $H_0$ distributions in question, i.e., that the two determinations are discrepant. 

Finally, we consider the Akaike Information Criterion (AIC) as a model comparison statistic that takes into account model complexity \citep{Akaike:1974vps}. The AIC itself is defined as 
\begin{equation}
    {\rm AIC} = \chi^2_{{\rm min}} + 2 N_{\rm param}\,,
\end{equation}
evaluated separately for each dataset combination. We can then define $\Delta {\rm AIC}$ as
\begin{equation}
    \Delta {\rm AIC} = {\rm AIC}_{\rm clumping} - {\rm AIC}_{\Lambda{\rm CDM}}.
\end{equation}
The recombination-only and full models add one and two parameters, so they must lower $\chi^2_{\rm min}$ by more than $2$ and $4$, respectively, to be preferred.

\subsection{Linear-response analysis}\label{sec:fisher}

Several of our results use a linearized treatment of the likelihood about a fiducial \lcdm\ model: the decomposition of the constraints by channel (Section~\ref{sec:degeneracy}), the forecast for future experiments (Section~\ref{sec:forecast}), the $\Delta\chi^2$ penalties of Section~\ref{sec:te_correlation}, and the systematic biases of Appendix~\ref{app:range_of_validity}. We collect the necessary machinery here, following \citet[][Eqs.~36--43]{Lee:2021bmn}. 

We assume a Gaussian likelihood with fixed covariance. For changes $X$ and $Y$ to the theory spectra ($TT$, $TE$, $EE$) we define the inner product
\begin{equation}\label{eqn:metric}
    \langle X, Y \rangle \equiv (\mathcal{B}X)^{\intercal}\, \mathsf{\Sigma}^{-1} (\mathcal{B}Y)\,,
\end{equation}
so that $\abs{X}^2 \equiv \langle X, X\rangle$ is the $\Delta\chi^2$ of $X$ for data equal to the fiducial. For an existing likelihood, $\mathcal{B}$ and $\mathsf{\Sigma}$ are its bandpower window functions and covariance. For a forecast, $\mathcal{B}$ is the identity and $\mathsf{\Sigma}$ is the Gaussian covariance of $(C_\ell^{TT}, C_\ell^{TE}, C_\ell^{EE})$ at each $\ell$, computed from $C_\ell^{XX}+N_\ell^{XX}$ with $(2\ell+1)f_{\rm sky}$ modes. The nuisance parameters $\theta_p$, the six \lcdm\ parameters and the calibrations, have responses $D_p \equiv \partial C_\ell/\partial\theta_p$ and Gaussian priors of width $\sigma_p$. These are $\sigma(\tau_{\rm reio}) = 0.0073$, standing in for low-$\ell$ polarization data, and the calibration priors of each likelihood. For a forecast we include a single calibration factor multiplying all spectra, with a prior of width $0.0025$. Minimizing $\abs{X - \sum_p\theta_pD_p}^2 + \sum_p\theta_p^2/\sigma_p^2$ over the $\theta_p$ gives the marginalized inner product
\begin{equation}\label{eqn:marg_metric}
    \langle X, Y \rangle_\perp \equiv \langle X, Y\rangle - \sum_{p,q}\langle X, D_p\rangle \left(F^{-1}\right)_{pq}\langle D_q, Y\rangle\,,
\end{equation}
with $F_{pq} \equiv \langle D_p, D_q\rangle + \delta_{pq}/\sigma_p^2$. The quantity $\abs{X}_\perp^2 \equiv \langle X, X\rangle_\perp$ is the part of the $\Delta\chi^2$ of $X$ that no shift of the nuisance parameters can absorb, and the marginal likelihood of a model predicting $X$ is $\propto \exp(-\abs{X}_\perp^2/2)$.

We use Equation~\ref{eqn:marg_metric} in three ways. For a parameter $\lambda$ with response $S \equiv \partial C_\ell/\partial\lambda$, the marginalized variance is $1/\abs{S}_\perp^2$. If the data contain $\lambda(S + U)$ but are fit with the template $S$ alone, minimizing $\abs{\lambda_0(S+U)-\lambda S}_\perp^2$ over $\lambda$ gives $\hat\lambda=\lambda_0(1+b)$, with
\begin{equation}\label{eqn:fisher_bias}
    b = \frac{\langle U, S\rangle_\perp}{\abs{S}_\perp^2}\,.
\end{equation}
Finally, to forecast a limit we cannot rely on the linear response alone, because at the values of $\sigma_{\rm e}^2$ near our limits the spectra depend on it non-linearly, through both the $\Delta X_{\rm e}$ table (Figure~\ref{fig:delta_xe}) and the shape functions (Equation~\ref{eqn:f_lognormal}). Following \citet{Lee:2021bmn}, we linearize only in the nuisance parameters and keep the full dependence of the spectra on $\sigma_{\rm e}^2$ at fixed $\Delta_c$, $\chi^2_\perp(\sigma_{\rm e}^2) \equiv \abs{C_\ell(\sigma_{\rm e}^2, \Delta_c) - C_\ell^{\rm fid}}_\perp^2$, and take the expected $95\%$ upper limit $\sigma_{95}^2$ to be the $95\%$ quantile of $\exp(-\chi^2_\perp/2)$ under a prior flat in $\sigma_{\rm e}^2 \geq 0$, the conditional prior of our limit curves (Section~\ref{sec:results}):
\begin{equation}\label{eqn:forecast_quantile}
    \int_0^{\sigma_{95}^2} e^{-\chi^2_\perp(s)/2}\,\dd s = 0.95\int_0^{\infty} e^{-\chi^2_\perp(s)/2}\,\dd s\,.
\end{equation}
If $C_\ell$ were linear in $\sigma_{\rm e}^2$, $\chi^2_\perp$ would be quadratic and this would reduce to the Fisher result for a half-Gaussian, $\sigma_{95}^2 = 1.96/\abs{S}_\perp$.

\section{Constraints on electron density fluctuations}\label{sec:constraints}
\label{sec:results}

We find no evidence for small-scale electron density fluctuations from CMB data alone: both the PR3 and SPA posteriors are consistent with $\sigma_{\rm e}^2 = 0$ (Table~\ref{tab:params}). With SPA we obtain $95\%$ upper limits of $\eavg{\delta_{\rm e}^2} < 0.21$ on the variance of the electron density and $\zeta_{\rm e} < 4.8\times10^{-3}$ on the transport parameter. Order-unity fluctuations at recombination are therefore excluded on the scales we probe: the rms density contrast is below $46\%$.

\begin{table*}[t]
    \caption{\label{tab:params} Constraints from CMB data: $95\%$ upper limits on the variance $\langle\delta_{\rm e}^2\rangle = e^{\sigma_{\rm e}^2}-1$ and on $\zeta_{\rm e}$, and marginalized means with $68\%$ intervals for $H_0$, $n_s$, and $\omega_{\rm b}$. Marginalized limits on $\Delta_c$ are not reported (see text); limits conditioned on $\Delta_c$ are shown in Figure~\ref{fig:power_spectrum_constraints}.}
    \begin{ruledtabular}
    \begin{tabular}{llccccc}
    Model & Dataset & $\langle\delta_{\rm e}^2\rangle$ & $\zeta_{\rm e}\,[10^{-3}]$\footnotemark[1] & $H_0\,[{\rm km}\,{\rm s}^{-1}\,{\rm Mpc}^{-1}]$ & $n_s$ & $\omega_{\rm b}$\\
    \colrule
    \multirow{1}{*}{Recombination-only} & PR3 & $<0.27$ & --- & $67.6\pm0.7$ & $0.9664\pm0.0048$ & $0.02243\pm0.00017$ \\
    \cline{1-7}\noalign{\vskip 1pt}
    \multirow{2}{*}{Full} & PR3 & $<0.51$ & $<16$ & $68.1\pm0.8$ & $0.9709\pm0.0056$ &  $0.02247\pm0.00019$ \\
                                           & SPA & $<0.21$ & $<4.8$ & $67.6\pm0.5$ & $0.9713\pm0.0039$ &   $0.02246\pm0.00011$ \\
    \cline{1-7}\noalign{\vskip 1pt}
    \multirow{1}{*}{Full, power law:} &  &  &  &  \\
    \multirow{1}{*}{$n=-3$} & PR3 & $<0.26$ & $<12$ & $67.7\pm0.7$ & $0.9722\pm0.0059$ & $0.02246\pm0.00017$ \\
    \multirow{1}{*}{$n=-4$} & PR3 & $<0.25$ & $<14$ & $67.7\pm0.7$ & $0.9721\pm0.0058$ & $0.02245\pm0.00017$\\
    \multirow{1}{*}{$n=-5$} & PR3 & $<0.27$ & $<15$ & $67.7\pm0.7$ & $0.9717\pm0.0058$ &  $0.02245\pm0.00017$ \\

    \end{tabular}
    \end{ruledtabular}
    \footnotetext[1]{For the power-law models, $\zeta_{\rm e}^{\rm ref}$ at $z_{\rm ref}=1100$ (Section~\ref{sec:new_params}).}
\end{table*}

\begin{figure}[t]
    \centering
    \includegraphics{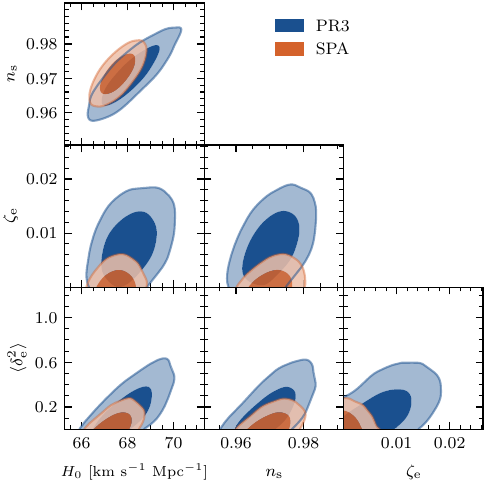}
    \caption{Marginalized 2D posteriors from the full model for the SPA and PR3 datasets. We only present a subset of \lcdm\ parameters that exhibit correlations with the extension parameters. We show the physical field variance $\eavg{\delta_{\rm e}^2} = e^{\sigma_{\rm e}^2} -1$ in place of the OU variance. }\label{fig:small_triangle}
\end{figure}

With PR3, opening the transport channel weakens the variance limit, from $\eavg{\delta_{\rm e}^2} < 0.27$ in the recombination-only model to $<0.51$ in the full model, and raises the means of $H_0$ and $n_s$ by $0.5\,{\rm km}\,{\rm s}^{-1}\,{\rm Mpc}^{-1}$ and $0.0045$ respectively (Figure~\ref{fig:small_triangle}). Once the \lcdm\ parameters are refit, the two channels leave partially cancelling imprints on the \planck\ spectra, so a larger variance is accommodated (Section~\ref{sec:degeneracy}). The high-$\ell$ data in SPA break this degeneracy: the full-model SPA limit, $\eavg{\delta_{\rm e}^2} < 0.21$, is tighter than even the recombination-only PR3 limit, and the limit on $\zeta_{\rm e}$ tightens by a factor of $3.3$ because ACT and SPT-3G measure the damping tail, where the transport corrections are largest.

\begin{figure*}[t]
    \centering
    \includegraphics{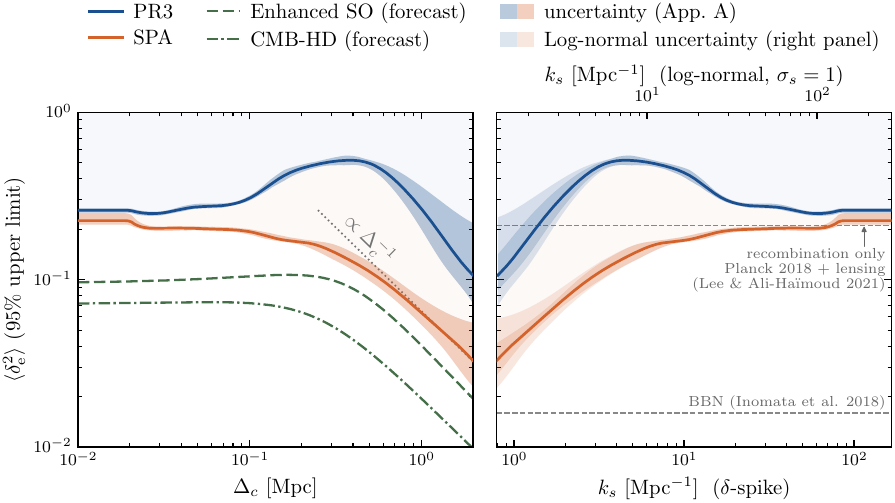}
    \caption{\label{fig:power_spectrum_constraints} Upper limits ($95\%$) on the variance of electron density fluctuations conditioned on the coherence length, from PR3 (blue) and SPA (orange), under a prior flat in $\sigma_{\rm e}^2$ and $\Delta_c$. Bands combine the systematic estimate of Appendix~\ref{app:error_budget} with the $68\%$ bootstrap interval of the curve. \textit{Left:} limit on $\eavg{\delta_{\rm e}^2}$ as a function of $\Delta_c$ (comoving, at $z_{\rm ref}=1100$), from $10$~kpc, below which our treatment of recombination is not valid (Appendix~\ref{app:range_of_validity}), to $2$~Mpc. The limits are flat where the recombination channel sets the constraint and approach $\eavg{\delta_{\rm e}^2}\propto\Delta_c^{-1}$ (dotted) where the transport channel does. Green curves are the expected limits of Section~\ref{sec:forecast} for enhanced SO (dashed) and CMB-HD (dash-dotted), without a systematic band. \textit{Right:} the same limits restated as limits on the integrated power $\eavg{\delta_{\rm e}^2}=\int\Delta_{\rm e}^2\,\dd\ln k$ of a $\delta$-function spike at $k_s$ (bottom axis) or a log-normal spectrum of width $\sigma_s=1$ centered on $k_s$ (top axis); for a given $\eta_c^{\rm phys}$ the two axes differ by $e^{\sigma_s^2/2}$. Darker bands are for the spike and lighter bands for the log-normal spectrum. Gray dashed lines show the recombination-only limit of \citet{Lee:2021bmn} from \planck\ 2018 temperature, polarization, and lensing data for baryon perturbations constant in time (equivalently, compensated isocurvature on these scales), and the deuterium bound of \citet{Inomata:2018htm}, which applies only to perturbations in the baryon-to-photon ratio already present at nucleosynthesis (Section~\ref{sec:comparison}).}
\end{figure*}

We do not quote marginalized limits on $\Delta_c$ (or $\tau_c$): as $\sigma_{\rm e}^2 \to 0$ both channels vanish and the data become insensitive to $\tau_c$, so the marginal posterior of $\Delta_c$ is dominated by the prior volume of this \lcdm-like region. Instead, Figure~\ref{fig:power_spectrum_constraints} shows $L(\Delta_c)$, the $95$th percentile of $\eavg{\delta_{\rm e}^2}$ conditioned on $\Delta_c$. We compute it using penalized quantile regression \citep[e.g.,][]{Koenker:1978}, with sample weights $w_i$ equal to the chain weights divided by $\sigma_{\rm e}^2$, which makes the prior flat in $\sigma_{\rm e}^2$ and $\Delta_c$. Writing $y_i = \ln\eavg{\delta_{\rm e}^2}_i$ and $x_i = \ln\Delta_{c,i}$ for the chain samples, we expand $\ln L(x) = \sum_j c_j B_j(x)$ in cubic B-splines with $14$ knots uniform in $\ln\Delta_c$ over $0.02$--$4$~Mpc. We choose coefficients $c_j$ that minimize
\begin{equation}\label{eqn:quantile_loss}
    \sum_i w_i\, \rho_q\big(y_i - \ln L(x_i)\big) + \sum_j \left(c_{j+1} - 2c_j + c_{j-1}\right)^2\,,
\end{equation}
with $q = 0.95$ and
\begin{equation}
    \rho_q(u) = \begin{cases} q\,u\,, & u \geq 0\,,\\ (q-1)\,u\,, & u < 0\,. \end{cases}
\end{equation}
The first term is minimized when a fraction $1-q$ of the weight lies above the curve, as desired. The second term penalizes curvature of $\ln L$ not required by the data, which would otherwise appear as oscillations where the chains are sparse \citep{Eilers:1996}. We additionally hold the curve at its plateau value below $\Delta_c = 0.02$~Mpc. The resulting limits agree with the percentile of the chain in narrow bins of $\Delta_c$ to within $5\%$ wherever the chains are well sampled. Finally, the band around each curve combines its $68\%$ bootstrap interval with the systematic uncertainty from the approximations of our model at large $\Delta_c$ (Appendix~\ref{app:error_budget}).

Both curves show the two regimes anticipated in Section~\ref{sec:constrained_quants}. At small $\Delta_c$ the transport corrections are negligible and the recombination channel sets a plateau, $\eavg{\delta_{\rm e}^2} < 0.26$ (PR3) and $0.22$ (SPA). We note that the PR3 value matches the recombination-only limit of Table~\ref{tab:params}. At large $\Delta_c$ the transport channel dominates, and since at leading order it constrains $\zeta_{\rm e} = \bar{\Gamma}\Delta_c\sigma_{\rm e}^2$, the limit approaches $\eavg{\delta_{\rm e}^2} \propto \Delta_c^{-1}$, reaching $0.11$ (PR3) and $0.033$ (SPA) at $\Delta_c = 2$~Mpc. For SPA the transition between the regimes is monotonic. However, for PR3 the limit instead rises to $0.52$ at $\Delta_c \approx 0.4$~Mpc, where the two channels contribute comparably and partially cancel, as we now discuss.

\subsection{Two-channel degeneracy}
\label{sec:degeneracy}

\begin{figure}[t]
    \centering
    \includegraphics{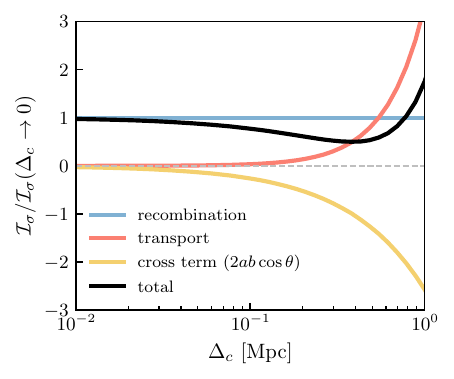}
    \caption{\label{fig:decomp} Decomposition of the marginalized Fisher information on $\sigma_{\rm e}^2$ (Equation~\ref{eqn:decomp}) as a function of $\Delta_c$, normalized to its $\Delta_c \to 0$ value, about the PR3 \lcdm\ best fit with the \texttt{plik\_lite} covariance. The blue recombination curve corresponds to the $a^2$ term in Equation~\ref{eqn:decomp}, and the red transport curve to the $b^2$ term. The minimum near $\Delta_c \approx 0.4$~Mpc coincides with the peak of the PR3 limit in Figure~\ref{fig:power_spectrum_constraints}.}
\end{figure}

The weakened PR3 bound at intermediate $\Delta_c$ stems from a degeneracy between the two channels. After marginalizing over the cosmological parameters, each channel leaves a residual imprint that no cosmological parameter shift can absorb. These two residuals are anti-aligned, so where they grow to comparable magnitudes they partially cancel and reduce the information on $\sigma_{\rm e}^2$. We quantify this with the linear-response formalism of Section~\ref{sec:fisher}, about the PR3 \lcdm\ best fit with the high-$\ell$ \texttt{plik\_lite} bandpower windows and covariance, marginalizing over the six \lcdm\ parameters and $A_{\rm Planck}$.

At fixed $\Delta_c$ we compute the response of the spectra to $\sigma_{\rm e}^2$ and split it by channel,
\begin{subequations}
    \begin{align}  
    S_{\rm full} &\equiv \frac{C_\ell - C_\ell^{\rm fid}}{\sigma_{\rm e}^2}\,, \\
    S_{\rm tr} &\equiv \frac{C_\ell^{\rm tr} - C_\ell^{\rm fid}}{\sigma_{\rm e}^2}\,,\\
    S_{X_{\rm e}} &\equiv S_{\rm full} - S_{\rm tr}\,,
    \end{align}
\end{subequations}
where $C_\ell^{\rm tr}$ includes the transport corrections but the standard $X_{\rm e}$, and all spectra are evaluated at $\sigma_{\rm e}^2 = 0.1$. The marginalized information on $\sigma_{\rm e}^2$, $\mathcal{I}_\sigma \equiv 1/\mathrm{Var}_{\rm marg}(\sigma_{\rm e}^2) = \abs{S_{\rm full}}_\perp^2$, then splits into channel auto- and cross-terms,
\begin{equation}\label{eqn:decomp}
\begin{split}
    \mathcal{I}_\sigma &= \abs{S_{X_{\rm e}}}_\perp^2 + 2\langle S_{X_{\rm e}}, S_{\rm tr}\rangle_\perp + \abs{S_{\rm tr}}_\perp^2 \\
    &\equiv a^2 + 2ab\cos\theta + b^2\,.
\end{split}
\end{equation}

Figure~\ref{fig:decomp} shows these terms as functions of $\Delta_c$, normalized so that $a^2 = 1$. The recombination term is independent of the coherence length and sets the plateau at small $\Delta_c$.  Most of the raw recombination response is absorbed by shifts in the cosmological parameters, with $\abs{S_{X_{\rm e}}}_\perp^2/\abs{S_{X_{\rm e}}}^2 \approx 0.03$. The transport response is less degenerate with the cosmological parameters ($\abs{S_{\rm tr}}_\perp^2/\abs{S_{\rm tr}}^2 \approx 0.5$), and since it scales with $\tau_c$, $b$ grows linearly with the coherence length, $b^2 \approx 3.3\,(\Delta_c/{\rm Mpc})^2$. The two residuals are anti-aligned, with $\cos\theta \approx -0.71$ nearly independent of $\Delta_c$.

Due to this anti-alignment, the total information is not monotonic in $\Delta_c$. It is smallest at $b = a\abs{\cos\theta}$, i.e., at $\Delta_c \approx 0.39$~Mpc, where the PR3 limit in Figure~\ref{fig:power_spectrum_constraints} peaks, and recovers its $\Delta_c \to 0$ value at $b = 2a\abs{\cos\theta}$, i.e., $\Delta_c \approx 0.78$~Mpc, close to where the PR3 limit returns to its plateau value in Figure~\ref{fig:power_spectrum_constraints} ($\Delta_c \approx 1$~Mpc). Therefore, while opening the transport channel tightens the bound at longer coherence lengths where the data are sensitive to the transport corrections, doing so costs information on $\sigma_{\rm e}^2$ at shorter coherence lengths.

Adding SPT-3G and ACT data to \planck\ data largely eliminates the degeneracy. Indeed, repeating the decomposition with the SPA bandpowers and covariance, $\cos\theta$ falls from $-0.71$ to $\approx -0.18$. The high-$\ell$ temperature and polarization data constrain the damping tail, where the rescaled scattering rates change photon diffusion and the generation of polarization, and so give the transport residual a component orthogonal to the recombination residual. 
Because a larger variance accelerates recombination and raises $H_0$, removing the degeneracy also narrows the range of $H_0$ opened by the small-scale fluctuations (Section~\ref{sec:hubble}).

\subsection{Projected constraints on the fluctuation power spectrum}\label{sec:projected_constraints}

The constraints of the previous sections are expressed in terms of the
parameters of the OU process. As discussed in Section~\ref{sec:constrained_quants}, these can instead be expressed as constraints on the power spectrum of the fluctuations, for different assumed spectral shapes. The right panel of Figure~\ref{fig:power_spectrum_constraints} shows the conditional limits for two such cases, a $\delta$-function spike and a log-normal spectrum of width $\sigma_s = 1$, using Equations~\ref{eqn:proj_var} and \ref{eqn:proj_corr} (see Appendix~\ref{app:projection} for explicit relations). Both are plotted as the integrated power $\eavg{\delta_{\rm e}^2}$, so a single curve serves for both, read against the bottom axis for the spike and the top axis for the log-normal spectrum. The limits span $0.8 \lesssim k_s \lesssim 170\,{\rm Mpc}^{-1}$ for the spike and $1.3 \lesssim k_s \lesssim 280\,{\rm Mpc}^{-1}$ for the log-normal spectrum, bounded by the truncation at $\Delta_c = 2$~Mpc and the validity limit at $10$~kpc. At large $k_s$ they reduce to the recombination plateau, $\eavg{\delta_{\rm e}^2} < 0.26$ (PR3) and $0.22$ (SPA), independent of spectral shape; at small $k_s$ the transport channel tightens them to $0.11$ (PR3) and $0.033$ (SPA). The PR3 peak lies at $k_s \approx 4.5\,{\rm Mpc}^{-1}$ for the spike and $7.5\,{\rm Mpc}^{-1}$ for the log-normal spectrum.

The systematic band widens toward small $k_s$, where the next-order correction to the transport channel, proportional to the mean separation $\bar{r}$ of the spectrum, grows (Appendix~\ref{app:error_budget}). It is wider for the log-normal spectrum than for the spike at the same $k_s$, because power at $k < k_s$ correlates the field over longer separations and so raises $\bar{r}$.

\subsection{Time-dependent coherence scales}\label{sec:timedep}

\begin{figure}[t]
    \centering  
    \includegraphics{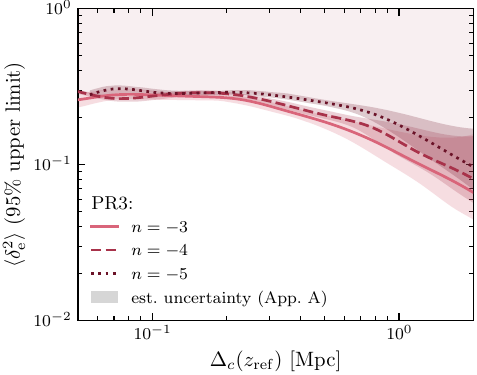}
    \caption{\label{fig:powerlaw_constraints} As the left panel of Figure~\ref{fig:power_spectrum_constraints}, for the power-law coherence lengths of Equation~\ref{eqn:time_dep_corr} with $n = -3$, $-4$, and $-5$, from PR3, with $\Delta_c$ evaluated at $z_{\rm ref} = 1100$. We show curves only where each chain retains an effective sample size of at least $200$ within $\pm0.2$ in $\ln\Delta_c$. Below this the chains sample the plateau too sparsely to locate the $95$th percentile. Bands are constructed as in Appendix~\ref{app:error_budget}, with $\epsilon_{\rm mean}$ taken from an explicit calculation of the next-order correction, which exceeds $\nu_\star\bar{r}$ by a factor of $1.2$--$2.1$ for these models. The omitted fluctuating power is negligible here ($\epsilon_{\rm inc} < 0.04$). At $2$~Mpc the total systematic error reaches $\epsilon_{\rm tot} = 0.66$, $0.54$, and $0.47$.}
\end{figure}

The constant-$\tau_c$ model implies $\Delta_c \propto \bar{\Gamma}^{-1}$, a redshift dependence unlikely to arise from a physical mechanism. We therefore repeat the PR3 analysis for the power-law coherence lengths of Equation~\ref{eqn:time_dep_corr}, with $n = -3$, $-4$, and $-5$ (Table~\ref{tab:params} and Figure~\ref{fig:powerlaw_constraints}). At fixed $\tau_c(z_{\rm ref})$ these models have larger $\tau_c$ before last scattering and hence more damping per unit $\zeta_{\rm e}^{\rm ref}$, so the transport channel becomes important at shorter coherence lengths.

The marginalized variance limits, $\eavg{\delta_{\rm e}^2} < 0.25$--$0.27$, equal the recombination-only PR3 limit: the weakening to $0.51$ seen for constant $\tau_c$ is absent. The conditional limits show no weakening at intermediate $\Delta_c$, and between $\Delta_c \approx 0.1$ and $1$~Mpc they lie below the constant-$\tau_c$ limit and its band, reaching $0.38$ ($n=-3$), $0.42$ ($n=-4$), and $0.50$ ($n=-5$) times that limit near $0.5$~Mpc. Because these models have a larger impact per unit $\zeta_{\rm e}^{\rm ref}$ at fixed $\tau_c(z_{\rm ref})$, PR3 data alone can distinguish between the two channels and the degeneracy is not as significant. At longer coherence lengths they remain below the constant-$\tau_c$ limit but within its systematic uncertainty, reaching $0.066$, $0.081$, and $0.097$ at $2$~Mpc, compared with $0.11$ in the constant case. At the shortest coherence lengths all models approach the same recombination plateau. Over the range where the transport channel matters, our constant-$\tau_c$ limits are therefore conservative relative to these redshift dependences.

\subsection{Comparison with previous limits}\label{sec:comparison}

Previous CMB limits on small-scale inhomogeneity rest on the recombination channel alone. \citet{Lee:2021bmn} computed the response of the mean ionization history to small-scale isocurvature perturbations at second order in $\delta_{\rm b}$, following their time and scale dependence, and constrained it with \planck\ 2018 temperature, polarization, and lensing data. For baryon perturbations constant in time, as in our separate-universe treatment, they found $\eavg{\delta_{\rm b}^2}<0.21$ ($95\%$), the same as their limit for compensated isocurvature, whose baryon perturbations remain constant through recombination on these scales. Under the identification $\sigma_{\rm b}^2=\sigma_{\rm e}^2$ of Section~\ref{sec:modified_rec}, our recombination-only PR3 limit of $0.27$ constrains the same quantity, and is consistent with theirs given that we omit lensing and average over a log-normal distribution of densities. For pure baryon isocurvature their limit is tighter, $\eavg{\delta_{\rm b}^2}<0.17$, because the baryons grow after kinematic decoupling.

The primordial deuterium abundance constrains baryon inhomogeneity on the same comoving scales at nucleosynthesis. Because the deuterium yield depends non-linearly on the local baryon-to-photon ratio, inhomogeneity shifts its mean, giving $\eavg{\delta_{\rm b}^2}<0.016$ ($2\sigma$) \citep{Inomata:2018htm}, or $<0.019$ ($95\%$) in the reanalysis of \citet{Lee:2021bmn}. This bound applies to the integrated power up to the neutron diffusion scale at nucleosynthesis, $k\sim10^8\,{\rm Mpc}^{-1}$ \citep{Inomata:2018htm}, and so to our scales. These lie outside the horizon at BBN, and after horizon entry Compton drag holds the baryons fixed until kinematic decoupling \citep{Lee:2021bmn}. On these scales the baryon density contrast is therefore the same at recombination as at nucleosynthesis, and the deuterium bound constrains the same quantity as our recombination channel, and is an order of magnitude stronger than both our plateau and the limit of \citet{Lee:2021bmn}.

The deuterium bound, however, applies only to inhomogeneity already present at nucleosynthesis. Baryon overdensities driven after horizon entry, by magnetic fields or by gravitational seeds, form later, and modulations of the ionization fraction leave the baryon density untouched. Our limits instead constrain fluctuations present at recombination, and in the transport regime they depend only on the statistics of the free electrons, however those fluctuations arose. There the transport channel also gives the tightest CMB constraint: at $\Delta_c=2$~Mpc the SPA limit is seven times below the plateau at small coherence lengths and about a factor of two above the deuterium bound.

\subsection{Forecast}\label{sec:forecast}

We consider forecasts for data from two future CMB experiments. The first is the enhanced Simons Observatory Large Aperture Telescope \citep{SimonsObservatory:2025wwn}, with minimum-variance ILC noise at its goal depths, $f_{\rm sky}=0.4$, $40\le\ell\le3000$ in $TT$ and $40\le\ell\le5000$ in $TE$ and $EE$. The second is CMB-HD \citep{Sehgal2019CMB-HD}, with its published noise including residual extragalactic foregrounds in $TT$, $f_{\rm sky}=0.59$, and $30\le\ell\le5000$ \citep{MacInnis2026CMB-HD,Cheslog:2026xyr}.\footnote{Available at \url{https://github.com/CMB-HD/hdMockData}. We use \code{v1.2}.} In both cases we include \planck-like data (white noise of $30$ and $60\,\mu{\rm K}$-arcmin in temperature and polarization, $7'$ beam) on $60\%$ of the sky. Where the two overlap, they are combined with the experiment's noise by inverse variance, and on the remaining sky they enter as an independent dataset for $30\le\ell\le2500$. The fiducial is the SPA \lcdm\ best fit, and the spectra are computed with \textsc{classito} at finite $\sigma_{\rm e}^2$ for the baseline constant-$\tau_c$ model. The data are taken equal to the fiducial.

Both forecasts are shown in Figure~\ref{fig:power_spectrum_constraints}. Enhanced SO would reach $\eavg{\delta_{\rm e}^2}<0.10$ on the plateau and $<0.019$ at $\Delta_c=2$~Mpc, a factor of $2$--$2.5$ below the limit the same calculation gives for SPA at every coherence length. CMB-HD would reach $0.072$ on the plateau and $0.010$ at $2$~Mpc, below the deuterium bound for $\Delta_c\gtrsim1.2$~Mpc. These CMB-HD limits are within $7\%$ (plateau) and $25\%$ ($\Delta_c\gtrsim1$~Mpc) of those of a cosmic-variance-limited survey with the same sky fraction and multipole range, so further improvement must come mainly from greater sky area.

\section{The Hubble tension and small-scale clumping}\label{sec:hubble}

\begin{table*}[t]
    \caption{\label{tab:hubble} The Hubble tension in the recombination-only and full models. For each dataset we give the marginalized mean and $68\%$ interval of $H_0$, and either the $95\%$ upper limit or, with SH0ES, the mean and $68\%$ interval of $\eavg{\delta_{\rm e}^2} = e^{\sigma_{\rm e}^2}-1$. The remaining columns compare each model with \lcdm\ fit to the same data: $\Delta\chi^2_{\rm CMB}$ and $\Delta\chi^2_{\rm ext}$ are the changes in the CMB and external (DESI DR2 or SH0ES) contributions to $\chi^2$ at the maximum-likelihood point, followed by the metrics of Section~\ref{sec:tension_metrics}. $\Delta_{\rm DMAP}$ is given for the CMB-only rows, and because $\Delta_{\rm shift}$ compares the $H_0$ posterior with the distance-ladder value, it is undefined when SH0ES is in the fit. For reference, \lcdm\ gives $\Delta_{\rm DMAP} = 5.59\,\sigma$ (PR3) and $6.06\,\sigma$ (SPA) and $H_0 = 67.27\pm 0.60\, {\rm km}\,{\rm s}^{-1}\,{\rm Mpc}^{-1}$ (PR3) and $67.26\pm 0.45\, {\rm km}\,{\rm s}^{-1}\,{\rm Mpc}^{-1}$ (SPA).}
    \begin{ruledtabular}
    \begin{tabular}{llccccccc}
    Model & Dataset & $H_0$ [${\rm km}\,{\rm s}^{-1}\,{\rm Mpc}^{-1}$] & $\eavg{\delta_{\rm e}^2}$ & $\Delta\chi^2_{\rm CMB}$(MLE) & $\Delta\chi^2_{\rm ext}$(MLE) & $\Delta\mathrm{AIC}$ & $\Delta_\mathrm{DMAP}$ & $\Delta_\mathrm{shift}$ \\
    \colrule
    \multirow{3}{*}{Recombination-only} & PR3 & $67.6\pm0.7$ & $<0.27$ & $0.00$ & --- & $2.00$ & $5.12$ & $4.95$ \\
    & PR3 + DESI DR2 & $68.8\pm0.4$ & $<0.34$ & $-0.17$ & $-0.14$ & $1.69$ & --- & $4.54$ \\
    & PR3 + SH0ES & $70.0\pm0.6$ & $0.32\pm0.15$ & $2.48$ & $-7.47$ & $-2.99$ & --- & --- \\
    \colrule
    \multirow{3}{*}{Full} & PR3 & $68.1\pm0.8$ & $<0.51$ & $-0.50$ & --- & $3.50$ & $4.11$ & $4.22$ \\
    & PR3 + DESI DR2 & $69.1\pm0.5$ & $<0.65$ & $-1.57$ & $-1.07$ & $1.36$ & --- & $4.16$ \\
    & PR3 + SH0ES & $71.1\pm0.7$ & $0.83\pm0.31$ & $0.96$ & $-15.65$ & $-10.68$ & --- & --- \\
    \colrule
    \multirow{3}{*}{Full} & SPA & $67.6\pm0.5$ & $<0.21$ & $-0.01$ & --- & $3.99$ & $5.79$ & $5.51$ \\
    & SPA + DESI DR2 & $68.6\pm0.3$ & $<0.30$ & $-0.53$ & $0.37$ & $3.84$ & --- & $4.94$ \\
    & SPA + SH0ES & $69.3\pm0.5$ & $0.28\pm0.15$ & $6.81$ & $-9.92$ & $0.89$ & --- & --- \\
    \end{tabular}
    \end{ruledtabular}
\end{table*}

We now reassess small-scale clumping as a resolution of the Hubble tension. As a representative of local distance-ladder measurements, we adopt the value of $H_0 = (73.17\pm 0.86)\,{\rm km}\,{\rm s}^{-1}\,{\rm Mpc}^{-1}$ from \citet{Breuval:2024lsv}, obtained from Cepheid-calibrated measurements of distances to Type Ia supernovae in the Hubble flow; however, we note that the tension remains robust even with other choices of distance indicators \citep{H0DN:2025lyy}. On the other hand, recent major CMB experiments such as \planck \citep{Planck:2018vyg}, ACT \citep{AtacamaCosmologyTelescope:2025blo,AtacamaCosmologyTelescope:2025vnj}, and SPT-3G \citep{SPT-3G:2025bzu,SPT-3G:2026krg} infer values near $H_0 \sim 67\,{\rm km}\,{\rm s}^{-1}\,{\rm Mpc}^{-1}$. The significance of the discrepancy between the \citet{Breuval:2024lsv} value and the CMB-SPA value from \citet{SPT-3G:2025bzu} is $6.4 \sigma$.

Inferences of $H_0$ from the CMB rely on a standard ruler, the comoving sound horizon $r_s^\star$, to judge the comoving distance to last scattering, $D_A^\star$. This inference flows through the angular scale of the sound horizon, $\theta_s^\star = r_s^\star / D_A^\star$, which is measured in a relatively model-independent manner at $0.02\%$ precision \citep[e.g.,][]{SPT-3G:2025bzu} via the location of the acoustic peaks in the CMB, $\ell_{\rm peak} \sim 1/\theta_s^\star \propto (H_0 r_s^\star)^{-1}$, where we have used $D_A^\star \propto H_0^{-1}$ at fixed $\Omega_{\rm m}$. The peak positions therefore constrain the product $H_0 r_s^\star$, so a smaller $r_s^\star$ admits a larger $H_0$ \citep[see][for details]{Knox:2019rjx}, although lowering $r_s^\star$ alone may not reconcile all datasets \citep{Jedamzik:2020zmd}.

However, solutions that lower $r_s^\star$ must do so while remaining compatible with measurements of the damping tail. The characteristic damping multipole is
\begin{equation}
    \ell_{\rm d} \equiv k_{\rm d} D_A^\star = \frac{\pi}{\theta_s^\star}\frac{r_s^\star}{r_{\rm d}^\star}\,, \qquad r_{\rm d}^\star \equiv \frac{\pi}{k_{\rm d}}\,,
\end{equation}
where $k_{\rm d}$ is the damping wavenumber from the second-order tight-coupling expansion \citep{Hu:1996mn}. At fixed $\theta_s^\star$, $\delta\ln\ell_{\rm d} = \delta\ln(r_s^\star/r_{\rm d}^\star)$, so a reduction in $r_s^\star$ must be matched by a proportional reduction in $r_{\rm d}^\star$. This is difficult to arrange with small-scale electron fluctuations. Overdense regions recombine earlier than underdense ones, so the visibility function broadens as well as shifting to higher redshift, and $r_{\rm d}^\star$, which accumulates mostly as tight coupling breaks down near last scattering, is sensitive to this change of shape \citep{Rashkovetskyi:2021rwg}. Clumping therefore does not preserve $r_s^\star/r_{\rm d}^\star$, a generic difficulty for pre-recombination solutions that change $r_s^\star$ \citep{Knox:2019rjx,Sekiguchi:2020teg}.

Clumping solutions do reduce $r_{\rm d}^\star$, but they overshoot, reducing it by a larger fraction than $r_s^\star$. \citet{Rashkovetskyi:2021rwg} show that, at fixed cosmological parameters, $r_{\rm d}^\star$ responds non-monotonically to the clumping factor $b$ (the baryon density variance, which we identify with $\eavg{\delta_{\rm e}^2}$), first increasing and then decreasing. Joint fits of clumping models to \planck\ and SH0ES have accordingly landed on less damping than in \lcdm\ \citep{Rashkovetskyi:2021rwg,Mirpoorian:2024fka}, which ACT and SPT-3G data, measuring these scales, disfavor \citep{Thiele:2021okz,Galli:2021mxk}. Model-independent reconstructions likewise find the damping scale to be an important degree of freedom for accommodating a high $H_0$ \citep{Lynch:2024gmp}.

Given that the transport channel acts directly on the damping scale, it is worth revisiting these conclusions. By lowering the effective scattering rates (Section~\ref{sec:modified_BH}) it lengthens the photon mean free path and increases damping, the opposite of the excess power at $\ell \gtrsim 2000$ that the recombination-only model leaves when refit to \planck+SH0ES. Because any physical clumping mechanism produces both effects, in this sense transport is not an optional extension but a part of the scenario omitted in previous analyses.

\subsection{Results with \planck data}\label{sec:planck_hubble}

\begin{figure}[t]
    \centering
    \includegraphics{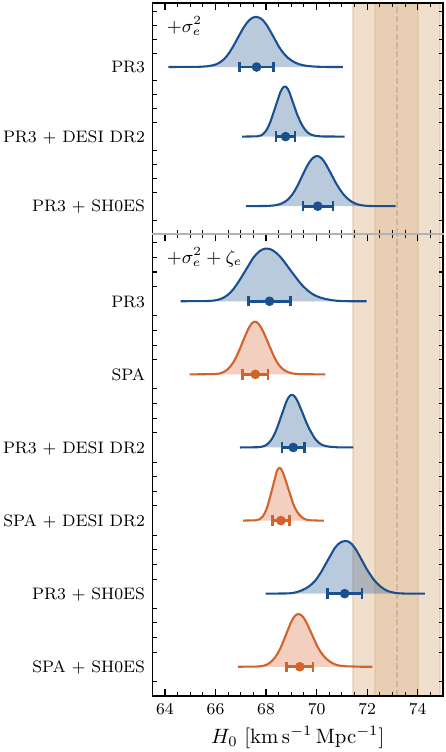}
    \caption{\label{fig:h0_forest}Marginalized constraints on $H_0$ for the recombination-only model (\textit{top}) and the full model including the
    transport channel (\textit{bottom}), for each dataset combination. Points and error bars are marginalized means with $68\%$ intervals; the vertical band shows the \citet{Breuval:2024lsv} distance-ladder measurement.}
\end{figure}

We first consider \planck\ PR3 alone, comparing the recombination-only and full models with \lcdm\ using the metrics of Section~\ref{sec:tension_metrics} (Table~\ref{tab:hubble}, Figure~\ref{fig:h0_forest}).

With PR3 alone neither model is preferred over \lcdm: the CMB fit improves by at most $\Delta\chi^2_{\rm CMB} = -0.50$ and $\Delta{\rm AIC}$ is positive. The marginalized $H_0$ moves only modestly, to $(67.6\pm0.7)\, {\rm km}\,{\rm s}^{-1}\,{\rm Mpc}^{-1}$ (recombination-only) and $(68.1\pm0.8)\,{\rm km}\,{\rm s}^{-1}\,{\rm Mpc}^{-1}$ (full), from $(67.27\pm 0.60)\, {\rm km}\,{\rm s}^{-1}\,{\rm Mpc}^{-1}$ in \lcdm. Additionally, the tension metrics fall as the two channels are opened in turn: $\Delta_{\rm DMAP}$ from $5.59\,\sigma$ (\lcdm) to $5.12\,\sigma$ and $4.11\,\sigma$ ($\Delta_{\rm shift}$: $4.95\,\sigma \to 4.22\,\sigma$). The fluctuations thus widen the range of $H_0$ the CMB tolerates more than they shift its central value.

Including a SH0ES prior on $H_0$ illustrates the payoff of this increased tolerance. The recombination-only model reaches $H_0 = (70.0\pm0.6)\, {\rm km}\,{\rm s}^{-1}\,{\rm Mpc}^{-1}$ at a CMB cost of $\Delta\chi^2_{\rm CMB} = +2.48$, with $\eavg{\delta_{\rm e}^2} = 0.32\pm0.15$. The full model reaches $H_0 = (71.1\pm0.7)\, {\rm km}\,{\rm s}^{-1}\,{\rm Mpc}^{-1}$ with a variance $2.6$ times larger, $\eavg{\delta_{\rm e}^2} = 0.83\pm0.31$, yet its CMB cost \emph{falls} to $\Delta\chi^2_{\rm CMB} = +0.96$. This reflects the degeneracy discussed in Section~\ref{sec:degeneracy}: the transport residual cancels part of the recombination residual that no \lcdm\ parameter shift absorbs, so a larger variance, and with it a smaller $r_s^\star$, costs less in the fit to \planck. Part of the added damping is also absorbed by the tilt of the primordial spectrum. We see that $n_s$ and $\zeta_{\rm e}$ are positively correlated (Figure~\ref{fig:small_triangle}), and with SH0ES the full model prefers $n_s = 0.9885\pm0.0051$, compared with $0.9796\pm 0.0044$ in the recombination-only model. Overall the full model improves on \lcdm\ by $\Delta{\rm AIC} = -10.68$ for PR3+SH0ES, entirely through the fit to the distance ladder ($\Delta\chi^2_{\rm ext} = -15.65$).

\subsection{Results with small-scale data}

Small-scale data from ACT and SPT-3G substantially restrict the allowed level of clumping and, correspondingly, the ceiling on acceptable values of $H_0$. With SPA alone the variance limit tightens to $\eavg{\delta_{\rm e}^2} < 0.21$ (from $0.51$ with PR3), and the marginalized $H_0 = (67.6\pm0.5)\, {\rm km}\,{\rm s}^{-1}\,{\rm Mpc}^{-1}$ lies barely above the \lcdm\ value of $(67.26\pm0.45)\, {\rm km}\,{\rm s}^{-1}\,{\rm Mpc}^{-1}$. In terms of the tension metrics, $\Delta_{\rm DMAP}$ falls only from $6.06\,\sigma$ to $5.79\,\sigma$, compared with $5.59\,\sigma \to 4.11\,\sigma$ for PR3. With the SH0ES prior added to the fit, the full model reaches $H_0 = (69.3\pm0.5)\, {\rm km}\,{\rm s}^{-1}\,{\rm Mpc}^{-1}$, below even the recombination-only PR3+SH0ES value, with $\eavg{\delta_{\rm e}^2} = 0.28\pm0.15$ and a CMB cost of $\Delta\chi^2_{\rm CMB} = +6.81$. $\Delta{\rm AIC}$ changes sign, from $-10.68$ to $+0.89$. These trends paint a consistent picture: the improved fit to the distance ladder no longer pays for the two added parameters and the worse fit to the small-scale CMB.

\begin{figure}[ht]
    \centering
    \includegraphics{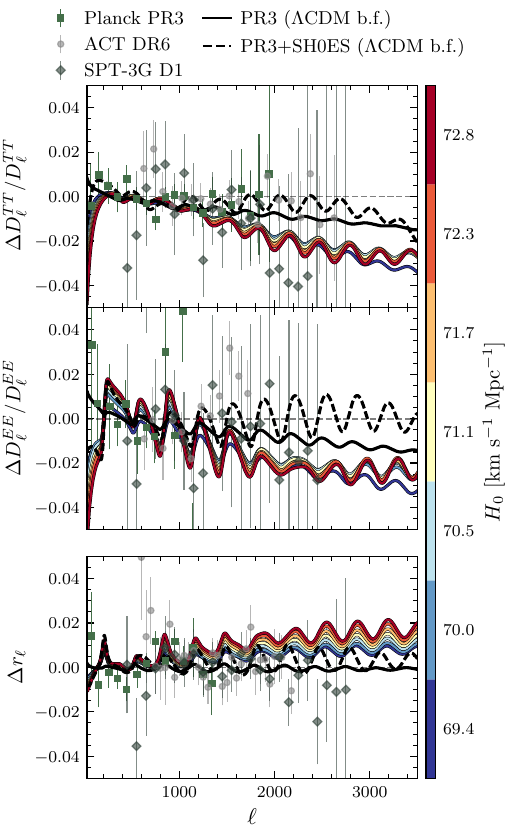}
    \caption{\label{fig:cl_residuals} Residuals of model power spectra with respect to the \lcdm\ best fit to SPA. Colored curves are spectra evaluated at the mean parameters of each of seven $H_0$ bins of the full-model PR3+SH0ES chain; black curves are the \lcdm\ best fits to PR3 (solid) and PR3+SH0ES (dashed). Points are rebinned bandpowers from the \planck\ PR3, ACT DR6, and SPT-3G D1 ``lite'' likelihoods. \textit{Top, middle:} fractional residuals $\Delta D_\ell / D_\ell$ in $TT$ and $EE$. \textit{Bottom:} change in the $T$--$E$ correlation coefficient $\Delta r_\ell$ (Equation~\ref{eqn:rell}). In the bottom panel, the data points are formed from the measured $TT$, $TE$, and $EE$ bandpowers in each bin, and the reference is binned with the same window functions before the ratio is taken. Neighboring bins in $TT$ are correlated due to foreground marginalization, so a coherent offset across several points is more significant than it appears.}
\end{figure}

As discussed in Section~\ref{sec:degeneracy}, ACT and SPT-3G break the degeneracy between the two channels that \planck\ allowed. To see where they exclude the high-$H_0$ models that fit \planck, we divide the full-model PR3+SH0ES chain into seven bins in $H_0$ spanning $\pm3\sigma$ about its mean (bin centers $69.4$--$72.8\,{\rm km}\,{\rm s}^{-1}\,{\rm Mpc}^{-1}$) and evaluate the spectra at the mean parameters of each bin. Figure~\ref{fig:cl_residuals} shows these relative to the \lcdm\ best fit to SPA, together with the \lcdm\ best fits to PR3 and PR3+SH0ES.

At $\ell \lesssim 1200$ the curves roughly coincide, since all are drawn from a posterior that fits \planck. At higher $\ell$ every model has less $TT$ and $EE$ power than the SPA reference, with little dependence on $H_0$. About half of this deficit is an offset of the reference: the PR3 \lcdm\ best fit itself lies $\approx1.5\%$ below it at small scales, in part because including ACT data raises $n_s$ \citep{AtacamaCosmologyTelescope:2025blo,Poulin:2025nfb}. The rest, a further $\approx1.5\%$ by $\ell \approx 3000$, is the added damping of the transport channel, which more than offsets the excess power left by the recombination channel. That the residual barely changes with $H_0$, although $\zeta_{\rm e}$ nearly doubles across the bins ($0.011$ to $0.019$), shows that the \lcdm\ parameters absorb the growth in damping. This deficit is therefore not what increasingly disfavors the higher-$H_0$ models. Instead, the dominant discriminator is the $T$--$E$ correlation, to which we now turn.

\subsection{The $T$--$E$ correlation}\label{sec:te_correlation}

\begin{figure}[t]
    \centering
    \includegraphics{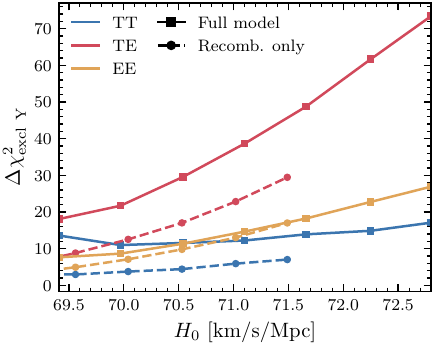}
    \caption{\label{fig:chi2_loo} Contribution of each spectrum to the ACT and SPT-3G penalty on the \planck-compatible high-$H_0$ models of Figure~\ref{fig:cl_residuals}. For each model, $\Delta\chi^2$ is computed relative to the SPA \lcdm\ best fit with the combined ACT DR6 and SPT-3G D1 bandpower covariance and calibration parameters fixed to unity. The penalty attributable to spectrum $Y$ is $\Delta\chi^2_{\text{excl }Y}$ (Equation~\ref{eqn:chi2_excl}), obtained by removing the $Y$ blocks from the difference vector and covariance. Solid lines are for the full model, and dashed for the recombination-only model along its own PR3+SH0ES posterior. $TE$ dominates at every $H_0$, and its contribution grows with $H_0$.}
\end{figure}

The bottom panel of Figure~\ref{fig:cl_residuals} shows the cross-spectrum not as a fractional change in $C_\ell^{TE}$ but through the $T$--$E$ correlation coefficient,
\begin{equation}\label{eqn:rell}
    r_\ell \equiv \frac{C_\ell^{TE}}{\sqrt{C_\ell^{TT}C_\ell^{EE}}}\,,\qquad \Delta r_\ell \equiv r_\ell - r_\ell^{\rm ref}\,.
\end{equation}
This quantity is advantageous to consider because, unlike $C_\ell^{TE}$, $r_\ell$ is nearly insensitive to the primordial power spectrum. Each spectrum is a projection
\begin{equation}
    C_\ell^{XY} = 4\pi\int \dd\ln k\, \mathcal{P}_{\mathcal{R}}(k)\,\Delta^X_\ell(k)\,\Delta^Y_\ell(k)\,,
\end{equation}
where $\Delta^X_\ell = \Theta_\ell^X/\mathcal{R}$ are the line-of-sight solutions of Equations~\ref{eqn:t_los} and \ref{eqn:e_los} per unit primordial curvature perturbation. The projection kernels are localized near $k \simeq \ell/D_A^\star$, so where $\mathcal{P}_{\mathcal{R}}$ is smooth the same factor multiplies the numerator and denominator of Equation~\ref{eqn:rell} and cancels. Changes to $A_s$, $n_s$, or a running of the spectral index therefore leave $r_\ell$ nearly unchanged ($\abs{\Delta r_\ell} < 10^{-3}$ over $1500 \le \ell \le 3500$ for $\Delta n_s = 0.02$ or $\alpha_s = 0.01$), as does an overall calibration, while they produce fractional residuals in $C_\ell^{TE}$ comparable to those in $TT$ and $EE$. Lensing, which scales with $A_s$, breaks the invariance at the level $\abs{\Delta r_\ell} \approx 4\times10^{-3}$ for $\Delta\ln(10^{10}A_s) = 0.05$, and a shift in $\theta_s^\star$ remaps $\ell$, producing an oscillatory $\Delta r_\ell$ with no net offset. What survives in $\Delta r_\ell$ are primarily changes to the transfer functions $\Delta^T_\ell$ and $\Delta^E_\ell$.

Every model from the chain shows a smooth, positive $\Delta r_\ell$ that grows with $\ell$ and, at fixed $\ell$, with $H_0$, reaching $\Delta r_\ell \approx 0.01$--$0.02$ at $\ell \simeq 3500$. The two \lcdm\ fits, which differ from the reference in $\theta_s^\star$, $n_s$, $A_s$, and $\omega_{\rm b}$, oscillate about zero with no net offset. The measured $\Delta \hat r_\ell$ (points; see the caption of Figure~\ref{fig:cl_residuals}) scatter about zero over $1500 \lesssim \ell \lesssim 2800$, where ACT and SPT-3G are most constraining, with no indication of the positive offset the high-$H_0$ models predict.

It is this signal in the cross-spectrum to which ACT and SPT-3G data are most sensitive. For each model we compute $\Delta\chi^2 = \langle X, X\rangle$ (Equation~\ref{eqn:metric}, with no nuisance parameters marginalized) with $X = C_\ell^{\rm model} - C_\ell^{\rm ref}$, using the combined ACT and SPT-3G bandpower windows and covariance with calibration parameters fixed to unity. This is the $\chi^2$ penalty the model would incur if the data matched the SPA \lcdm\ fit. To isolate the contribution of each spectrum $Y \in \{TT, TE, EE\}$, we remove the $Y$ blocks from $X$ and the covariance and define
\begin{equation}\label{eqn:chi2_excl}
    \Delta \chi^2_{\text{excl }Y} \equiv \Delta \chi^2_{\rm all} - \Delta \chi^2_{\text{w/o }Y}\,,
\end{equation}
the penalty that $Y$ adds given the other two spectra. Figure~\ref{fig:chi2_loo} shows the result. For the full model $\Delta \chi^2_{\text{excl }TE}$ rises from $18$ to $73$ across the $H_0$ bins and exceeds the $TT$ and $EE$ contributions at every $H_0$. The high-$H_0$ models that fit \planck\ are therefore disfavored by ACT and SPT-3G primarily through $TE$, and increasingly so as $H_0$ rises.

For the recombination-only model (dashed lines), the trends are similar but $\Delta\chi^2_{\text{excl }TE}$ is smaller by a factor of $1.5$--$2$ at matched $H_0$: the transport channel strengthens the $TE$ penalty but does not create it. Because the combined covariance treats ACT and SPT-3G as independent, their individual $\Delta\chi^2_{\text{excl }TE}$ sum to the total. We find that the $TE$ penalty splits nearly evenly between them ($\sim55\%$ ACT) at every $H_0$, so neither experiment drives the exclusion.

We have not isolated a single ingredient responsible for the transport contribution to this residual. It persists when the corrections to the line-of-sight sources (Equations~\ref{eqn:t_los} and \ref{eqn:e_los}) are dropped, so it arises from the modified hierarchy rather than the modified projection, and it is carried mainly by the Sachs--Wolfe and Doppler terms of the temperature source. Removing the differential dipole correction ($f_1 \to f_3$) or the polarization-source correction ($f_2^{\rm P} \to f_3$, which restores the \lcdm\ ratio of the polarization relaxation rate to the scattering rate) reduces the Sachs--Wolfe part but eliminates neither, and the Doppler part is less affected. The residual therefore reflects the combined, $\ell$-dependent rescaling of the scattering rates in Equations~\ref{eqn:temperature_hierarchy}--\ref{eqn:polarization_hierarchy} rather than any one term.

A signal of this general character is nonetheless expected. The $E$-mode is sourced by the photon quadrupole, which is generated from the dipole by free streaming and relaxes, in the absence of further sourcing, at the rate $(3/10)\bar{\Gamma}$ (Section~\ref{sec:modified_BH}). Because the quadrupole retains a memory of the dipole over this relaxation time, polarization samples the acoustic phase from slightly earlier than the temperature, and its effective visibility peaks before that of the temperature \citep{Zaldarriaga:1995gi}. Changes to the ionization history, which reshape the visibility function, and changes to the scattering rates in the hierarchy, which alter the relaxation of the polarization source, both shift this relative timing. Because the acoustic phase advances at a rate $k c_s$ (approximating $c_s$ as constant near decoupling), a fixed shift in relative timing changes $r_\ell$ smoothly and by more at smaller scales, the shape seen in Figure~\ref{fig:cl_residuals}. ACT and SPT-3G now measure the $T$--$E$ correlation in the damping tail precisely enough that these details of how polarization is generated at last scattering are important.

\subsection{Impact of DESI DR2 data}

Combining \planck\ with DESI DR1 BAO data and allowing free-form departures from the standard ionization history, \citet{Lynch:2024hzh} found $H_0 = 70.0\pm1.1\,{\rm km}\,{\rm s}^{-1}\,{\rm Mpc}^{-1}$ together with a clumping-like ionization history. This preference was data-driven: the {\tt ModRec} model used there can represent far more general departures than the family of Figure~\ref{fig:delta_xe}. \citet{Mirpoorian:2024fka} found a similar result with a lower-dimensional phenomenological model, and subsequently extended it to DESI DR2 \citep{Mirpoorian:2025rfp}. We therefore ask whether DESI DR2 favors our full model.

Adding DESI DR2 raises the inferred $H_0$ by $\approx 1\,{\rm km}\,{\rm s}^{-1}\,{\rm Mpc}^{-1}$ in all three model and dataset combinations, loosens the variance limits (to $0.34$, $0.65$, and $0.30$), and reduces $\Delta_{\rm shift}$ (Table~\ref{tab:hubble}). In \lcdm, DESI prefers a lower $\Omega_{\rm m}$ than the CMB \citep{Loverde:2024nfi}. With $\omega_{\rm m}$ fixed by the CMB this pulls $H_0$ upward, and a model that lowers $r_s^\star$ can compensate for that and keep $\theta_s^\star$ fixed. In our model, however, the shift is largely the response of the \lcdm\ parameters: within \lcdm, $H_0$ moves from $67.27\pm0.60$ to $68.44\pm 0.30$ (PR3) and from $67.26\pm0.45$ to $68.27\pm 0.26$ (SPA) when DESI is added, and the fluctuations improve the fit to DESI by at most $\Delta\chi^2_{\rm ext} = -1.07$. Every combination including DESI retains $\Delta_{\rm shift} > 4\,\sigma$ and $\Delta{\rm AIC} > 0$, so there is no preference over \lcdm. That the preference found with {\tt ModRec} is not reproduced may reflect both the more restricted family of ionization histories that small-scale fluctuations generate and the greater precision of DR2.

\section{Discussion and conclusion}\label{sec:conclusion}

We have implemented the transport corrections of CVB, together with a separate-universe model of inhomogeneous recombination, in the Boltzmann code \textsc{classito}, and used them to constrain small-scale electron density fluctuations at last scattering with CMB data from \planck, ACT DR6, and \sptthreeg\ D1. We find no evidence for such fluctuations. From SPA we obtain $\eavg{\delta_{\rm e}^2} < 0.21$ and $\zeta_{\rm e} < 4.8\times10^{-3}$ at $95\%$ confidence, excluding order-unity inhomogeneity on the scales our model describes. Because the transport channel responds to the coherence length of the fluctuations, we also obtain limits on their variance conditioned on $\Delta_c$, over $10\,{\rm kpc}\lesssim\Delta_c\lesssim2\,{\rm Mpc}$, which constrain the fluctuations directly at last scattering. These are flat at short coherence lengths, where the recombination channel sets the limit at essentially its marginalized value, and tighten as $\Delta_c^{-1}$ at long ones, where the transport channel constrains $\eavg{\delta_{\rm e}^2}\eta_c^{\rm phys}$, reaching $\eavg{\delta_{\rm e}^2}<0.033$ at $2$~Mpc. Since in the optically thin limit the data depend on the fluctuation power spectrum only through two of its integrals, the same limits apply to the integrated power of other spectral shapes, covering $1\lesssim k\lesssim10^2\,{\rm Mpc}^{-1}$ for the two we consider. Allowing the coherence length to grow as a power law in redshift tightens the limits between $0.1$ and $1$~Mpc, so our baseline limits are conservative there.

With \planck\ alone, including the transport channel weakens the bound on the variance, from $\eavg{\delta_{\rm e}^2}<0.27$ in the recombination-only model to $<0.51$. Once the \lcdm\ parameters are marginalized, the residual imprints of the two channels on the \planck\ spectra are anti-aligned, and where the channels contribute comparably, near $\Delta_c\approx0.4$~Mpc, the conditional limit is twice its plateau value. An analysis of \planck-like data that includes only the recombination channel therefore overstates the constraining power of the data. ACT and \sptthreeg\ break this degeneracy through their measurement of the damping tail, and the full-model SPA limit is tighter than even the recombination-only \planck\ limit. Future CMB experiments will measure $TE$ and $EE$ in the damping tail with several times the current sensitivity. We forecast that the enhanced Simons Observatory \citep{SimonsObservatory:2025wwn} would tighten our limits by a factor of $2$--$2.5$ at every coherence length, and CMB-HD \citep{Sehgal2019CMB-HD} by a further factor of two in the transport regime (Section~\ref{sec:forecast}).

Including the transport channel also changes the assessment of clumping as a resolution of the Hubble tension. With \planck\ alone neither model is preferred over \lcdm, but the fluctuations widen the range of $H_0$ the data tolerate, lowering $\Delta_{\rm DMAP}$ from $5.59\,\sigma$ to $5.12\,\sigma$ (recombination-only) and $4.11\,\sigma$ (full). With the distance-ladder prior, the full model reaches $H_0 = 71.1\pm0.7\,{\rm km}\,{\rm s}^{-1}\,{\rm Mpc}^{-1}$ with $2.6$ times the variance preferred by the recombination-only model, yet a smaller CMB penalty, $\Delta\chi^2_{\rm CMB}=+0.96$ against $+2.48$, because the transport residual conceals part of the recombination residual that \lcdm\ cannot absorb. It improves on \lcdm\ by $\Delta{\rm AIC}=-10.68$, entirely through the fit to the distance ladder. ACT and \sptthreeg\ remove this freedom: for SPA$+$SH0ES the model reaches only $H_0=69.3\pm0.5\,{\rm km}\,{\rm s}^{-1}\,{\rm Mpc}^{-1}$ at $\Delta\chi^2_{\rm CMB}=+6.81$, $\Delta{\rm AIC}$ becomes $+0.89$, and $\Delta_{\rm DMAP}$ falls only from $6.06\,\sigma$ to $5.79\,\sigma$. Adding DESI DR2 raises $H_0$ mainly through the response of the \lcdm\ parameters and produces no preference for the model.

The high-$H_0$ solutions allowed by \planck\ are excluded mainly through the $T$--$E$ correlation coefficient $r_\ell$, in which they predict a positive offset at $\ell\gtrsim1500$ that the data do not show. The penalty grows with $H_0$, is shared nearly equally between ACT and \sptthreeg, and is present in the recombination-only model, with the transport channel strengthening it. Because $r_\ell$ is nearly insensitive to the primordial power spectrum (Section~\ref{sec:te_correlation}), we do not expect a running of the spectral index to relax the SPA bound, although we have not tested this. This may explain why, in the comparison of \citet{Schoneberg:2026eys}, running aids dark-radiation-like solutions more than those that modify recombination: the former leave residuals mainly in the damping tails of $TT$ and $EE$, which a scale-dependent tilt can partly repair, whereas changes to the visibility function or the scattering rates alter how polarization is generated at last scattering, and with it the correlation of $T$ with $E$. Residuals of this form are visible in other models that reshape the visibility function \citep[][Fig.~14]{Lynch:2024gmp} and \citep[][Fig.~5]{Mirpoorian:2024fka}. We suggest that $\Delta r_\ell$ be reported alongside the $TT$, $TE$, and $EE$ residuals whenever modified-recombination models are confronted with small-scale data.

The approximations used in our analysis affect the two channels differently. For the transport channel, we estimated the systematic error incurred by these approximations. The CVB formalism requires the coherence length to be short compared with the scales on which the photon perturbations evolve, and at long coherence lengths it neglects both the decay of the perturbation sourced by one fluctuation before the photons meet a correlated one, and the power in the fluctuating part of the photon field, which appears as an unresolved, Poisson-like component of the spectra. We estimate both in Appendix~\ref{app:validity} and include them in the bands of Figures~\ref{fig:power_spectrum_constraints} and \ref{fig:powerlaw_constraints}. For the baseline model they amount to $\lesssim25\%$ of the transport corrections for $\Delta_c\lesssim1$~Mpc and $\approx40\%$ at $2$~Mpc, where we truncate the limits. Reaching longer coherence lengths will require going beyond first order in the separation of scales, particularly near the end of tight coupling.

The recombination channel sets the plateau, and with it our marginalized variance limits and the conclusions on $H_0$, but its uncertainties are not quantified. Our separate-universe average over static, independent patches neglects baryon dynamics below the Jeans scale, which restricts it to $\Delta_c\gtrsim10$~kpc. Studies that include these dynamics or non-local Lyman-$\alpha$ transport find that recombination is accelerated further \citep{Lee:2021bmn,Jedamzik:2023rfd,Schiff:2025vjf}, so in these respects our limits are conservative. We also neglect heating associated with the fluctuations, which alters the mapping from baryon to electron density in a mechanism-dependent way that we have not assessed. Moreover, we have identified the variance that drives recombination with that of the electrons at last scattering, rather than the baryons. Were the two to differ, the transport-regime limits would still apply, approximately, to the electron field and the plateau to the baryon density, and only the regime in which the channels compete would depend on their relation. The resulting family of ionization histories is narrower than the range the CMB allows \citep{Lynch:2024gmp}, so our conclusions on $H_0$ apply to small-scale fluctuations as modeled here. Finally, we have applied the stochastic formalism to photon transport but not to recombination itself, whose mean we compute from an ensemble of deterministic histories. An \ito\ treatment of recombination is the natural next step.

Looking ahead, measurements of the CMB frequency spectrum would complement our limits most directly. The cosmological recombination radiation (CRR), the line emission from the recombination of hydrogen and helium, is produced by the same electrons as our recombination channel. Inhomogeneous recombination broadens and shifts the CRR features at second order in the fluctuations, an effect enhanced at the levels of clumping relevant to the Hubble tension \citep{Lucca2024CRR}, so a measurement of the CRR would test the recombination channel through the spectrum rather than through its impact on the anisotropies. Spectral distortions from the dissipation of acoustic modes are sensitive to primordial power on $1\lesssim k\lesssim10^4\,{\rm Mpc}^{-1}$ \citep{Chluba:2012we,Chluba:2019kpb}, overlapping and extending the range constrained here. The proposed FOSSIL mission \citep{Aghanim:2026fossil} would improve on the sensitivity of COBE/FIRAS by roughly three orders of magnitude and aims to detect the $\mu\simeq2\times10^{-8}$ expected from this dissipation in \lcdm\ \citep{Cyr:2026pic}.

Other probes constrain different components of the small-scale universe at other epochs: the deuterium abundance constrains baryon inhomogeneity at nucleosynthesis (Section~\ref{sec:comparison}), and 21-cm observations of the dark ages and cosmic dawn will probe small-scale matter power and the thermal state of the gas after recombination \citep{Munoz:2019hjh,Cole:2019zhu,Zhu:2026mbp}. With \textsc{classito} and its \textsc{cobaya} interface public, any mechanism that predicts the variance and coherence length of the electron density can be confronted with CMB data directly. The CMB does not resolve these scales, but it records how photons scattered through them. With the damping tail now measured precisely, that record has become a quantitative probe of the small-scale universe at last scattering.

\begin{acknowledgments}
    GPL thanks Jonathan Schiff for a helpful discussion regarding non-local recombination, and Srini Raghunathan for providing noise curves for forecasting. GPL was supported by DOE Office of Science award DE-SC0009999 and the Michael and Ester Vaida Endowed Chair in Cosmology and Astrophysics.
    The authors acknowledge the use of Claude Code (Claude Opus versions 4.6--5) to: prepare figures and tables, write the \textsc{cobaya} wrapper, and document code. All scientific decisions rest with the authors, who accept full responsibility for the results. This work made use of the following software packages: \textsc{numpy}~\citep{Harris:2020xlr}, \textsc{scipy}~\citep{Virtanen:2019joe}, \textsc{matplotlib}~\citep{Hunter:2007ouj}, \textsc{getdist}~\citep{Lewis:2019xzd}, and \textsc{cmasher}~\citep{cmasher}.
\end{acknowledgments}

\bibliography{main}

\begin{thebibliography}{103}%
\makeatletter
\providecommand \@ifxundefined [1]{%
 \@ifx{#1\undefined}
}%
\providecommand \@ifnum [1]{%
 \ifnum #1\expandafter \@firstoftwo
 \else \expandafter \@secondoftwo
 \fi
}%
\providecommand \@ifx [1]{%
 \ifx #1\expandafter \@firstoftwo
 \else \expandafter \@secondoftwo
 \fi
}%
\providecommand \natexlab [1]{#1}%
\providecommand \enquote  [1]{``#1''}%
\providecommand \bibnamefont  [1]{#1}%
\providecommand \bibfnamefont [1]{#1}%
\providecommand \citenamefont [1]{#1}%
\providecommand \href@noop [0]{\@secondoftwo}%
\providecommand \href [0]{\begingroup \@sanitize@url \@href}%
\providecommand \@href[1]{\@@startlink{#1}\@@href}%
\providecommand \@@href[1]{\endgroup#1\@@endlink}%
\providecommand \@sanitize@url [0]{\catcode `\\12\catcode `\$12\catcode
  `\&12\catcode `\#12\catcode `\^12\catcode `\_12\catcode `\%12\relax}%
\providecommand \@@startlink[1]{}%
\providecommand \@@endlink[0]{}%
\providecommand \url  [0]{\begingroup\@sanitize@url \@url }%
\providecommand \@url [1]{\endgroup\@href {#1}{\urlprefix }}%
\providecommand \urlprefix  [0]{URL }%
\providecommand \Eprint [0]{\href }%
\providecommand \doibase [0]{https://doi.org/}%
\providecommand \selectlanguage [0]{\@gobble}%
\providecommand \bibinfo  [0]{\@secondoftwo}%
\providecommand \bibfield  [0]{\@secondoftwo}%
\providecommand \translation [1]{[#1]}%
\providecommand \BibitemOpen [0]{}%
\providecommand \bibitemStop [0]{}%
\providecommand \bibitemNoStop [0]{.\EOS\space}%
\providecommand \EOS [0]{\spacefactor3000\relax}%
\providecommand \BibitemShut  [1]{\csname bibitem#1\endcsname}%
\let\auto@bib@innerbib\@empty
\bibitem [{\citenamefont {Chluba}\ \emph {et~al.}(2012)\citenamefont {Chluba},
  \citenamefont {Erickcek},\ and\ \citenamefont {Ben-Dayan}}]{Chluba:2012we}%
  \BibitemOpen
  \bibfield  {author} {\bibinfo {author} {\bibfnamefont {J.}~\bibnamefont
  {Chluba}}, \bibinfo {author} {\bibfnamefont {A.~L.}\ \bibnamefont
  {Erickcek}},\ and\ \bibinfo {author} {\bibfnamefont {I.}~\bibnamefont
  {Ben-Dayan}},\ }\bibfield  {title} {\bibinfo {title} {{Probing the inflaton:
  Small-scale power spectrum constraints from measurements of the CMB energy
  spectrum}},\ }\href {https://doi.org/10.1088/0004-637X/758/2/76} {\bibfield
  {journal} {\bibinfo  {journal} {Astrophys. J.}\ }\textbf {\bibinfo {volume}
  {758}},\ \bibinfo {pages} {76} (\bibinfo {year} {2012})},\ \Eprint
  {https://arxiv.org/abs/1203.2681} {arXiv:1203.2681 [astro-ph.CO]}
  \BibitemShut {NoStop}%
\bibitem [{\citenamefont {Chluba}\ and\ \citenamefont
  {Grin}(2013)}]{Chluba:2013dna}%
  \BibitemOpen
  \bibfield  {author} {\bibinfo {author} {\bibfnamefont {J.}~\bibnamefont
  {Chluba}}\ and\ \bibinfo {author} {\bibfnamefont {D.}~\bibnamefont {Grin}},\
  }\bibfield  {title} {\bibinfo {title} {{CMB spectral distortions from
  small-scale isocurvature fluctuations}},\ }\href
  {https://doi.org/10.1093/mnras/stt1129} {\bibfield  {journal} {\bibinfo
  {journal} {Mon. Not. Roy. Astron. Soc.}\ }\textbf {\bibinfo {volume} {434}},\
  \bibinfo {pages} {1619} (\bibinfo {year} {2013})},\ \Eprint
  {https://arxiv.org/abs/1304.4596} {arXiv:1304.4596 [astro-ph.CO]}
  \BibitemShut {NoStop}%
\bibitem [{\citenamefont {Nakama}\ \emph {et~al.}(2014)\citenamefont {Nakama},
  \citenamefont {Suyama},\ and\ \citenamefont {Yokoyama}}]{Nakama:2014vla}%
  \BibitemOpen
  \bibfield  {author} {\bibinfo {author} {\bibfnamefont {T.}~\bibnamefont
  {Nakama}}, \bibinfo {author} {\bibfnamefont {T.}~\bibnamefont {Suyama}},\
  and\ \bibinfo {author} {\bibfnamefont {J.}~\bibnamefont {Yokoyama}},\
  }\bibfield  {title} {\bibinfo {title} {{Reheating the Universe Once More: The
  Dissipation of Acoustic Waves as a Novel Probe of Primordial Inhomogeneities
  on Even Smaller Scales}},\ }\href
  {https://doi.org/10.1103/PhysRevLett.113.061302} {\bibfield  {journal}
  {\bibinfo  {journal} {Phys. Rev. Lett.}\ }\textbf {\bibinfo {volume} {113}},\
  \bibinfo {pages} {061302} (\bibinfo {year} {2014})},\ \Eprint
  {https://arxiv.org/abs/1403.5407} {arXiv:1403.5407 [astro-ph.CO]}
  \BibitemShut {NoStop}%
\bibitem [{\citenamefont {Chluba}\ \emph {et~al.}(2019)\citenamefont {Chluba}
  \emph {et~al.}}]{Chluba:2019kpb}%
  \BibitemOpen
  \bibfield  {author} {\bibinfo {author} {\bibfnamefont {J.}~\bibnamefont
  {Chluba}} \emph {et~al.},\ }\bibfield  {title} {\bibinfo {title} {{Spectral
  Distortions of the CMB as a Probe of Inflation, Recombination, Structure
  Formation and Particle Physics}: {Astro2020 Science White Paper}},\
  }\href@noop {} {\bibfield  {journal} {\bibinfo  {journal} {Bull. Am. Astron.
  Soc.}\ }\textbf {\bibinfo {volume} {51}},\ \bibinfo {pages} {184} (\bibinfo
  {year} {2019})},\ \Eprint {https://arxiv.org/abs/1903.04218}
  {arXiv:1903.04218 [astro-ph.CO]} \BibitemShut {NoStop}%
\bibitem [{\citenamefont {Barrow}\ and\ \citenamefont
  {Scherrer}(2018)}]{Barrow:2018yyg}%
  \BibitemOpen
  \bibfield  {author} {\bibinfo {author} {\bibfnamefont {J.~D.}\ \bibnamefont
  {Barrow}}\ and\ \bibinfo {author} {\bibfnamefont {R.~J.}\ \bibnamefont
  {Scherrer}},\ }\bibfield  {title} {\bibinfo {title} {Constraining {{Density
  Fluctuations}} with {{Big Bang Nucleosynthesis}} in the {{Era}} of
  {{Precision Cosmology}}},\ }\href
  {https://doi.org/10.1103/PhysRevD.98.043534} {\bibfield  {journal} {\bibinfo
  {journal} {Physical Review D}\ }\textbf {\bibinfo {volume} {98}},\ \bibinfo
  {pages} {043534} (\bibinfo {year} {2018})},\ \Eprint
  {https://arxiv.org/abs/1803.02383} {arXiv:1803.02383 [astro-ph.CO]}
  \BibitemShut {NoStop}%
\bibitem [{\citenamefont {Inomata}\ \emph {et~al.}(2018)\citenamefont
  {Inomata}, \citenamefont {Kawasaki}, \citenamefont {Kusenko},\ and\
  \citenamefont {Yang}}]{Inomata:2018htm}%
  \BibitemOpen
  \bibfield  {author} {\bibinfo {author} {\bibfnamefont {K.}~\bibnamefont
  {Inomata}}, \bibinfo {author} {\bibfnamefont {M.}~\bibnamefont {Kawasaki}},
  \bibinfo {author} {\bibfnamefont {A.}~\bibnamefont {Kusenko}},\ and\ \bibinfo
  {author} {\bibfnamefont {L.}~\bibnamefont {Yang}},\ }\bibfield  {title}
  {\bibinfo {title} {{Big Bang Nucleosynthesis Constraint on Baryonic
  Isocurvature Perturbations}},\ }\href
  {https://doi.org/10.1088/1475-7516/2018/12/003} {\bibfield  {journal}
  {\bibinfo  {journal} {JCAP}\ }\textbf {\bibinfo {volume} {12}},\ \bibinfo
  {pages} {003}},\ \Eprint {https://arxiv.org/abs/1806.00123} {arXiv:1806.00123
  [astro-ph.CO]} \BibitemShut {NoStop}%
\bibitem [{\citenamefont {Bagherian}\ \emph {et~al.}(2025)\citenamefont
  {Bagherian}, \citenamefont {Ekhterachian},\ and\ \citenamefont
  {Stelzl}}]{Bagherian:2025puf}%
  \BibitemOpen
  \bibfield  {author} {\bibinfo {author} {\bibfnamefont {H.}~\bibnamefont
  {Bagherian}}, \bibinfo {author} {\bibfnamefont {M.}~\bibnamefont
  {Ekhterachian}},\ and\ \bibinfo {author} {\bibfnamefont {S.}~\bibnamefont
  {Stelzl}},\ }\href {https://doi.org/10.48550/arXiv.2505.15904} {\bibinfo
  {title} {The {{Bearable Inhomogeneity}} of the {{Baryon Asymmetry}}}}
  (\bibinfo {year} {2025}),\ \Eprint {https://arxiv.org/abs/2505.15904}
  {arXiv:2505.15904 [hep-ph]} \BibitemShut {NoStop}%
\bibitem [{\citenamefont {Co}\ and\ \citenamefont
  {Tadepalli}(2026)}]{Co:2026klr}%
  \BibitemOpen
  \bibfield  {author} {\bibinfo {author} {\bibfnamefont {R.~T.}\ \bibnamefont
  {Co}}\ and\ \bibinfo {author} {\bibfnamefont {S.~C.}\ \bibnamefont
  {Tadepalli}},\ }\href@noop {} {\bibinfo {title} {{New Isocurvature
  Constraints from JWST UV Luminosity Function}}} (\bibinfo {year} {2026}),\
  \Eprint {https://arxiv.org/abs/2605.11079} {arXiv:2605.11079 [astro-ph.CO]}
  \BibitemShut {NoStop}%
\bibitem [{\citenamefont {Jedamzik}\ and\ \citenamefont
  {Abel}(2013)}]{Jedamzik:2013gua}%
  \BibitemOpen
  \bibfield  {author} {\bibinfo {author} {\bibfnamefont {K.}~\bibnamefont
  {Jedamzik}}\ and\ \bibinfo {author} {\bibfnamefont {T.}~\bibnamefont
  {Abel}},\ }\bibfield  {title} {\bibinfo {title} {Small-scale primordial
  magnetic fields and anisotropies in the cosmic microwave background
  radiation},\ }\href {https://doi.org/10.1088/1475-7516/2013/10/050}
  {\bibfield  {journal} {\bibinfo  {journal} {Journal of Cosmology and
  Astroparticle Physics}\ }\textbf {\bibinfo {volume} {2013}}\bibinfo  {number}
  { (10)},\ \bibinfo {pages} {050}}\BibitemShut {NoStop}%
\bibitem [{\citenamefont {Lee}\ and\ \citenamefont
  {{Ali-Ha{\"i}moud}}(2021)}]{Lee:2021bmn}%
  \BibitemOpen
\bibfield  {number} {  }\bibfield  {author} {\bibinfo {author} {\bibfnamefont
  {N.}~\bibnamefont {Lee}}\ and\ \bibinfo {author} {\bibfnamefont
  {Y.}~\bibnamefont {{Ali-Ha{\"i}moud}}},\ }\bibfield  {title} {\bibinfo
  {title} {Probing small-scale baryon and dark matter isocurvature
  perturbations with cosmic microwave background anisotropies},\ }\href
  {https://doi.org/10.1103/PhysRevD.104.103509} {\bibfield  {journal} {\bibinfo
   {journal} {Physical Review D}\ }\textbf {\bibinfo {volume} {104}},\ \bibinfo
  {pages} {103509} (\bibinfo {year} {2021})}\BibitemShut {NoStop}%
\bibitem [{\citenamefont {Chluba}\ \emph {et~al.}(2025)\citenamefont {Chluba},
  \citenamefont {Vasil},\ and\ \citenamefont {Battye}}]{Chluba:2025jqt}%
  \BibitemOpen
  \bibfield  {author} {\bibinfo {author} {\bibfnamefont {J.}~\bibnamefont
  {Chluba}}, \bibinfo {author} {\bibfnamefont {G.}~\bibnamefont {Vasil}},\ and\
  \bibinfo {author} {\bibfnamefont {R.}~\bibnamefont {Battye}},\ }\href
  {https://doi.org/10.48550/arXiv.2505.22242} {\bibinfo {title} {Ito calculus
  meets the {{Hubble}} tension: {{Effects}} of small-scale electron density
  fluctuations on the {{CMB}} anisotropies}} (\bibinfo {year} {2025}),\ \Eprint
  {https://arxiv.org/abs/2505.22242} {arXiv:2505.22242 [astro-ph]} \BibitemShut
  {NoStop}%
\bibitem [{\citenamefont {Affleck}\ and\ \citenamefont
  {Dine}(1985)}]{Affleck:1984fy}%
  \BibitemOpen
  \bibfield  {author} {\bibinfo {author} {\bibfnamefont {I.}~\bibnamefont
  {Affleck}}\ and\ \bibinfo {author} {\bibfnamefont {M.}~\bibnamefont {Dine}},\
  }\bibfield  {title} {\bibinfo {title} {A new mechanism for baryogenesis},\
  }\href {https://doi.org/10.1016/0550-3213(85)90021-5} {\bibfield  {journal}
  {\bibinfo  {journal} {Nuclear Physics B}\ }\textbf {\bibinfo {volume}
  {249}},\ \bibinfo {pages} {361} (\bibinfo {year} {1985})}\BibitemShut
  {NoStop}%
\bibitem [{\citenamefont {Enqvist}\ and\ \citenamefont
  {McDonald}(1999)}]{Enqvist:1998pf}%
  \BibitemOpen
  \bibfield  {author} {\bibinfo {author} {\bibfnamefont {K.}~\bibnamefont
  {Enqvist}}\ and\ \bibinfo {author} {\bibfnamefont {J.}~\bibnamefont
  {McDonald}},\ }\bibfield  {title} {\bibinfo {title} {{Observable isocurvature
  fluctuations from the Affleck-Dine condensate}},\ }\href
  {https://doi.org/10.1103/PhysRevLett.83.2510} {\bibfield  {journal} {\bibinfo
   {journal} {Phys. Rev. Lett.}\ }\textbf {\bibinfo {volume} {83}},\ \bibinfo
  {pages} {2510} (\bibinfo {year} {1999})},\ \Eprint
  {https://arxiv.org/abs/hep-ph/9811412} {arXiv:hep-ph/9811412} \BibitemShut
  {NoStop}%
\bibitem [{\citenamefont {Kasuya}\ \emph {et~al.}(2008)\citenamefont {Kasuya},
  \citenamefont {Kawasaki},\ and\ \citenamefont {Takahashi}}]{Kasuya:2008xp}%
  \BibitemOpen
  \bibfield  {author} {\bibinfo {author} {\bibfnamefont {S.}~\bibnamefont
  {Kasuya}}, \bibinfo {author} {\bibfnamefont {M.}~\bibnamefont {Kawasaki}},\
  and\ \bibinfo {author} {\bibfnamefont {F.}~\bibnamefont {Takahashi}},\
  }\bibfield  {title} {\bibinfo {title} {{Isocurvature fluctuations in
  Affleck-Dine mechanism and constraints on inflation models}},\ }\href
  {https://doi.org/10.1088/1475-7516/2008/10/017} {\bibfield  {journal}
  {\bibinfo  {journal} {JCAP}\ }\textbf {\bibinfo {volume} {10}},\ \bibinfo
  {pages} {017}},\ \Eprint {https://arxiv.org/abs/0805.4245} {arXiv:0805.4245
  [hep-ph]} \BibitemShut {NoStop}%
\bibitem [{\citenamefont {Akrami}\ \emph {et~al.}(2020)\citenamefont {Akrami}
  \emph {et~al.}}]{Planck:2018jri}%
  \BibitemOpen
  \bibfield  {author} {\bibinfo {author} {\bibfnamefont {Y.}~\bibnamefont
  {Akrami}} \emph {et~al.} (\bibinfo {collaboration} {Planck}),\ }\bibfield
  {title} {\bibinfo {title} {{Planck 2018 results. X. Constraints on
  inflation}},\ }\href {https://doi.org/10.1051/0004-6361/201833887} {\bibfield
   {journal} {\bibinfo  {journal} {Astron. Astrophys.}\ }\textbf {\bibinfo
  {volume} {641}},\ \bibinfo {pages} {A10} (\bibinfo {year} {2020})},\ \Eprint
  {https://arxiv.org/abs/1807.06211} {arXiv:1807.06211 [astro-ph.CO]}
  \BibitemShut {NoStop}%
\bibitem [{\citenamefont {Jedamzik}\ \emph {et~al.}(2025)\citenamefont
  {Jedamzik}, \citenamefont {Abel},\ and\ \citenamefont
  {Ali-Haimoud}}]{Jedamzik:2023rfd}%
  \BibitemOpen
  \bibfield  {author} {\bibinfo {author} {\bibfnamefont {K.}~\bibnamefont
  {Jedamzik}}, \bibinfo {author} {\bibfnamefont {T.}~\bibnamefont {Abel}},\
  and\ \bibinfo {author} {\bibfnamefont {Y.}~\bibnamefont {Ali-Haimoud}},\
  }\bibfield  {title} {\bibinfo {title} {{Cosmic recombination in the presence
  of primordial magnetic fields}},\ }\href
  {https://doi.org/10.1088/1475-7516/2025/03/012} {\bibfield  {journal}
  {\bibinfo  {journal} {JCAP}\ }\textbf {\bibinfo {volume} {03}},\ \bibinfo
  {pages} {012}},\ \Eprint {https://arxiv.org/abs/2312.11448} {arXiv:2312.11448
  [astro-ph.CO]} \BibitemShut {NoStop}%
\bibitem [{\citenamefont {Ralegankar}(2023)}]{Ralegankar:2023pyx}%
  \BibitemOpen
  \bibfield  {author} {\bibinfo {author} {\bibfnamefont {P.}~\bibnamefont
  {Ralegankar}},\ }\bibfield  {title} {\bibinfo {title} {{Dark Matter Minihalos
  from Primordial Magnetic Fields}},\ }\href
  {https://doi.org/10.1103/PhysRevLett.131.231002} {\bibfield  {journal}
  {\bibinfo  {journal} {Phys. Rev. Lett.}\ }\textbf {\bibinfo {volume} {131}},\
  \bibinfo {pages} {231002} (\bibinfo {year} {2023})},\ \Eprint
  {https://arxiv.org/abs/2303.11861} {arXiv:2303.11861 [astro-ph.CO]}
  \BibitemShut {NoStop}%
\bibitem [{\citenamefont {Ali-Ha{\"\i}moud}\ and\ \citenamefont
  {Kamionkowski}(2017)}]{Ali-Haimoud:2016mbv}%
  \BibitemOpen
  \bibfield  {author} {\bibinfo {author} {\bibfnamefont {Y.}~\bibnamefont
  {Ali-Ha{\"\i}moud}}\ and\ \bibinfo {author} {\bibfnamefont {M.}~\bibnamefont
  {Kamionkowski}},\ }\bibfield  {title} {\bibinfo {title} {{Cosmic microwave
  background limits on accreting primordial black holes}},\ }\href
  {https://doi.org/10.1103/PhysRevD.95.043534} {\bibfield  {journal} {\bibinfo
  {journal} {Phys. Rev. D}\ }\textbf {\bibinfo {volume} {95}},\ \bibinfo
  {pages} {043534} (\bibinfo {year} {2017})},\ \Eprint
  {https://arxiv.org/abs/1612.05644} {arXiv:1612.05644 [astro-ph.CO]}
  \BibitemShut {NoStop}%
\bibitem [{\citenamefont {Ricotti}\ and\ \citenamefont
  {Gould}(2009)}]{Ricotti:2009bs}%
  \BibitemOpen
  \bibfield  {author} {\bibinfo {author} {\bibfnamefont {M.}~\bibnamefont
  {Ricotti}}\ and\ \bibinfo {author} {\bibfnamefont {A.}~\bibnamefont
  {Gould}},\ }\bibfield  {title} {\bibinfo {title} {{A New Probe of Dark Matter
  and High-Energy Universe Using Microlensing}},\ }\href
  {https://doi.org/10.1088/0004-637X/707/2/979} {\bibfield  {journal} {\bibinfo
   {journal} {Astrophys. J.}\ }\textbf {\bibinfo {volume} {707}},\ \bibinfo
  {pages} {979} (\bibinfo {year} {2009})},\ \Eprint
  {https://arxiv.org/abs/0908.0735} {arXiv:0908.0735 [astro-ph.CO]}
  \BibitemShut {NoStop}%
\bibitem [{\citenamefont {Inman}\ and\ \citenamefont
  {{Ali-Ha{\"i}moud}}(2019)}]{Inman:2019wvr}%
  \BibitemOpen
  \bibfield  {author} {\bibinfo {author} {\bibfnamefont {D.}~\bibnamefont
  {Inman}}\ and\ \bibinfo {author} {\bibfnamefont {Y.}~\bibnamefont
  {{Ali-Ha{\"i}moud}}},\ }\bibfield  {title} {\bibinfo {title} {Early
  {{Structure Formation}} in \${{$\Lambda$}}\${{PBH Cosmologies}}},\ }\href
  {https://doi.org/10.1103/PhysRevD.100.083528} {\bibfield  {journal} {\bibinfo
   {journal} {Physical Review D}\ }\textbf {\bibinfo {volume} {100}},\ \bibinfo
  {pages} {083528} (\bibinfo {year} {2019})},\ \Eprint
  {https://arxiv.org/abs/1907.08129} {arXiv:1907.08129 [astro-ph]} \BibitemShut
  {NoStop}%
\bibitem [{\citenamefont {Cyr}(2025)}]{Cyr:2025tkt}%
  \BibitemOpen
  \bibfield  {author} {\bibinfo {author} {\bibfnamefont {B.}~\bibnamefont
  {Cyr}},\ }\href@noop {} {\bibinfo {title} {{Not-quite-primordial black holes
  seeded by cosmic string loops}}} (\bibinfo {year} {2025}),\ \Eprint
  {https://arxiv.org/abs/2507.17833} {arXiv:2507.17833 [astro-ph.CO]}
  \BibitemShut {NoStop}%
\bibitem [{\citenamefont {Hirano}\ \emph {et~al.}(2015)\citenamefont {Hirano},
  \citenamefont {Zhu}, \citenamefont {Yoshida}, \citenamefont {Spergel},\ and\
  \citenamefont {Yorke}}]{Hirano:2015wla}%
  \BibitemOpen
  \bibfield  {author} {\bibinfo {author} {\bibfnamefont {S.}~\bibnamefont
  {Hirano}}, \bibinfo {author} {\bibfnamefont {N.}~\bibnamefont {Zhu}},
  \bibinfo {author} {\bibfnamefont {N.}~\bibnamefont {Yoshida}}, \bibinfo
  {author} {\bibfnamefont {D.}~\bibnamefont {Spergel}},\ and\ \bibinfo {author}
  {\bibfnamefont {H.~W.}\ \bibnamefont {Yorke}},\ }\bibfield  {title} {\bibinfo
  {title} {{Early structure formation from primordial density fluctuations with
  a blue-tilted power spectrum}},\ }\href
  {https://doi.org/10.1088/0004-637X/814/1/18} {\bibfield  {journal} {\bibinfo
  {journal} {Astrophys. J.}\ }\textbf {\bibinfo {volume} {814}},\ \bibinfo
  {pages} {18} (\bibinfo {year} {2015})},\ \Eprint
  {https://arxiv.org/abs/1504.05186} {arXiv:1504.05186 [astro-ph.CO]}
  \BibitemShut {NoStop}%
\bibitem [{\citenamefont {Qin}\ \emph {et~al.}(2026)\citenamefont {Qin},
  \citenamefont {Kumar}, \citenamefont {Natarajan},\ and\ \citenamefont
  {Weiner}}]{Qin:2025ymc}%
  \BibitemOpen
  \bibfield  {author} {\bibinfo {author} {\bibfnamefont {W.}~\bibnamefont
  {Qin}}, \bibinfo {author} {\bibfnamefont {S.}~\bibnamefont {Kumar}}, \bibinfo
  {author} {\bibfnamefont {P.}~\bibnamefont {Natarajan}},\ and\ \bibinfo
  {author} {\bibfnamefont {N.}~\bibnamefont {Weiner}},\ }\bibfield  {title}
  {\bibinfo {title} {{Not-Quite-Primordial Black Holes}},\ }\href
  {https://doi.org/10.1103/zfjm-8fnt} {\bibfield  {journal} {\bibinfo
  {journal} {Phys. Rev. Lett.}\ }\textbf {\bibinfo {volume} {137}},\ \bibinfo
  {pages} {111004} (\bibinfo {year} {2026})},\ \Eprint
  {https://arxiv.org/abs/2506.13858} {arXiv:2506.13858 [astro-ph.CO]}
  \BibitemShut {NoStop}%
\bibitem [{\citenamefont {Dvorkin}\ \emph {et~al.}(2013)\citenamefont
  {Dvorkin}, \citenamefont {Blum},\ and\ \citenamefont
  {Zaldarriaga}}]{Dvorkin:2013cga}%
  \BibitemOpen
  \bibfield  {author} {\bibinfo {author} {\bibfnamefont {C.}~\bibnamefont
  {Dvorkin}}, \bibinfo {author} {\bibfnamefont {K.}~\bibnamefont {Blum}},\ and\
  \bibinfo {author} {\bibfnamefont {M.}~\bibnamefont {Zaldarriaga}},\
  }\bibfield  {title} {\bibinfo {title} {{Perturbed Recombination from Dark
  Matter Annihilation}},\ }\href {https://doi.org/10.1103/PhysRevD.87.103522}
  {\bibfield  {journal} {\bibinfo  {journal} {Phys. Rev. D}\ }\textbf {\bibinfo
  {volume} {87}},\ \bibinfo {pages} {103522} (\bibinfo {year} {2013})},\
  \Eprint {https://arxiv.org/abs/1302.4753} {arXiv:1302.4753 [astro-ph.CO]}
  \BibitemShut {NoStop}%
\bibitem [{\citenamefont {Jensen}\ and\ \citenamefont
  {Ali-Ha{\"\i}moud}(2021)}]{Jensen:2021mik}%
  \BibitemOpen
  \bibfield  {author} {\bibinfo {author} {\bibfnamefont {T.~W.}\ \bibnamefont
  {Jensen}}\ and\ \bibinfo {author} {\bibfnamefont {Y.}~\bibnamefont
  {Ali-Ha{\"\i}moud}},\ }\bibfield  {title} {\bibinfo {title} {{Perturbed
  recombination from inhomogeneous photon injection and application to
  accreting primordial black holes}},\ }\href
  {https://doi.org/10.1103/PhysRevD.104.063534} {\bibfield  {journal} {\bibinfo
   {journal} {Phys. Rev. D}\ }\textbf {\bibinfo {volume} {104}},\ \bibinfo
  {pages} {063534} (\bibinfo {year} {2021})},\ \Eprint
  {https://arxiv.org/abs/2106.10266} {arXiv:2106.10266 [astro-ph.CO]}
  \BibitemShut {NoStop}%
\bibitem [{\citenamefont {Smith}\ \emph {et~al.}(2019)\citenamefont {Smith},
  \citenamefont {Grin}, \citenamefont {Robinson},\ and\ \citenamefont
  {Qi}}]{Smith:2018rnu}%
  \BibitemOpen
  \bibfield  {author} {\bibinfo {author} {\bibfnamefont {T.~L.}\ \bibnamefont
  {Smith}}, \bibinfo {author} {\bibfnamefont {D.}~\bibnamefont {Grin}},
  \bibinfo {author} {\bibfnamefont {D.}~\bibnamefont {Robinson}},\ and\
  \bibinfo {author} {\bibfnamefont {D.}~\bibnamefont {Qi}},\ }\bibfield
  {title} {\bibinfo {title} {{Probing spatial variation of the fine-structure
  constant using the CMB}},\ }\href
  {https://doi.org/10.1103/PhysRevD.99.043531} {\bibfield  {journal} {\bibinfo
  {journal} {Phys. Rev. D}\ }\textbf {\bibinfo {volume} {99}},\ \bibinfo
  {pages} {043531} (\bibinfo {year} {2019})},\ \Eprint
  {https://arxiv.org/abs/1808.07486} {arXiv:1808.07486 [astro-ph.CO]}
  \BibitemShut {NoStop}%
\bibitem [{\citenamefont {Jedamzik}\ and\ \citenamefont
  {Saveliev}(2019)}]{Jedamzik:2018itu}%
  \BibitemOpen
  \bibfield  {author} {\bibinfo {author} {\bibfnamefont {K.}~\bibnamefont
  {Jedamzik}}\ and\ \bibinfo {author} {\bibfnamefont {A.}~\bibnamefont
  {Saveliev}},\ }\bibfield  {title} {\bibinfo {title} {{Stringent Limit on
  Primordial Magnetic Fields from the Cosmic Microwave Background Radiation}},\
  }\href {https://doi.org/10.1103/PhysRevLett.123.021301} {\bibfield  {journal}
  {\bibinfo  {journal} {Phys. Rev. Lett.}\ }\textbf {\bibinfo {volume} {123}},\
  \bibinfo {pages} {021301} (\bibinfo {year} {2019})},\ \Eprint
  {https://arxiv.org/abs/1804.06115} {arXiv:1804.06115 [astro-ph.CO]}
  \BibitemShut {NoStop}%
\bibitem [{\citenamefont {Di~Valentino}\ \emph {et~al.}(2025)\citenamefont
  {Di~Valentino} \emph {et~al.}}]{CosmoVerseNetwork:2025alb}%
  \BibitemOpen
  \bibfield  {author} {\bibinfo {author} {\bibfnamefont {E.}~\bibnamefont
  {Di~Valentino}} \emph {et~al.} (\bibinfo {collaboration} {CosmoVerse
  Network}),\ }\bibfield  {title} {\bibinfo {title} {{The CosmoVerse White
  Paper: Addressing observational tensions in cosmology with systematics and
  fundamental physics}},\ }\href {https://doi.org/10.1016/j.dark.2025.101965}
  {\bibfield  {journal} {\bibinfo  {journal} {Phys. Dark Univ.}\ }\textbf
  {\bibinfo {volume} {49}},\ \bibinfo {pages} {101965} (\bibinfo {year}
  {2025})},\ \Eprint {https://arxiv.org/abs/2504.01669} {arXiv:2504.01669
  [astro-ph.CO]} \BibitemShut {NoStop}%
\bibitem [{\citenamefont {Chluba}\ and\ \citenamefont
  {Thomas}(2011)}]{Chluba:2010ca}%
  \BibitemOpen
  \bibfield  {author} {\bibinfo {author} {\bibfnamefont {J.}~\bibnamefont
  {Chluba}}\ and\ \bibinfo {author} {\bibfnamefont {R.~M.}\ \bibnamefont
  {Thomas}},\ }\bibfield  {title} {\bibinfo {title} {Towards a complete
  treatment of the cosmological recombination problem},\ }\href
  {https://doi.org/10.1111/j.1365-2966.2010.17940.x} {\bibfield  {journal}
  {\bibinfo  {journal} {Monthly Notices of the Royal Astronomical Society}\
  }\textbf {\bibinfo {volume} {412}},\ \bibinfo {pages} {748} (\bibinfo {year}
  {2011})}\BibitemShut {NoStop}%
\bibitem [{\citenamefont {Ali-Ha\"{\i}moud}\ and\ \citenamefont
  {Hirata}(2011)}]{PhysRevD.83.043513}%
  \BibitemOpen
  \bibfield  {author} {\bibinfo {author} {\bibfnamefont {Y.}~\bibnamefont
  {Ali-Ha\"{\i}moud}}\ and\ \bibinfo {author} {\bibfnamefont {C.~M.}\
  \bibnamefont {Hirata}},\ }\bibfield  {title} {\bibinfo {title} {Hyrec: A fast
  and highly accurate primordial hydrogen and helium recombination code},\
  }\href {https://doi.org/10.1103/PhysRevD.83.043513} {\bibfield  {journal}
  {\bibinfo  {journal} {Phys. Rev. D}\ }\textbf {\bibinfo {volume} {83}},\
  \bibinfo {pages} {043513} (\bibinfo {year} {2011})}\BibitemShut {NoStop}%
\bibitem [{\citenamefont {Chiang}\ and\ \citenamefont
  {Slosar}(2018)}]{Chiang:2018xpn}%
  \BibitemOpen
  \bibfield  {author} {\bibinfo {author} {\bibfnamefont {C.-T.}\ \bibnamefont
  {Chiang}}\ and\ \bibinfo {author} {\bibfnamefont {A.}~\bibnamefont
  {Slosar}},\ }\href@noop {} {\bibinfo {title} {{Inferences of $H_0$ in
  presence of a non-standard recombination}}} (\bibinfo {year} {2018}),\
  \Eprint {https://arxiv.org/abs/1811.03624} {arXiv:1811.03624 [astro-ph.CO]}
  \BibitemShut {NoStop}%
\bibitem [{\citenamefont {Hart}\ and\ \citenamefont
  {Chluba}(2020)}]{Hart:2019dxi}%
  \BibitemOpen
  \bibfield  {author} {\bibinfo {author} {\bibfnamefont {L.}~\bibnamefont
  {Hart}}\ and\ \bibinfo {author} {\bibfnamefont {J.}~\bibnamefont {Chluba}},\
  }\bibfield  {title} {\bibinfo {title} {Updated fundamental constant
  constraints from {{Planck}} 2018 data and possible relations to the
  {{Hubble}} tension},\ }\href {https://doi.org/10.1093/mnras/staa412}
  {\bibfield  {journal} {\bibinfo  {journal} {Monthly Notices of the Royal
  Astronomical Society}\ }\textbf {\bibinfo {volume} {493}},\ \bibinfo {pages}
  {3255} (\bibinfo {year} {2020})},\ \Eprint {https://arxiv.org/abs/1912.03986}
  {arXiv:1912.03986} \BibitemShut {NoStop}%
\bibitem [{\citenamefont {Sekiguchi}\ and\ \citenamefont
  {Takahashi}(2021)}]{Sekiguchi:2020teg}%
  \BibitemOpen
  \bibfield  {author} {\bibinfo {author} {\bibfnamefont {T.}~\bibnamefont
  {Sekiguchi}}\ and\ \bibinfo {author} {\bibfnamefont {T.}~\bibnamefont
  {Takahashi}},\ }\bibfield  {title} {\bibinfo {title} {Early recombination as
  a solution to the {$H_0$} tension},\ }\href
  {https://doi.org/10.1103/PhysRevD.103.083507} {\bibfield  {journal} {\bibinfo
   {journal} {Physical Review D}\ }\textbf {\bibinfo {volume} {103}},\ \bibinfo
  {pages} {083507} (\bibinfo {year} {2021})},\ \Eprint
  {https://arxiv.org/abs/2007.03381} {arXiv:2007.03381} \BibitemShut {NoStop}%
\bibitem [{\citenamefont {Knox}\ and\ \citenamefont
  {Millea}(2020)}]{Knox:2019rjx}%
  \BibitemOpen
  \bibfield  {author} {\bibinfo {author} {\bibfnamefont {L.}~\bibnamefont
  {Knox}}\ and\ \bibinfo {author} {\bibfnamefont {M.}~\bibnamefont {Millea}},\
  }\bibfield  {title} {\bibinfo {title} {{Hubble constant
  hunter\textquoteright{}s guide}},\ }\href
  {https://doi.org/10.1103/PhysRevD.101.043533} {\bibfield  {journal} {\bibinfo
   {journal} {Phys. Rev. D}\ }\textbf {\bibinfo {volume} {101}},\ \bibinfo
  {pages} {043533} (\bibinfo {year} {2020})},\ \Eprint
  {https://arxiv.org/abs/1908.03663} {arXiv:1908.03663 [astro-ph.CO]}
  \BibitemShut {NoStop}%
\bibitem [{\citenamefont {Lee}\ \emph {et~al.}(2023)\citenamefont {Lee},
  \citenamefont {Ali-Ha{\"\i}moud}, \citenamefont {Sch{\"o}neberg},\ and\
  \citenamefont {Poulin}}]{Lee:2022gzh}%
  \BibitemOpen
  \bibfield  {author} {\bibinfo {author} {\bibfnamefont {N.}~\bibnamefont
  {Lee}}, \bibinfo {author} {\bibfnamefont {Y.}~\bibnamefont
  {Ali-Ha{\"\i}moud}}, \bibinfo {author} {\bibfnamefont {N.}~\bibnamefont
  {Sch{\"o}neberg}},\ and\ \bibinfo {author} {\bibfnamefont {V.}~\bibnamefont
  {Poulin}},\ }\bibfield  {title} {\bibinfo {title} {{What It Takes to Solve
  the Hubble Tension through Modifications of Cosmological Recombination}},\
  }\href {https://doi.org/10.1103/PhysRevLett.130.161003} {\bibfield  {journal}
  {\bibinfo  {journal} {Phys. Rev. Lett.}\ }\textbf {\bibinfo {volume} {130}},\
  \bibinfo {pages} {161003} (\bibinfo {year} {2023})},\ \Eprint
  {https://arxiv.org/abs/2212.04494} {arXiv:2212.04494 [astro-ph.CO]}
  \BibitemShut {NoStop}%
\bibitem [{\citenamefont {Lynch}\ \emph
  {et~al.}(2024{\natexlab{a}})\citenamefont {Lynch}, \citenamefont {Knox},\
  and\ \citenamefont {Chluba}}]{Lynch:2024gmp}%
  \BibitemOpen
  \bibfield  {author} {\bibinfo {author} {\bibfnamefont {G.~P.}\ \bibnamefont
  {Lynch}}, \bibinfo {author} {\bibfnamefont {L.}~\bibnamefont {Knox}},\ and\
  \bibinfo {author} {\bibfnamefont {J.}~\bibnamefont {Chluba}},\ }\bibfield
  {title} {\bibinfo {title} {Reconstructing the recombination history by
  combining early and late cosmological probes},\ }\href
  {https://doi.org/10.1103/PhysRevD.110.063518} {\bibfield  {journal} {\bibinfo
   {journal} {Physical Review D}\ }\textbf {\bibinfo {volume} {110}},\ \bibinfo
  {pages} {063518} (\bibinfo {year} {2024}{\natexlab{a}})}\BibitemShut
  {NoStop}%
\bibitem [{\citenamefont {Mirpoorian}\ \emph
  {et~al.}(2025{\natexlab{a}})\citenamefont {Mirpoorian}, \citenamefont
  {Jedamzik},\ and\ \citenamefont {Pogosian}}]{Mirpoorian:2024fka}%
  \BibitemOpen
  \bibfield  {author} {\bibinfo {author} {\bibfnamefont {S.~H.}\ \bibnamefont
  {Mirpoorian}}, \bibinfo {author} {\bibfnamefont {K.}~\bibnamefont
  {Jedamzik}},\ and\ \bibinfo {author} {\bibfnamefont {L.}~\bibnamefont
  {Pogosian}},\ }\bibfield  {title} {\bibinfo {title} {{Modified recombination
  and the Hubble tension}},\ }\href
  {https://doi.org/10.1103/PhysRevD.111.083519} {\bibfield  {journal} {\bibinfo
   {journal} {Phys. Rev. D}\ }\textbf {\bibinfo {volume} {111}},\ \bibinfo
  {pages} {083519} (\bibinfo {year} {2025}{\natexlab{a}})},\ \Eprint
  {https://arxiv.org/abs/2411.16678} {arXiv:2411.16678 [astro-ph.CO]}
  \BibitemShut {NoStop}%
\bibitem [{\citenamefont {Lee}\ and\ \citenamefont {Zhou}(2026)}]{Lee:2026hlo}%
  \BibitemOpen
  \bibfield  {author} {\bibinfo {author} {\bibfnamefont {N.}~\bibnamefont
  {Lee}}\ and\ \bibinfo {author} {\bibfnamefont {T.}~\bibnamefont {Zhou}},\
  }\href@noop {} {\bibinfo {title} {{What it takes to solve the Hubble tension
  through Modifications of Cosmological Recombination II: in light of ACT DR6
  and DESI DR2}}} (\bibinfo {year} {2026}),\ \Eprint
  {https://arxiv.org/abs/2606.06495} {arXiv:2606.06495 [astro-ph.CO]}
  \BibitemShut {NoStop}%
\bibitem [{\citenamefont {Jedamzik}\ and\ \citenamefont
  {Pogosian}(2020)}]{Jedamzik:2020krr}%
  \BibitemOpen
  \bibfield  {author} {\bibinfo {author} {\bibfnamefont {K.}~\bibnamefont
  {Jedamzik}}\ and\ \bibinfo {author} {\bibfnamefont {L.}~\bibnamefont
  {Pogosian}},\ }\bibfield  {title} {\bibinfo {title} {{Relieving the Hubble
  tension with primordial magnetic fields}},\ }\href
  {https://doi.org/10.1103/PhysRevLett.125.181302} {\bibfield  {journal}
  {\bibinfo  {journal} {Phys. Rev. Lett.}\ }\textbf {\bibinfo {volume} {125}},\
  \bibinfo {pages} {181302} (\bibinfo {year} {2020})},\ \Eprint
  {https://arxiv.org/abs/2004.09487} {arXiv:2004.09487 [astro-ph.CO]}
  \BibitemShut {NoStop}%
\bibitem [{\citenamefont {Rashkovetskyi}\ \emph {et~al.}(2021)\citenamefont
  {Rashkovetskyi}, \citenamefont {Mu{\~n}oz}, \citenamefont {Eisenstein},\ and\
  \citenamefont {Dvorkin}}]{Rashkovetskyi:2021rwg}%
  \BibitemOpen
  \bibfield  {author} {\bibinfo {author} {\bibfnamefont {M.}~\bibnamefont
  {Rashkovetskyi}}, \bibinfo {author} {\bibfnamefont {J.~B.}\ \bibnamefont
  {Mu{\~n}oz}}, \bibinfo {author} {\bibfnamefont {D.~J.}\ \bibnamefont
  {Eisenstein}},\ and\ \bibinfo {author} {\bibfnamefont {C.}~\bibnamefont
  {Dvorkin}},\ }\bibfield  {title} {\bibinfo {title} {{Small-scale clumping at
  recombination and the Hubble tension}},\ }\href
  {https://doi.org/10.1103/PhysRevD.104.103517} {\bibfield  {journal} {\bibinfo
   {journal} {Phys. Rev. D}\ }\textbf {\bibinfo {volume} {104}},\ \bibinfo
  {pages} {103517} (\bibinfo {year} {2021})},\ \Eprint
  {https://arxiv.org/abs/2108.02747} {arXiv:2108.02747 [astro-ph.CO]}
  \BibitemShut {NoStop}%
\bibitem [{\citenamefont {Thiele}\ \emph {et~al.}(2021)\citenamefont {Thiele},
  \citenamefont {Guan}, \citenamefont {Hill}, \citenamefont {Kosowsky},\ and\
  \citenamefont {Spergel}}]{Thiele:2021okz}%
  \BibitemOpen
  \bibfield  {author} {\bibinfo {author} {\bibfnamefont {L.}~\bibnamefont
  {Thiele}}, \bibinfo {author} {\bibfnamefont {Y.}~\bibnamefont {Guan}},
  \bibinfo {author} {\bibfnamefont {J.~C.}\ \bibnamefont {Hill}}, \bibinfo
  {author} {\bibfnamefont {A.}~\bibnamefont {Kosowsky}},\ and\ \bibinfo
  {author} {\bibfnamefont {D.~N.}\ \bibnamefont {Spergel}},\ }\bibfield
  {title} {\bibinfo {title} {Can small-scale baryon inhomogeneities resolve the
  {{Hubble}} tension? {{An}} investigation with {{ACT DR4}}},\ }\href
  {https://doi.org/10.1103/PhysRevD.104.063535} {\bibfield  {journal} {\bibinfo
   {journal} {Physical Review D}\ }\textbf {\bibinfo {volume} {104}},\ \bibinfo
  {pages} {063535} (\bibinfo {year} {2021})},\ \Eprint
  {https://arxiv.org/abs/2105.03003} {arXiv:2105.03003 [astro-ph]} \BibitemShut
  {NoStop}%
\bibitem [{\citenamefont {Galli}\ \emph {et~al.}(2022)\citenamefont {Galli},
  \citenamefont {Pogosian}, \citenamefont {Jedamzik},\ and\ \citenamefont
  {Balkenhol}}]{Galli:2021mxk}%
  \BibitemOpen
  \bibfield  {author} {\bibinfo {author} {\bibfnamefont {S.}~\bibnamefont
  {Galli}}, \bibinfo {author} {\bibfnamefont {L.}~\bibnamefont {Pogosian}},
  \bibinfo {author} {\bibfnamefont {K.}~\bibnamefont {Jedamzik}},\ and\
  \bibinfo {author} {\bibfnamefont {L.}~\bibnamefont {Balkenhol}},\ }\bibfield
  {title} {\bibinfo {title} {{Consistency of Planck, ACT, and SPT constraints
  on magnetically assisted recombination and forecasts for future
  experiments}},\ }\href {https://doi.org/10.1103/PhysRevD.105.023513}
  {\bibfield  {journal} {\bibinfo  {journal} {Phys. Rev. D}\ }\textbf {\bibinfo
  {volume} {105}},\ \bibinfo {pages} {023513} (\bibinfo {year} {2022})},\
  \Eprint {https://arxiv.org/abs/2109.03816} {arXiv:2109.03816 [astro-ph.CO]}
  \BibitemShut {NoStop}%
\bibitem [{\citenamefont {Schiff}\ and\ \citenamefont
  {Venumadhav}(2025)}]{Schiff:2025vjf}%
  \BibitemOpen
  \bibfield  {author} {\bibinfo {author} {\bibfnamefont {J.}~\bibnamefont
  {Schiff}}\ and\ \bibinfo {author} {\bibfnamefont {T.}~\bibnamefont
  {Venumadhav}},\ }\href {https://doi.org/10.48550/arXiv.2506.16517} {\bibinfo
  {title} {Primordial magnetic fields and modified recombination histories}}
  (\bibinfo {year} {2025}),\ \Eprint {https://arxiv.org/abs/2506.16517}
  {arXiv:2506.16517 [astro-ph]} \BibitemShut {NoStop}%
\bibitem [{\citenamefont {Sch{\"o}neberg}\ \emph {et~al.}(2022)\citenamefont
  {Sch{\"o}neberg}, \citenamefont {Franco~Abell{\'a}n}, \citenamefont
  {P{\'e}rez~S{\'a}nchez}, \citenamefont {Witte}, \citenamefont {Poulin},\ and\
  \citenamefont {Lesgourgues}}]{Schoneberg:2021qvd}%
  \BibitemOpen
  \bibfield  {author} {\bibinfo {author} {\bibfnamefont {N.}~\bibnamefont
  {Sch{\"o}neberg}}, \bibinfo {author} {\bibfnamefont {G.}~\bibnamefont
  {Franco~Abell{\'a}n}}, \bibinfo {author} {\bibfnamefont {A.}~\bibnamefont
  {P{\'e}rez~S{\'a}nchez}}, \bibinfo {author} {\bibfnamefont {S.~J.}\
  \bibnamefont {Witte}}, \bibinfo {author} {\bibfnamefont {V.}~\bibnamefont
  {Poulin}},\ and\ \bibinfo {author} {\bibfnamefont {J.}~\bibnamefont
  {Lesgourgues}},\ }\bibfield  {title} {\bibinfo {title} {{The H0 Olympics: A
  fair ranking of proposed models}},\ }\href
  {https://doi.org/10.1016/j.physrep.2022.07.001} {\bibfield  {journal}
  {\bibinfo  {journal} {Phys. Rept.}\ }\textbf {\bibinfo {volume} {984}},\
  \bibinfo {pages} {1} (\bibinfo {year} {2022})},\ \Eprint
  {https://arxiv.org/abs/2107.10291} {arXiv:2107.10291 [astro-ph.CO]}
  \BibitemShut {NoStop}%
\bibitem [{\citenamefont {Calabrese}\ \emph {et~al.}(2025)\citenamefont
  {Calabrese} \emph {et~al.}}]{AtacamaCosmologyTelescope:2025nti}%
  \BibitemOpen
  \bibfield  {author} {\bibinfo {author} {\bibfnamefont {E.}~\bibnamefont
  {Calabrese}} \emph {et~al.} (\bibinfo {collaboration} {Atacama Cosmology
  Telescope}),\ }\bibfield  {title} {\bibinfo {title} {{The Atacama Cosmology
  Telescope: DR6 constraints on extended cosmological models}},\ }\href
  {https://doi.org/10.1088/1475-7516/2025/11/063} {\bibfield  {journal}
  {\bibinfo  {journal} {JCAP}\ }\textbf {\bibinfo {volume} {11}},\ \bibinfo
  {pages} {063}},\ \Eprint {https://arxiv.org/abs/2503.14454} {arXiv:2503.14454
  [astro-ph.CO]} \BibitemShut {NoStop}%
\bibitem [{\citenamefont {Camphuis}\ \emph {et~al.}(2026)\citenamefont
  {Camphuis} \emph {et~al.}}]{SPT-3G:2025bzu}%
  \BibitemOpen
  \bibfield  {author} {\bibinfo {author} {\bibfnamefont {E.}~\bibnamefont
  {Camphuis}} \emph {et~al.},\ }\bibfield  {title} {\bibinfo {title} {{{SPT-3G
  D1}}: {{CMB}} temperature and polarization power spectra and cosmology from
  2019 and 2020 observations of the {{SPT-3G}} main field},\ }\href
  {https://doi.org/10.1103/7wt3-9v2y} {\bibfield  {journal} {\bibinfo
  {journal} {Phys. Rev. D}\ }\textbf {\bibinfo {volume} {113}},\ \bibinfo
  {pages} {083504} (\bibinfo {year} {2026})}\BibitemShut {NoStop}%
\bibitem [{\citenamefont {Sch{\"o}neberg}\ \emph {et~al.}(2026)\citenamefont
  {Sch{\"o}neberg}, \citenamefont {Poulin}, \citenamefont {Ferrari},
  \citenamefont {Finelli}, \citenamefont {Lesgourgues}, \citenamefont
  {Morelli}, \citenamefont {Mosbech}, \citenamefont {Sharma},\ and\
  \citenamefont {Simon}}]{Schoneberg:2026eys}%
  \BibitemOpen
  \bibfield  {author} {\bibinfo {author} {\bibfnamefont {N.}~\bibnamefont
  {Sch{\"o}neberg}}, \bibinfo {author} {\bibfnamefont {V.}~\bibnamefont
  {Poulin}}, \bibinfo {author} {\bibfnamefont {A.~G.}\ \bibnamefont {Ferrari}},
  \bibinfo {author} {\bibfnamefont {F.}~\bibnamefont {Finelli}}, \bibinfo
  {author} {\bibfnamefont {J.}~\bibnamefont {Lesgourgues}}, \bibinfo {author}
  {\bibfnamefont {L.}~\bibnamefont {Morelli}}, \bibinfo {author} {\bibfnamefont
  {M.~R.}\ \bibnamefont {Mosbech}}, \bibinfo {author} {\bibfnamefont {R.~K.}\
  \bibnamefont {Sharma}},\ and\ \bibinfo {author} {\bibfnamefont
  {T.}~\bibnamefont {Simon}},\ }\href@noop {} {\bibinfo {title} {{The $H_0$
  World Cup. I. Summary of the baseline group stage results}}} (\bibinfo {year}
  {2026}),\ \Eprint {https://arxiv.org/abs/2607.13282} {arXiv:2607.13282
  [astro-ph.CO]} \BibitemShut {NoStop}%
\bibitem [{\citenamefont {Smith}\ and\ \citenamefont
  {Sch{\"o}neberg}(2025)}]{Smith:2025zsg}%
  \BibitemOpen
  \bibfield  {author} {\bibinfo {author} {\bibfnamefont {T.~L.}\ \bibnamefont
  {Smith}}\ and\ \bibinfo {author} {\bibfnamefont {N.}~\bibnamefont
  {Sch{\"o}neberg}},\ }\bibfield  {title} {\bibinfo {title} {{Predictions for
  new physics in the CMB damping tail}},\ }\href
  {https://doi.org/10.1103/sp34-j91r} {\bibfield  {journal} {\bibinfo
  {journal} {Phys. Rev. D}\ }\textbf {\bibinfo {volume} {112}},\ \bibinfo
  {pages} {083559} (\bibinfo {year} {2025})},\ \Eprint
  {https://arxiv.org/abs/2503.20002} {arXiv:2503.20002 [astro-ph.CO]}
  \BibitemShut {NoStop}%
\bibitem [{\citenamefont {Blas}\ \emph {et~al.}(2011)\citenamefont {Blas},
  \citenamefont {Lesgourgues},\ and\ \citenamefont {Tram}}]{Blas:2011rf}%
  \BibitemOpen
  \bibfield  {author} {\bibinfo {author} {\bibfnamefont {D.}~\bibnamefont
  {Blas}}, \bibinfo {author} {\bibfnamefont {J.}~\bibnamefont {Lesgourgues}},\
  and\ \bibinfo {author} {\bibfnamefont {T.}~\bibnamefont {Tram}},\ }\bibfield
  {title} {\bibinfo {title} {The {{Cosmic Linear Anisotropy Solving System}}
  ({{CLASS}}) {{II}}: {{Approximation}} schemes},\ }\href
  {https://doi.org/10.1088/1475-7516/2011/07/034} {\bibfield  {journal}
  {\bibinfo  {journal} {Journal of Cosmology and Astroparticle Physics}\
  }\textbf {\bibinfo {volume} {2011}}\bibfield  {number} {\bibinfo  {number} {
  (07)},\ \bibinfo {pages} {034}},\ }\Eprint {https://arxiv.org/abs/1104.2933}
  {arXiv:1104.2933} \BibitemShut {NoStop}%
\bibitem [{\citenamefont {{Sehgal}}\ \emph {et~al.}(2019)\citenamefont
  {{Sehgal}}, \citenamefont {{Aiola}}, \citenamefont {{Akrami}}, \citenamefont
  {{Basu}}, \citenamefont {{Boylan-Kolchin}}, \citenamefont {{Bryan}},
  \citenamefont {{Clesse}}, \citenamefont {{Cyr-Racine}}, \citenamefont {{Di
  Mascolo}}, \citenamefont {{Dicker}}, \citenamefont {{Essinger-Hileman}},
  \citenamefont {{Ferraro}}, \citenamefont {{Fuller}}, \citenamefont {{Han}},
  \citenamefont {{Hasselfield}}, \citenamefont {{Holder}}, \citenamefont
  {{Jain}}, \citenamefont {{Johnson}}, \citenamefont {{Johnson}}, \citenamefont
  {{Klaassen}}, \citenamefont {{Madhavacheril}}, \citenamefont {{Mauskopf}},
  \citenamefont {{Meerburg}}, \citenamefont {{Meyers}}, \citenamefont
  {{Mroczkowski}}, \citenamefont {{M{\"u}nchmeyer}}, \citenamefont {{Naess}},
  \citenamefont {{Nagai}}, \citenamefont {{Namikawa}}, \citenamefont
  {{Newburgh}}, \citenamefont {{Nguyen}}, \citenamefont {{Niemack}},
  \citenamefont {{Oppenheimer}}, \citenamefont {{Pierpaoli}}, \citenamefont
  {{Schaan}}, \citenamefont {{Slosar}}, \citenamefont {{Spergel}},
  \citenamefont {{Switzer}}, \citenamefont {{van Engelen}},\ and\ \citenamefont
  {{Wollack}}}]{Sehgal2019CMB-HD}%
  \BibitemOpen
  \bibfield  {author} {\bibinfo {author} {\bibfnamefont {N.}~\bibnamefont
  {{Sehgal}}}, \bibinfo {author} {\bibfnamefont {S.}~\bibnamefont {{Aiola}}},
  \bibinfo {author} {\bibfnamefont {Y.}~\bibnamefont {{Akrami}}}, \bibinfo
  {author} {\bibfnamefont {K.}~\bibnamefont {{Basu}}}, \bibinfo {author}
  {\bibfnamefont {M.}~\bibnamefont {{Boylan-Kolchin}}}, \bibinfo {author}
  {\bibfnamefont {S.}~\bibnamefont {{Bryan}}}, \bibinfo {author} {\bibfnamefont
  {S.}~\bibnamefont {{Clesse}}}, \bibinfo {author} {\bibfnamefont {F.-Y.}\
  \bibnamefont {{Cyr-Racine}}}, \bibinfo {author} {\bibfnamefont
  {L.}~\bibnamefont {{Di Mascolo}}}, \bibinfo {author} {\bibfnamefont
  {S.}~\bibnamefont {{Dicker}}}, \bibinfo {author} {\bibfnamefont
  {T.}~\bibnamefont {{Essinger-Hileman}}}, \bibinfo {author} {\bibfnamefont
  {S.}~\bibnamefont {{Ferraro}}}, \bibinfo {author} {\bibfnamefont
  {G.}~\bibnamefont {{Fuller}}}, \bibinfo {author} {\bibfnamefont
  {D.}~\bibnamefont {{Han}}}, \bibinfo {author} {\bibfnamefont
  {M.}~\bibnamefont {{Hasselfield}}}, \bibinfo {author} {\bibfnamefont
  {G.}~\bibnamefont {{Holder}}}, \bibinfo {author} {\bibfnamefont
  {B.}~\bibnamefont {{Jain}}}, \bibinfo {author} {\bibfnamefont {B.~R.}\
  \bibnamefont {{Johnson}}}, \bibinfo {author} {\bibfnamefont {M.}~\bibnamefont
  {{Johnson}}}, \bibinfo {author} {\bibfnamefont {P.}~\bibnamefont
  {{Klaassen}}}, \bibinfo {author} {\bibfnamefont {M.}~\bibnamefont
  {{Madhavacheril}}}, \bibinfo {author} {\bibfnamefont {P.}~\bibnamefont
  {{Mauskopf}}}, \bibinfo {author} {\bibfnamefont {D.}~\bibnamefont
  {{Meerburg}}}, \bibinfo {author} {\bibfnamefont {J.}~\bibnamefont
  {{Meyers}}}, \bibinfo {author} {\bibfnamefont {T.}~\bibnamefont
  {{Mroczkowski}}}, \bibinfo {author} {\bibfnamefont {M.}~\bibnamefont
  {{M{\"u}nchmeyer}}}, \bibinfo {author} {\bibfnamefont {S.~K.}\ \bibnamefont
  {{Naess}}}, \bibinfo {author} {\bibfnamefont {D.}~\bibnamefont {{Nagai}}},
  \bibinfo {author} {\bibfnamefont {T.}~\bibnamefont {{Namikawa}}}, \bibinfo
  {author} {\bibfnamefont {L.}~\bibnamefont {{Newburgh}}}, \bibinfo {author}
  {\bibfnamefont {N.}~\bibnamefont {{Nguyen}}}, \bibinfo {author}
  {\bibfnamefont {M.}~\bibnamefont {{Niemack}}}, \bibinfo {author}
  {\bibfnamefont {B.~D.}\ \bibnamefont {{Oppenheimer}}}, \bibinfo {author}
  {\bibfnamefont {E.}~\bibnamefont {{Pierpaoli}}}, \bibinfo {author}
  {\bibfnamefont {E.}~\bibnamefont {{Schaan}}}, \bibinfo {author}
  {\bibfnamefont {A.}~\bibnamefont {{Slosar}}}, \bibinfo {author}
  {\bibfnamefont {D.}~\bibnamefont {{Spergel}}}, \bibinfo {author}
  {\bibfnamefont {E.}~\bibnamefont {{Switzer}}}, \bibinfo {author}
  {\bibfnamefont {A.}~\bibnamefont {{van Engelen}}},\ and\ \bibinfo {author}
  {\bibfnamefont {E.}~\bibnamefont {{Wollack}}},\ }\bibfield  {title} {\bibinfo
  {title} {{CMB-HD: An Ultra-Deep, High-Resolution Millimeter-Wave Survey Over
  Half the Sky}},\ }in\ \href {https://doi.org/10.48550/arXiv.1906.10134}
  {\emph {\bibinfo {booktitle} {Bulletin of the American Astronomical
  Society}}},\ Vol.~\bibinfo {volume} {51}\ (\bibinfo  {publisher} {AIP},\
  \bibinfo {year} {2019})\ p.~\bibinfo {pages} {6},\ \Eprint
  {https://arxiv.org/abs/1906.10134} {arXiv:1906.10134 [astro-ph.CO]}
  \BibitemShut {NoStop}%
\bibitem [{\citenamefont {{MacInnis}}\ \emph {et~al.}(2026)\citenamefont
  {{MacInnis}}, \citenamefont {{Ange}}, \citenamefont {{Sehgal}}, \citenamefont
  {{Kable}},\ and\ \citenamefont {{Blackstad}}}]{MacInnis2026CMB-HD}%
  \BibitemOpen
  \bibfield  {author} {\bibinfo {author} {\bibfnamefont {A.}~\bibnamefont
  {{MacInnis}}}, \bibinfo {author} {\bibfnamefont {J.}~\bibnamefont {{Ange}}},
  \bibinfo {author} {\bibfnamefont {N.}~\bibnamefont {{Sehgal}}}, \bibinfo
  {author} {\bibfnamefont {J.~A.}\ \bibnamefont {{Kable}}},\ and\ \bibinfo
  {author} {\bibfnamefont {I.}~\bibnamefont {{Blackstad}}},\ }\bibfield
  {title} {\bibinfo {title} {{CMB-HD Foregrounds: Simulations, Source
  Detection, and Foreground Removal}},\ }\href
  {https://doi.org/10.48550/arXiv.2609.16128} {\bibfield  {journal} {\bibinfo
  {journal} {arXiv e-prints}\ ,\ \bibinfo {eid} {arXiv:2609.16128}} (\bibinfo
  {year} {2026})},\ \Eprint {https://arxiv.org/abs/2609.16128}
  {arXiv:2609.16128 [astro-ph.CO]} \BibitemShut {NoStop}%
\bibitem [{\citenamefont {Ma}\ and\ \citenamefont
  {Bertschinger}(1995)}]{Ma:1995ey}%
  \BibitemOpen
  \bibfield  {author} {\bibinfo {author} {\bibfnamefont {C.-P.}\ \bibnamefont
  {Ma}}\ and\ \bibinfo {author} {\bibfnamefont {E.}~\bibnamefont
  {Bertschinger}},\ }\bibfield  {title} {\bibinfo {title} {Cosmological
  {{Perturbation Theory}} in the {{Synchronous}} and {{Conformal Newtonian
  Gauges}}},\ }\href {https://doi.org/10.1086/176550} {\bibfield  {journal}
  {\bibinfo  {journal} {The Astrophysical Journal}\ }\textbf {\bibinfo {volume}
  {455}},\ \bibinfo {pages} {7} (\bibinfo {year} {1995})},\ \Eprint
  {https://arxiv.org/abs/astro-ph/9506072} {arXiv:astro-ph/9506072}
  \BibitemShut {NoStop}%
\bibitem [{\citenamefont {Uhlenbeck}\ and\ \citenamefont
  {Ornstein}(1930)}]{Uhlenbeck:1930zz}%
  \BibitemOpen
  \bibfield  {author} {\bibinfo {author} {\bibfnamefont {G.~E.}\ \bibnamefont
  {Uhlenbeck}}\ and\ \bibinfo {author} {\bibfnamefont {L.~S.}\ \bibnamefont
  {Ornstein}},\ }\bibfield  {title} {\bibinfo {title} {On the {{Theory}} of the
  {{Brownian Motion}}},\ }\href {https://doi.org/10.1103/PhysRev.36.823}
  {\bibfield  {journal} {\bibinfo  {journal} {Physical Review}\ }\textbf
  {\bibinfo {volume} {36}},\ \bibinfo {pages} {823} (\bibinfo {year}
  {1930})}\BibitemShut {NoStop}%
\bibitem [{\citenamefont {Aghanim}\ \emph
  {et~al.}(2020{\natexlab{a}})\citenamefont {Aghanim} \emph
  {et~al.}}]{Planck:2018vyg}%
  \BibitemOpen
  \bibfield  {author} {\bibinfo {author} {\bibfnamefont {N.}~\bibnamefont
  {Aghanim}} \emph {et~al.} (\bibinfo {collaboration} {Planck}),\ }\bibfield
  {title} {\bibinfo {title} {{Planck 2018 results. VI. Cosmological
  parameters}},\ }\href {https://doi.org/10.1051/0004-6361/201833910}
  {\bibfield  {journal} {\bibinfo  {journal} {Astron. Astrophys.}\ }\textbf
  {\bibinfo {volume} {641}},\ \bibinfo {pages} {A6} (\bibinfo {year}
  {2020}{\natexlab{a}})},\ \bibinfo {note} {[Erratum: Astron.Astrophys. 652, C4
  (2021)]},\ \Eprint {https://arxiv.org/abs/1807.06209} {arXiv:1807.06209
  [astro-ph.CO]} \BibitemShut {NoStop}%
\bibitem [{\citenamefont {Peebles}(1968)}]{Peebles:1968ja}%
  \BibitemOpen
  \bibfield  {author} {\bibinfo {author} {\bibfnamefont {P.~J.~E.}\
  \bibnamefont {Peebles}},\ }\bibfield  {title} {\bibinfo {title}
  {Recombination of the {{Primeval Plasma}}},\ }\href
  {https://doi.org/10.1086/149628} {\bibfield  {journal} {\bibinfo  {journal}
  {The Astrophysical Journal}\ }\textbf {\bibinfo {volume} {153}},\ \bibinfo
  {pages} {1} (\bibinfo {year} {1968})}\BibitemShut {NoStop}%
\bibitem [{\citenamefont {Zel'dovich}\ \emph {et~al.}(1969)\citenamefont
  {Zel'dovich}, \citenamefont {Kurt},\ and\ \citenamefont
  {Syunyaev}}]{Zeldovich:1969qmq}%
  \BibitemOpen
  \bibfield  {author} {\bibinfo {author} {\bibfnamefont {{\relax Ya}.~B.}\
  \bibnamefont {Zel'dovich}}, \bibinfo {author} {\bibfnamefont {V.~G.}\
  \bibnamefont {Kurt}},\ and\ \bibinfo {author} {\bibfnamefont {R.~A.}\
  \bibnamefont {Syunyaev}},\ }\bibfield  {title} {\bibinfo {title}
  {Recombination of {{Hydrogen}} in the {{Hot Model}} of the {{Universe}}},\
  }\href@noop {} {\bibfield  {journal} {\bibinfo  {journal} {Sov. Phys. JETP}\
  }\textbf {\bibinfo {volume} {28}},\ \bibinfo {pages} {146} (\bibinfo {year}
  {1969})}\BibitemShut {NoStop}%
\bibitem [{\citenamefont {{Chluba}}\ \emph {et~al.}(2015)\citenamefont
  {{Chluba}}, \citenamefont {{Paoletti}}, \citenamefont {{Finelli}},\ and\
  \citenamefont {{Rubi{\~n}o-Mart{\'\i}n}}}]{Chluba2015PMF}%
  \BibitemOpen
  \bibfield  {author} {\bibinfo {author} {\bibfnamefont {J.}~\bibnamefont
  {{Chluba}}}, \bibinfo {author} {\bibfnamefont {D.}~\bibnamefont
  {{Paoletti}}}, \bibinfo {author} {\bibfnamefont {F.}~\bibnamefont
  {{Finelli}}},\ and\ \bibinfo {author} {\bibfnamefont {J.~A.}\ \bibnamefont
  {{Rubi{\~n}o-Mart{\'\i}n}}},\ }\bibfield  {title} {\bibinfo {title} {{Effect
  of primordial magnetic fields on the ionization history}},\ }\href
  {https://doi.org/10.1093/mnras/stv1096} {\bibfield  {journal} {\bibinfo
  {journal} {\mnras}\ }\textbf {\bibinfo {volume} {451}},\ \bibinfo {pages}
  {2244} (\bibinfo {year} {2015})},\ \Eprint {https://arxiv.org/abs/1503.04827}
  {arXiv:1503.04827 [astro-ph.CO]} \BibitemShut {NoStop}%
\bibitem [{\citenamefont {Chluba}\ and\ \citenamefont
  {Ali-Haïmoud}(2016)}]{Chluba:2015gta}%
  \BibitemOpen
  \bibfield  {author} {\bibinfo {author} {\bibfnamefont {J.}~\bibnamefont
  {Chluba}}\ and\ \bibinfo {author} {\bibfnamefont {Y.}~\bibnamefont
  {Ali-Haïmoud}},\ }\bibfield  {title} {\bibinfo {title} {Cosmospec: Fast and
  detailed computation of the cosmological recombination radiation from
  hydrogen and helium},\ }\href {https://doi.org/10.1093/mnras/stv2691}
  {\bibfield  {journal} {\bibinfo  {journal} {Monthly Notices of the Royal
  Astronomical Society}\ }\textbf {\bibinfo {volume} {456}},\ \bibinfo {pages}
  {3494} (\bibinfo {year} {2016})}\BibitemShut {NoStop}%
\bibitem [{\citenamefont {{Sethi}}\ and\ \citenamefont
  {{Subramanian}}(2005)}]{Sethi2005}%
  \BibitemOpen
  \bibfield  {author} {\bibinfo {author} {\bibfnamefont {S.~K.}\ \bibnamefont
  {{Sethi}}}\ and\ \bibinfo {author} {\bibfnamefont {K.}~\bibnamefont
  {{Subramanian}}},\ }\bibfield  {title} {\bibinfo {title} {{Primordial
  magnetic fields in the post-recombination era and early reionization}},\
  }\href {https://doi.org/10.1111/j.1365-2966.2004.08520.x} {\bibfield
  {journal} {\bibinfo  {journal} {\mnras}\ }\textbf {\bibinfo {volume} {356}},\
  \bibinfo {pages} {778} (\bibinfo {year} {2005})},\ \Eprint
  {https://arxiv.org/abs/astro-ph/0405413} {astro-ph/0405413} \BibitemShut
  {NoStop}%
\bibitem [{\citenamefont {Senatore}\ \emph {et~al.}(2009)\citenamefont
  {Senatore}, \citenamefont {Tassev},\ and\ \citenamefont
  {Zaldarriaga}}]{Senatore:2008vi}%
  \BibitemOpen
  \bibfield  {author} {\bibinfo {author} {\bibfnamefont {L.}~\bibnamefont
  {Senatore}}, \bibinfo {author} {\bibfnamefont {S.}~\bibnamefont {Tassev}},\
  and\ \bibinfo {author} {\bibfnamefont {M.}~\bibnamefont {Zaldarriaga}},\
  }\bibfield  {title} {\bibinfo {title} {Cosmological {{Perturbations}} at
  {{Second Order}} and {{Recombination Perturbed}}},\ }\href
  {https://doi.org/10.1088/1475-7516/2009/08/031} {\bibfield  {journal}
  {\bibinfo  {journal} {Journal of Cosmology and Astroparticle Physics}\
  }\textbf {\bibinfo {volume} {2009}}\bibfield  {number} {\bibinfo  {number} {
  (08)},\ \bibinfo {pages} {031}},\ }\Eprint {https://arxiv.org/abs/0812.3652}
  {arXiv:0812.3652 [astro-ph]} \BibitemShut {NoStop}%
\bibitem [{\citenamefont
  {Hummer}(1994)}]{hummerTotalRecombinationEnergyloss1994}%
  \BibitemOpen
  \bibfield  {author} {\bibinfo {author} {\bibfnamefont {D.~G.}\ \bibnamefont
  {Hummer}},\ }\bibfield  {title} {\bibinfo {title} {Total recombination and
  energy-loss coefficients for hydrogenic ions at low density for {{$10 \le
  T_\mathrm{e} / Z^2 \le 10^7 K$}}},\ }\href
  {https://doi.org/10.1093/mnras/268.1.109} {\bibfield  {journal} {\bibinfo
  {journal} {Monthly Notices of the Royal Astronomical Society}\ }\textbf
  {\bibinfo {volume} {268}},\ \bibinfo {pages} {109} (\bibinfo {year}
  {1994})}\BibitemShut {NoStop}%
\bibitem [{\citenamefont {Seager}\ \emph {et~al.}(2000)\citenamefont {Seager},
  \citenamefont {Sasselov},\ and\ \citenamefont {Scott}}]{Seager:1999km}%
  \BibitemOpen
  \bibfield  {author} {\bibinfo {author} {\bibfnamefont {S.}~\bibnamefont
  {Seager}}, \bibinfo {author} {\bibfnamefont {D.~D.}\ \bibnamefont
  {Sasselov}},\ and\ \bibinfo {author} {\bibfnamefont {D.}~\bibnamefont
  {Scott}},\ }\bibfield  {title} {\bibinfo {title} {How {{Exactly Did}} the
  {{Universe Become Neutral}}?},\ }\href {https://doi.org/10.1086/313388}
  {\bibfield  {journal} {\bibinfo  {journal} {The Astrophysical Journal
  Supplement Series}\ }\textbf {\bibinfo {volume} {128}},\ \bibinfo {pages}
  {407} (\bibinfo {year} {2000})}\BibitemShut {NoStop}%
\bibitem [{\citenamefont {Seljak}\ and\ \citenamefont
  {Zaldarriaga}(1996)}]{Seljak:1996is}%
  \BibitemOpen
  \bibfield  {author} {\bibinfo {author} {\bibfnamefont {U.}~\bibnamefont
  {Seljak}}\ and\ \bibinfo {author} {\bibfnamefont {M.}~\bibnamefont
  {Zaldarriaga}},\ }\bibfield  {title} {\bibinfo {title} {{A Line of sight
  integration approach to cosmic microwave background anisotropies}},\ }\href
  {https://doi.org/10.1086/177793} {\bibfield  {journal} {\bibinfo  {journal}
  {Astrophys. J.}\ }\textbf {\bibinfo {volume} {469}},\ \bibinfo {pages} {437}
  (\bibinfo {year} {1996})},\ \Eprint {https://arxiv.org/abs/astro-ph/9603033}
  {arXiv:astro-ph/9603033} \BibitemShut {NoStop}%
\bibitem [{\citenamefont {Aghanim}\ \emph
  {et~al.}(2020{\natexlab{b}})\citenamefont {Aghanim} \emph
  {et~al.}}]{Planck:2019nip}%
  \BibitemOpen
  \bibfield  {author} {\bibinfo {author} {\bibfnamefont {N.}~\bibnamefont
  {Aghanim}} \emph {et~al.} (\bibinfo {collaboration} {Planck Collaboration}),\
  }\bibfield  {title} {\bibinfo {title} {Planck 2018 results. {{V}}. {{CMB}}
  power spectra and likelihoods},\ }\href
  {https://doi.org/10.1051/0004-6361/201936386} {\bibfield  {journal} {\bibinfo
   {journal} {Astronomy \& Astrophysics}\ }\textbf {\bibinfo {volume} {641}},\
  \bibinfo {pages} {A5} (\bibinfo {year} {2020}{\natexlab{b}})},\ \Eprint
  {https://arxiv.org/abs/1907.12875} {arXiv:1907.12875} \BibitemShut {NoStop}%
\bibitem [{\citenamefont {Naess}\ \emph {et~al.}(2025)\citenamefont {Naess}
  \emph {et~al.}}]{AtacamaCosmologyTelescope:2025vnj}%
  \BibitemOpen
  \bibfield  {author} {\bibinfo {author} {\bibfnamefont {S.}~\bibnamefont
  {Naess}} \emph {et~al.} (\bibinfo {collaboration} {Atacama Cosmology
  Telescope}),\ }\bibfield  {title} {\bibinfo {title} {{The Atacama Cosmology
  Telescope: DR6 maps}},\ }\href
  {https://doi.org/10.1088/1475-7516/2025/11/061} {\bibfield  {journal}
  {\bibinfo  {journal} {JCAP}\ }\textbf {\bibinfo {volume} {11}},\ \bibinfo
  {pages} {061}},\ \Eprint {https://arxiv.org/abs/2503.14451} {arXiv:2503.14451
  [astro-ph.CO]} \BibitemShut {NoStop}%
\bibitem [{\citenamefont {Louis}\ \emph {et~al.}(2025)\citenamefont {Louis}
  \emph {et~al.}}]{AtacamaCosmologyTelescope:2025blo}%
  \BibitemOpen
  \bibfield  {author} {\bibinfo {author} {\bibfnamefont {T.}~\bibnamefont
  {Louis}} \emph {et~al.} (\bibinfo {collaboration} {Atacama Cosmology
  Telescope}),\ }\bibfield  {title} {\bibinfo {title} {{The Atacama Cosmology
  Telescope: DR6 power spectra, likelihoods and {\ensuremath{\Lambda}}CDM
  parameters}},\ }\href {https://doi.org/10.1088/1475-7516/2025/11/062}
  {\bibfield  {journal} {\bibinfo  {journal} {JCAP}\ }\textbf {\bibinfo
  {volume} {11}},\ \bibinfo {pages} {062}},\ \Eprint
  {https://arxiv.org/abs/2503.14452} {arXiv:2503.14452 [astro-ph.CO]}
  \BibitemShut {NoStop}%
\bibitem [{\citenamefont {Abdul~Karim}\ \emph {et~al.}(2025)\citenamefont
  {Abdul~Karim} \emph {et~al.}}]{DESI:2025zgx}%
  \BibitemOpen
  \bibfield  {author} {\bibinfo {author} {\bibfnamefont {M.}~\bibnamefont
  {Abdul~Karim}} \emph {et~al.},\ }\bibfield  {title} {\bibinfo {title} {{{DESI
  DR2}} results. {{II}}. {{Measurements}} of baryon acoustic oscillations and
  cosmological constraints},\ }\href {https://doi.org/10.1103/tr6y-kpc6}
  {\bibfield  {journal} {\bibinfo  {journal} {Phys. Rev. D}\ }\textbf {\bibinfo
  {volume} {112}},\ \bibinfo {pages} {083515} (\bibinfo {year}
  {2025})}\BibitemShut {NoStop}%
\bibitem [{\citenamefont {Breuval}\ \emph {et~al.}(2024)\citenamefont
  {Breuval}, \citenamefont {Riess}, \citenamefont {Casertano}, \citenamefont
  {Yuan}, \citenamefont {Macri}, \citenamefont {Romaniello}, \citenamefont
  {Murakami}, \citenamefont {Scolnic}, \citenamefont {Anand},\ and\
  \citenamefont {Soszy{\'n}ski}}]{Breuval:2024lsv}%
  \BibitemOpen
  \bibfield  {author} {\bibinfo {author} {\bibfnamefont {L.}~\bibnamefont
  {Breuval}}, \bibinfo {author} {\bibfnamefont {A.~G.}\ \bibnamefont {Riess}},
  \bibinfo {author} {\bibfnamefont {S.}~\bibnamefont {Casertano}}, \bibinfo
  {author} {\bibfnamefont {W.}~\bibnamefont {Yuan}}, \bibinfo {author}
  {\bibfnamefont {L.~M.}\ \bibnamefont {Macri}}, \bibinfo {author}
  {\bibfnamefont {M.}~\bibnamefont {Romaniello}}, \bibinfo {author}
  {\bibfnamefont {Y.~S.}\ \bibnamefont {Murakami}}, \bibinfo {author}
  {\bibfnamefont {D.}~\bibnamefont {Scolnic}}, \bibinfo {author} {\bibfnamefont
  {G.~S.}\ \bibnamefont {Anand}},\ and\ \bibinfo {author} {\bibfnamefont
  {I.}~\bibnamefont {Soszy{\'n}ski}},\ }\bibfield  {title} {\bibinfo {title}
  {Small {{Magellanic Cloud Cepheids Observed}} with the {{Hubble Space
  Telescope Provide}} a {{New Anchor}} for the {{SH0ES Distance Ladder}}},\
  }\href {https://doi.org/10.3847/1538-4357/ad630e} {\bibfield  {journal}
  {\bibinfo  {journal} {Astrophys. J.}\ }\textbf {\bibinfo {volume} {973}},\
  \bibinfo {pages} {30} (\bibinfo {year} {2024})}\BibitemShut {NoStop}%
\bibitem [{\citenamefont {{Neal}}(2005)}]{Neal_dragging}%
  \BibitemOpen
  \bibfield  {author} {\bibinfo {author} {\bibfnamefont {R.~M.}\ \bibnamefont
  {{Neal}}},\ }\bibfield  {title} {\bibinfo {title} {{Taking Bigger Metropolis
  Steps by Dragging Fast Variables}},\ }\href
  {https://doi.org/10.48550/arXiv.math/0502099} {\bibfield  {journal} {\bibinfo
   {journal} {arXiv Mathematics e-prints}\ ,\ \bibinfo {eid} {math/0502099}}
  (\bibinfo {year} {2005})},\ \Eprint {https://arxiv.org/abs/math/0502099}
  {arXiv:math/0502099 [math.ST]} \BibitemShut {NoStop}%
\bibitem [{\citenamefont {Lewis}\ and\ \citenamefont
  {Bridle}(2002)}]{Lewis:2002ah}%
  \BibitemOpen
  \bibfield  {author} {\bibinfo {author} {\bibfnamefont {A.}~\bibnamefont
  {Lewis}}\ and\ \bibinfo {author} {\bibfnamefont {S.}~\bibnamefont {Bridle}},\
  }\bibfield  {title} {\bibinfo {title} {{Cosmological parameters from CMB and
  other data: A Monte Carlo approach}},\ }\href
  {https://doi.org/10.1103/PhysRevD.66.103511} {\bibfield  {journal} {\bibinfo
  {journal} {Phys. Rev. D}\ }\textbf {\bibinfo {volume} {66}},\ \bibinfo
  {pages} {103511} (\bibinfo {year} {2002})},\ \Eprint
  {https://arxiv.org/abs/astro-ph/0205436} {arXiv:astro-ph/0205436}
  \BibitemShut {NoStop}%
\bibitem [{\citenamefont {Lewis}(2013)}]{Lewis:2013hha}%
  \BibitemOpen
  \bibfield  {author} {\bibinfo {author} {\bibfnamefont {A.}~\bibnamefont
  {Lewis}},\ }\bibfield  {title} {\bibinfo {title} {{Efficient sampling of fast
  and slow cosmological parameters}},\ }\href
  {https://doi.org/10.1103/PhysRevD.87.103529} {\bibfield  {journal} {\bibinfo
  {journal} {Phys. Rev. D}\ }\textbf {\bibinfo {volume} {87}},\ \bibinfo
  {pages} {103529} (\bibinfo {year} {2013})},\ \Eprint
  {https://arxiv.org/abs/1304.4473} {arXiv:1304.4473 [astro-ph.CO]}
  \BibitemShut {NoStop}%
\bibitem [{\citenamefont {Torrado}\ and\ \citenamefont
  {Lewis}(2021)}]{Torrado:2020dgo}%
  \BibitemOpen
  \bibfield  {author} {\bibinfo {author} {\bibfnamefont {J.}~\bibnamefont
  {Torrado}}\ and\ \bibinfo {author} {\bibfnamefont {A.}~\bibnamefont
  {Lewis}},\ }\bibfield  {title} {\bibinfo {title} {{Cobaya: Code for Bayesian
  Analysis of hierarchical physical models}},\ }\href
  {https://doi.org/10.1088/1475-7516/2021/05/057} {\bibfield  {journal}
  {\bibinfo  {journal} {JCAP}\ }\textbf {\bibinfo {volume} {05}},\ \bibinfo
  {pages} {057}},\ \Eprint {https://arxiv.org/abs/2005.05290} {arXiv:2005.05290
  [astro-ph.IM]} \BibitemShut {NoStop}%
\bibitem [{\citenamefont {G{\"u}nther}\ \emph {et~al.}(2025)\citenamefont
  {G{\"u}nther}, \citenamefont {Balkenhol}, \citenamefont {Fidler},
  \citenamefont {Khalife}, \citenamefont {Lesgourgues}, \citenamefont
  {Mosbech},\ and\ \citenamefont {Sharma}}]{Gunther:2025xrq}%
  \BibitemOpen
  \bibfield  {author} {\bibinfo {author} {\bibfnamefont {S.}~\bibnamefont
  {G{\"u}nther}}, \bibinfo {author} {\bibfnamefont {L.}~\bibnamefont
  {Balkenhol}}, \bibinfo {author} {\bibfnamefont {C.}~\bibnamefont {Fidler}},
  \bibinfo {author} {\bibfnamefont {A.~R.}\ \bibnamefont {Khalife}}, \bibinfo
  {author} {\bibfnamefont {J.}~\bibnamefont {Lesgourgues}}, \bibinfo {author}
  {\bibfnamefont {M.~R.}\ \bibnamefont {Mosbech}},\ and\ \bibinfo {author}
  {\bibfnamefont {R.~K.}\ \bibnamefont {Sharma}},\ }\bibfield  {title}
  {\bibinfo {title} {{{OL\'E}} --- {{Online Learning Emulation}} in
  cosmology},\ }\href {https://doi.org/10.1088/1475-7516/2025/09/059}
  {\bibfield  {journal} {\bibinfo  {journal} {JCAP}\ }\textbf {\bibinfo
  {volume} {09}},\ \bibinfo {pages} {059}}\BibitemShut {NoStop}%
\bibitem [{\citenamefont {Raveri}\ and\ \citenamefont
  {Doux}(2021)}]{Raveri:2021wfz}%
  \BibitemOpen
  \bibfield  {author} {\bibinfo {author} {\bibfnamefont {M.}~\bibnamefont
  {Raveri}}\ and\ \bibinfo {author} {\bibfnamefont {C.}~\bibnamefont {Doux}},\
  }\bibfield  {title} {\bibinfo {title} {{Non-Gaussian estimates of tensions in
  cosmological parameters}},\ }\href
  {https://doi.org/10.1103/PhysRevD.104.043504} {\bibfield  {journal} {\bibinfo
   {journal} {Phys. Rev. D}\ }\textbf {\bibinfo {volume} {104}},\ \bibinfo
  {pages} {043504} (\bibinfo {year} {2021})},\ \Eprint
  {https://arxiv.org/abs/2105.03324} {arXiv:2105.03324 [astro-ph.CO]}
  \BibitemShut {NoStop}%
\bibitem [{\citenamefont {Leizerovich}\ \emph {et~al.}(2024)\citenamefont
  {Leizerovich}, \citenamefont {Landau},\ and\ \citenamefont
  {Sc{\'o}ccola}}]{Leizerovich:2023qqt}%
  \BibitemOpen
  \bibfield  {author} {\bibinfo {author} {\bibfnamefont {M.}~\bibnamefont
  {Leizerovich}}, \bibinfo {author} {\bibfnamefont {S.~J.}\ \bibnamefont
  {Landau}},\ and\ \bibinfo {author} {\bibfnamefont {C.~G.}\ \bibnamefont
  {Sc{\'o}ccola}},\ }\bibfield  {title} {\bibinfo {title} {{Tensions in
  cosmology: A discussion of statistical tools to determine inconsistencies}},\
  }\href {https://doi.org/10.1016/j.physletb.2024.138844} {\bibfield  {journal}
  {\bibinfo  {journal} {Phys. Lett. B}\ }\textbf {\bibinfo {volume} {855}},\
  \bibinfo {pages} {138844} (\bibinfo {year} {2024})},\ \Eprint
  {https://arxiv.org/abs/2312.08542} {arXiv:2312.08542 [astro-ph.CO]}
  \BibitemShut {NoStop}%
\bibitem [{\citenamefont {Khalife}\ \emph {et~al.}(2024)\citenamefont
  {Khalife}, \citenamefont {Zanjani}, \citenamefont {Galli}, \citenamefont
  {G{\"u}nther}, \citenamefont {Lesgourgues},\ and\ \citenamefont
  {Benabed}}]{Khalife:2023qbu}%
  \BibitemOpen
  \bibfield  {author} {\bibinfo {author} {\bibfnamefont {A.~R.}\ \bibnamefont
  {Khalife}}, \bibinfo {author} {\bibfnamefont {M.~B.}\ \bibnamefont
  {Zanjani}}, \bibinfo {author} {\bibfnamefont {S.}~\bibnamefont {Galli}},
  \bibinfo {author} {\bibfnamefont {S.}~\bibnamefont {G{\"u}nther}}, \bibinfo
  {author} {\bibfnamefont {J.}~\bibnamefont {Lesgourgues}},\ and\ \bibinfo
  {author} {\bibfnamefont {K.}~\bibnamefont {Benabed}},\ }\bibfield  {title}
  {\bibinfo {title} {Review of {{Hubble}} tension solutions with new {{SH0ES}}
  and {{SPT-3G}} data},\ }\href {https://doi.org/10.1088/1475-7516/2024/04/059}
  {\bibfield  {journal} {\bibinfo  {journal} {Journal of Cosmology and
  Astroparticle Physics}\ }\textbf {\bibinfo {volume} {2024}}\bibinfo  {number}
  { (04)},\ \bibinfo {pages} {059}}\BibitemShut {NoStop}%
\bibitem [{\citenamefont {Akaike}(1974)}]{Akaike:1974vps}%
  \BibitemOpen
\bibfield  {number} {  }\bibfield  {author} {\bibinfo {author} {\bibfnamefont
  {H.}~\bibnamefont {Akaike}},\ }\bibfield  {title} {\bibinfo {title} {{A new
  look at the statistical model identification}},\ }\href
  {https://doi.org/10.1109/TAC.1974.1100705} {\bibfield  {journal} {\bibinfo
  {journal} {IEEE Trans. Automatic Control}\ }\textbf {\bibinfo {volume}
  {19}},\ \bibinfo {pages} {716} (\bibinfo {year} {1974})}\BibitemShut
  {NoStop}%
\bibitem [{\citenamefont {Koenker}\ and\ \citenamefont
  {Bassett}(1978)}]{Koenker:1978}%
  \BibitemOpen
  \bibfield  {author} {\bibinfo {author} {\bibfnamefont {R.}~\bibnamefont
  {Koenker}}\ and\ \bibinfo {author} {\bibfnamefont {G.}~\bibnamefont
  {Bassett}},\ }\bibfield  {title} {\bibinfo {title} {Regression quantiles},\
  }\href {http://www.jstor.org/stable/1913643} {\bibfield  {journal} {\bibinfo
  {journal} {Econometrica}\ }\textbf {\bibinfo {volume} {46}},\ \bibinfo
  {pages} {33} (\bibinfo {year} {1978})}\BibitemShut {NoStop}%
\bibitem [{\citenamefont {Eilers}\ and\ \citenamefont
  {Marx}(1996)}]{Eilers:1996}%
  \BibitemOpen
  \bibfield  {author} {\bibinfo {author} {\bibfnamefont {P.~H.~C.}\
  \bibnamefont {Eilers}}\ and\ \bibinfo {author} {\bibfnamefont {B.~D.}\
  \bibnamefont {Marx}},\ }\bibfield  {title} {\bibinfo {title} {Flexible
  smoothing with $b$-splines and penalties},\ }\href
  {http://www.jstor.org/stable/2246049} {\bibfield  {journal} {\bibinfo
  {journal} {Statistical Science}\ }\textbf {\bibinfo {volume} {11}},\ \bibinfo
  {pages} {89} (\bibinfo {year} {1996})}\BibitemShut {NoStop}%
\bibitem [{\citenamefont {Abitbol}\ \emph {et~al.}(2025)\citenamefont {Abitbol}
  \emph {et~al.}}]{SimonsObservatory:2025wwn}%
  \BibitemOpen
  \bibfield  {author} {\bibinfo {author} {\bibfnamefont {M.}~\bibnamefont
  {Abitbol}} \emph {et~al.} (\bibinfo {collaboration} {The Simons Observatory
  Collaboration}),\ }\href {https://doi.org/10.48550/arXiv.2503.00636}
  {\bibinfo {title} {The {{Simons Observatory}}: {{Science Goals}} and
  {{Forecasts}} for the {{Enhanced Large Aperture Telescope}}}} (\bibinfo
  {year} {2025}),\ \Eprint {https://arxiv.org/abs/2503.00636} {arXiv:2503.00636
  [astro-ph]} \BibitemShut {NoStop}%
\bibitem [{\citenamefont {Cheslog}\ \emph {et~al.}(2026)\citenamefont
  {Cheslog}, \citenamefont {Finson}, \citenamefont {MacInnis}, \citenamefont
  {Sehgal}, \citenamefont {Afshordi}, \citenamefont {Nerval},\ and\
  \citenamefont {Hlo{\v{z}}ek}}]{Cheslog:2026xyr}%
  \BibitemOpen
  \bibfield  {author} {\bibinfo {author} {\bibfnamefont {Z.}~\bibnamefont
  {Cheslog}}, \bibinfo {author} {\bibfnamefont {E.}~\bibnamefont {Finson}},
  \bibinfo {author} {\bibfnamefont {A.}~\bibnamefont {MacInnis}}, \bibinfo
  {author} {\bibfnamefont {N.}~\bibnamefont {Sehgal}}, \bibinfo {author}
  {\bibfnamefont {N.}~\bibnamefont {Afshordi}}, \bibinfo {author}
  {\bibfnamefont {S.~K.}\ \bibnamefont {Nerval}},\ and\ \bibinfo {author}
  {\bibfnamefont {R.}~\bibnamefont {Hlo{\v{z}}ek}},\ }\href@noop {} {\bibinfo
  {title} {{Current and Future Constraints on the Primordial Power Spectrum}}}
  (\bibinfo {year} {2026}),\ \Eprint {https://arxiv.org/abs/2609.35964}
  {arXiv:2609.35964 [astro-ph.CO]} \BibitemShut {NoStop}%
\bibitem [{\citenamefont {Casertano}\ \emph {et~al.}(2026)\citenamefont
  {Casertano} \emph {et~al.}}]{H0DN:2025lyy}%
  \BibitemOpen
  \bibfield  {author} {\bibinfo {author} {\bibfnamefont {S.}~\bibnamefont
  {Casertano}} \emph {et~al.} (\bibinfo {collaboration} {H0DN Collaboration}),\
  }\bibfield  {title} {\bibinfo {title} {The {{Local Distance Network}}: A
  community consensus report on the measurement of the {{Hubble}} constant at
  1\% precision},\ }\href {https://doi.org/10.1051/0004-6361/202557993}
  {\bibfield  {journal} {\bibinfo  {journal} {Astronomy \& Astrophysics}\
  }\textbf {\bibinfo {volume} {708}},\ \bibinfo {pages} {A166} (\bibinfo {year}
  {2026})},\ \Eprint {https://arxiv.org/abs/2510.23823} {arXiv:2510.23823
  [astro-ph.CO]} \BibitemShut {NoStop}%
\bibitem [{\citenamefont {Quan}\ \emph {et~al.}(2026)\citenamefont {Quan} \emph
  {et~al.}}]{SPT-3G:2026krg}%
  \BibitemOpen
  \bibfield  {author} {\bibinfo {author} {\bibfnamefont {W.}~\bibnamefont
  {Quan}} \emph {et~al.} (\bibinfo {collaboration} {SPT-3G}),\ }\href@noop {}
  {\bibinfo {title} {{SPT-3G D1: Maps of the millimeter-wave sky from 2019 and
  2020 observations of the SPT-3G Main field}}} (\bibinfo {year} {2026}),\
  \Eprint {https://arxiv.org/abs/2603.20163} {arXiv:2603.20163 [astro-ph.CO]}
  \BibitemShut {NoStop}%
\bibitem [{\citenamefont {Jedamzik}\ \emph {et~al.}(2021)\citenamefont
  {Jedamzik}, \citenamefont {Pogosian},\ and\ \citenamefont
  {Zhao}}]{Jedamzik:2020zmd}%
  \BibitemOpen
  \bibfield  {author} {\bibinfo {author} {\bibfnamefont {K.}~\bibnamefont
  {Jedamzik}}, \bibinfo {author} {\bibfnamefont {L.}~\bibnamefont {Pogosian}},\
  and\ \bibinfo {author} {\bibfnamefont {G.-B.}\ \bibnamefont {Zhao}},\
  }\bibfield  {title} {\bibinfo {title} {{Why reducing the cosmic sound horizon
  alone can not fully resolve the Hubble tension}},\ }\href
  {https://doi.org/10.1038/s42005-021-00628-x} {\bibfield  {journal} {\bibinfo
  {journal} {Commun. in Phys.}\ }\textbf {\bibinfo {volume} {4}},\ \bibinfo
  {pages} {123} (\bibinfo {year} {2021})},\ \Eprint
  {https://arxiv.org/abs/2010.04158} {arXiv:2010.04158 [astro-ph.CO]}
  \BibitemShut {NoStop}%
\bibitem [{\citenamefont {Hu}\ and\ \citenamefont {White}(1997)}]{Hu:1996mn}%
  \BibitemOpen
  \bibfield  {author} {\bibinfo {author} {\bibfnamefont {W.}~\bibnamefont
  {Hu}}\ and\ \bibinfo {author} {\bibfnamefont {M.~J.}\ \bibnamefont {White}},\
  }\bibfield  {title} {\bibinfo {title} {{The Damping tail of CMB
  anisotropies}},\ }\href {https://doi.org/10.1086/303928} {\bibfield
  {journal} {\bibinfo  {journal} {Astrophys. J.}\ }\textbf {\bibinfo {volume}
  {479}},\ \bibinfo {pages} {568} (\bibinfo {year} {1997})},\ \Eprint
  {https://arxiv.org/abs/astro-ph/9609079} {arXiv:astro-ph/9609079}
  \BibitemShut {NoStop}%
\bibitem [{\citenamefont {Poulin}\ \emph {et~al.}(2026)\citenamefont {Poulin},
  \citenamefont {Smith}, \citenamefont {Calder{\'o}n},\ and\ \citenamefont
  {Simon}}]{Poulin:2025nfb}%
  \BibitemOpen
  \bibfield  {author} {\bibinfo {author} {\bibfnamefont {V.}~\bibnamefont
  {Poulin}}, \bibinfo {author} {\bibfnamefont {T.~L.}\ \bibnamefont {Smith}},
  \bibinfo {author} {\bibfnamefont {R.}~\bibnamefont {Calder{\'o}n}},\ and\
  \bibinfo {author} {\bibfnamefont {T.}~\bibnamefont {Simon}},\ }\bibfield
  {title} {\bibinfo {title} {{Impact of ACT DR6 and DESI DR2 for early dark
  energy and the Hubble tension}},\ }\href {https://doi.org/10.1103/bx25-1g5d}
  {\bibfield  {journal} {\bibinfo  {journal} {Phys. Rev. D}\ }\textbf {\bibinfo
  {volume} {113}},\ \bibinfo {pages} {063519} (\bibinfo {year} {2026})},\
  \Eprint {https://arxiv.org/abs/2505.08051} {arXiv:2505.08051 [astro-ph.CO]}
  \BibitemShut {NoStop}%
\bibitem [{\citenamefont {Zaldarriaga}\ and\ \citenamefont
  {Harari}(1995)}]{Zaldarriaga:1995gi}%
  \BibitemOpen
  \bibfield  {author} {\bibinfo {author} {\bibfnamefont {M.}~\bibnamefont
  {Zaldarriaga}}\ and\ \bibinfo {author} {\bibfnamefont {D.~D.}\ \bibnamefont
  {Harari}},\ }\bibfield  {title} {\bibinfo {title} {Analytic approach to the
  polarization of the cosmic microwave background in flat and open universes},\
  }\href {https://doi.org/10.1103/PhysRevD.52.3276} {\bibfield  {journal}
  {\bibinfo  {journal} {Physical Review D}\ }\textbf {\bibinfo {volume} {52}},\
  \bibinfo {pages} {3276} (\bibinfo {year} {1995})},\ \Eprint
  {https://arxiv.org/abs/astro-ph/9504085} {arXiv:astro-ph/9504085}
  \BibitemShut {NoStop}%
\bibitem [{\citenamefont {Lynch}\ \emph
  {et~al.}(2024{\natexlab{b}})\citenamefont {Lynch}, \citenamefont {Knox},\
  and\ \citenamefont {Chluba}}]{Lynch:2024hzh}%
  \BibitemOpen
  \bibfield  {author} {\bibinfo {author} {\bibfnamefont {G.~P.}\ \bibnamefont
  {Lynch}}, \bibinfo {author} {\bibfnamefont {L.}~\bibnamefont {Knox}},\ and\
  \bibinfo {author} {\bibfnamefont {J.}~\bibnamefont {Chluba}},\ }\href
  {https://doi.org/10.48550/arXiv.2406.10202} {\bibinfo {title} {{{DESI}} and
  the {{Hubble}} tension in light of modified recombination}} (\bibinfo {year}
  {2024}{\natexlab{b}}),\ \Eprint {https://arxiv.org/abs/2406.10202}
  {arXiv:2406.10202 [astro-ph]} \BibitemShut {NoStop}%
\bibitem [{\citenamefont {Mirpoorian}\ \emph
  {et~al.}(2025{\natexlab{b}})\citenamefont {Mirpoorian}, \citenamefont
  {Jedamzik},\ and\ \citenamefont {Pogosian}}]{Mirpoorian:2025rfp}%
  \BibitemOpen
  \bibfield  {author} {\bibinfo {author} {\bibfnamefont {S.~H.}\ \bibnamefont
  {Mirpoorian}}, \bibinfo {author} {\bibfnamefont {K.}~\bibnamefont
  {Jedamzik}},\ and\ \bibinfo {author} {\bibfnamefont {L.}~\bibnamefont
  {Pogosian}},\ }\bibfield  {title} {\bibinfo {title} {{Is dynamical dark
  energy necessary? DESI BAO and modified recombination}},\ }\href
  {https://doi.org/10.1088/1475-7516/2025/12/050} {\bibfield  {journal}
  {\bibinfo  {journal} {JCAP}\ }\textbf {\bibinfo {volume} {12}},\ \bibinfo
  {pages} {050}},\ \Eprint {https://arxiv.org/abs/2504.15274} {arXiv:2504.15274
  [astro-ph.CO]} \BibitemShut {NoStop}%
\bibitem [{\citenamefont {Loverde}\ and\ \citenamefont
  {Weiner}(2024)}]{Loverde:2024nfi}%
  \BibitemOpen
  \bibfield  {author} {\bibinfo {author} {\bibfnamefont {M.}~\bibnamefont
  {Loverde}}\ and\ \bibinfo {author} {\bibfnamefont {Z.~J.}\ \bibnamefont
  {Weiner}},\ }\bibfield  {title} {\bibinfo {title} {{Massive neutrinos and
  cosmic composition}},\ }\href {https://doi.org/10.1088/1475-7516/2024/12/048}
  {\bibfield  {journal} {\bibinfo  {journal} {JCAP}\ }\textbf {\bibinfo
  {volume} {12}},\ \bibinfo {pages} {048}},\ \Eprint
  {https://arxiv.org/abs/2410.00090} {arXiv:2410.00090 [astro-ph.CO]}
  \BibitemShut {NoStop}%
\bibitem [{\citenamefont {{Lucca}}\ \emph {et~al.}(2024)\citenamefont
  {{Lucca}}, \citenamefont {{Chluba}},\ and\ \citenamefont
  {{Rotti}}}]{Lucca2024CRR}%
  \BibitemOpen
  \bibfield  {author} {\bibinfo {author} {\bibfnamefont {M.}~\bibnamefont
  {{Lucca}}}, \bibinfo {author} {\bibfnamefont {J.}~\bibnamefont {{Chluba}}},\
  and\ \bibinfo {author} {\bibfnamefont {A.}~\bibnamefont {{Rotti}}},\
  }\bibfield  {title} {\bibinfo {title} {{CRRfast: an emulator for the
  cosmological recombination radiation with effects from inhomogeneous
  recombination}},\ }\href {https://doi.org/10.1093/mnras/stae915} {\bibfield
  {journal} {\bibinfo  {journal} {\mnras}\ }\textbf {\bibinfo {volume} {530}},\
  \bibinfo {pages} {668} (\bibinfo {year} {2024})},\ \Eprint
  {https://arxiv.org/abs/2306.08085} {arXiv:2306.08085 [astro-ph.CO]}
  \BibitemShut {NoStop}%
\bibitem [{\citenamefont {Aghanim}\ \emph {et~al.}(2026)\citenamefont
  {Aghanim}, \citenamefont {Maffei} \emph {et~al.}}]{Aghanim:2026fossil}%
  \BibitemOpen
  \bibfield  {author} {\bibinfo {author} {\bibfnamefont {N.}~\bibnamefont
  {Aghanim}}, \bibinfo {author} {\bibfnamefont {B.}~\bibnamefont {Maffei}},
  \emph {et~al.},\ }\href@noop {} {\bibinfo {title} {{FOSSIL: A future mission
  for CMB spectral distortion measurements}}} (\bibinfo {year} {2026}),\
  \Eprint {https://arxiv.org/abs/2609.17618} {arXiv:2609.17618 [astro-ph.CO]}
  \BibitemShut {NoStop}%
\bibitem [{\citenamefont {Cyr}\ \emph {et~al.}(2026)\citenamefont {Cyr} \emph
  {et~al.}}]{Cyr:2026pic}%
  \BibitemOpen
  \bibfield  {author} {\bibinfo {author} {\bibfnamefont {B.}~\bibnamefont
  {Cyr}} \emph {et~al.},\ }\href@noop {} {\bibinfo {title} {{Peering Beyond the
  Veil of Last Scattering: A View of the Universe with CMB Spectral
  Distortions}}} (\bibinfo {year} {2026}),\ \Eprint
  {https://arxiv.org/abs/2609.16194} {arXiv:2609.16194 [astro-ph.CO]}
  \BibitemShut {NoStop}%
\bibitem [{\citenamefont {Mu{\~n}oz}\ \emph {et~al.}(2020)\citenamefont
  {Mu{\~n}oz}, \citenamefont {Dvorkin},\ and\ \citenamefont
  {Cyr-Racine}}]{Munoz:2019hjh}%
  \BibitemOpen
  \bibfield  {author} {\bibinfo {author} {\bibfnamefont {J.~B.}\ \bibnamefont
  {Mu{\~n}oz}}, \bibinfo {author} {\bibfnamefont {C.}~\bibnamefont {Dvorkin}},\
  and\ \bibinfo {author} {\bibfnamefont {F.-Y.}\ \bibnamefont {Cyr-Racine}},\
  }\bibfield  {title} {\bibinfo {title} {{Probing the Small-Scale Matter Power
  Spectrum with Large-Scale 21-cm Data}},\ }\href
  {https://doi.org/10.1103/PhysRevD.101.063526} {\bibfield  {journal} {\bibinfo
   {journal} {Phys. Rev. D}\ }\textbf {\bibinfo {volume} {101}},\ \bibinfo
  {pages} {063526} (\bibinfo {year} {2020})},\ \Eprint
  {https://arxiv.org/abs/1911.11144} {arXiv:1911.11144 [astro-ph.CO]}
  \BibitemShut {NoStop}%
\bibitem [{\citenamefont {Cole}\ and\ \citenamefont
  {Silk}(2021)}]{Cole:2019zhu}%
  \BibitemOpen
  \bibfield  {author} {\bibinfo {author} {\bibfnamefont {P.~S.}\ \bibnamefont
  {Cole}}\ and\ \bibinfo {author} {\bibfnamefont {J.}~\bibnamefont {Silk}},\
  }\bibfield  {title} {\bibinfo {title} {{Small-scale primordial fluctuations
  in the 21 cm Dark Ages signal}},\ }\href
  {https://doi.org/10.1093/mnras/staa3638} {\bibfield  {journal} {\bibinfo
  {journal} {Mon. Not. Roy. Astron. Soc.}\ }\textbf {\bibinfo {volume} {501}},\
  \bibinfo {pages} {2627} (\bibinfo {year} {2021})},\ \Eprint
  {https://arxiv.org/abs/1912.02171} {arXiv:1912.02171 [astro-ph.CO]}
  \BibitemShut {NoStop}%
\bibitem [{\citenamefont {Zhu}\ \emph {et~al.}(2026)\citenamefont {Zhu},
  \citenamefont {Yue}, \citenamefont {Xu}, \citenamefont {Chen},\ and\
  \citenamefont {Huang}}]{Zhu:2026mbp}%
  \BibitemOpen
  \bibfield  {author} {\bibinfo {author} {\bibfnamefont {H.}~\bibnamefont
  {Zhu}}, \bibinfo {author} {\bibfnamefont {B.}~\bibnamefont {Yue}}, \bibinfo
  {author} {\bibfnamefont {Y.}~\bibnamefont {Xu}}, \bibinfo {author}
  {\bibfnamefont {X.}~\bibnamefont {Chen}},\ and\ \bibinfo {author}
  {\bibfnamefont {Z.}~\bibnamefont {Huang}},\ }\bibfield  {title} {\bibinfo
  {title} {{Probing Power Spectrum Enhancement at Small Scales with the SKA}},\
  }\href {https://doi.org/10.3847/1538-4357/ae4aa1} {\bibfield  {journal}
  {\bibinfo  {journal} {Astrophys. J.}\ }\textbf {\bibinfo {volume} {1001}},\
  \bibinfo {pages} {25} (\bibinfo {year} {2026})},\ \Eprint
  {https://arxiv.org/abs/2602.24117} {arXiv:2602.24117 [astro-ph.CO]}
  \BibitemShut {NoStop}%
\bibitem [{\citenamefont {Harris}\ \emph {et~al.}(2020)\citenamefont {Harris}
  \emph {et~al.}}]{Harris:2020xlr}%
  \BibitemOpen
  \bibfield  {author} {\bibinfo {author} {\bibfnamefont {C.~R.}\ \bibnamefont
  {Harris}} \emph {et~al.},\ }\bibfield  {title} {\bibinfo {title} {{Array
  programming with NumPy}},\ }\href {https://doi.org/10.1038/s41586-020-2649-2}
  {\bibfield  {journal} {\bibinfo  {journal} {Nature}\ }\textbf {\bibinfo
  {volume} {585}},\ \bibinfo {pages} {357} (\bibinfo {year} {2020})},\ \Eprint
  {https://arxiv.org/abs/2006.10256} {arXiv:2006.10256 [cs.MS]} \BibitemShut
  {NoStop}%
\bibitem [{\citenamefont {Virtanen}\ \emph {et~al.}(2020)\citenamefont
  {Virtanen} \emph {et~al.}}]{Virtanen:2019joe}%
  \BibitemOpen
  \bibfield  {author} {\bibinfo {author} {\bibfnamefont {P.}~\bibnamefont
  {Virtanen}} \emph {et~al.},\ }\bibfield  {title} {\bibinfo {title} {{SciPy
  1.0--Fundamental Algorithms for Scientific Computing in Python}},\ }\href
  {https://doi.org/10.1038/s41592-019-0686-2} {\bibfield  {journal} {\bibinfo
  {journal} {Nature Meth.}\ }\textbf {\bibinfo {volume} {17}},\ \bibinfo
  {pages} {261} (\bibinfo {year} {2020})},\ \Eprint
  {https://arxiv.org/abs/1907.10121} {arXiv:1907.10121 [cs.MS]} \BibitemShut
  {NoStop}%
\bibitem [{\citenamefont {Hunter}(2007)}]{Hunter:2007ouj}%
  \BibitemOpen
  \bibfield  {author} {\bibinfo {author} {\bibfnamefont {J.~D.}\ \bibnamefont
  {Hunter}},\ }\bibfield  {title} {\bibinfo {title} {{Matplotlib: A 2D Graphics
  Environment}},\ }\href {https://doi.org/10.1109/MCSE.2007.55} {\bibfield
  {journal} {\bibinfo  {journal} {Comput. Sci. Eng.}\ }\textbf {\bibinfo
  {volume} {9}},\ \bibinfo {pages} {90} (\bibinfo {year} {2007})}\BibitemShut
  {NoStop}%
\bibitem [{\citenamefont {Lewis}(2025)}]{Lewis:2019xzd}%
  \BibitemOpen
  \bibfield  {author} {\bibinfo {author} {\bibfnamefont {A.}~\bibnamefont
  {Lewis}},\ }\bibfield  {title} {\bibinfo {title} {{GetDist: a Python package
  for analysing Monte Carlo samples}},\ }\href
  {https://doi.org/10.1088/1475-7516/2025/08/025} {\bibfield  {journal}
  {\bibinfo  {journal} {JCAP}\ }\textbf {\bibinfo {volume} {08}},\ \bibinfo
  {pages} {025}},\ \Eprint {https://arxiv.org/abs/1910.13970} {arXiv:1910.13970
  [astro-ph.IM]} \BibitemShut {NoStop}%
\bibitem [{\citenamefont {{van der Velden}}(2020)}]{cmasher}%
  \BibitemOpen
  \bibfield  {author} {\bibinfo {author} {\bibfnamefont {E.}~\bibnamefont {{van
  der Velden}}},\ }\bibfield  {title} {\bibinfo {title} {{CMasher: Scientific
  colormaps for making accessible, informative and 'cmashing' plots}},\ }\href
  {https://doi.org/10.21105/joss.02004} {\bibfield  {journal} {\bibinfo
  {journal} {The Journal of Open Source Software}\ }\textbf {\bibinfo {volume}
  {5}},\ \bibinfo {eid} {2004} (\bibinfo {year} {2020})},\ \Eprint
  {https://arxiv.org/abs/2003.01069} {arXiv:2003.01069 [eess.IV]} \BibitemShut
  {NoStop}%
\bibitem [{\citenamefont {Dunkley}\ \emph {et~al.}(2013)\citenamefont {Dunkley}
  \emph {et~al.}}]{Dunkley:2013vu}%
  \BibitemOpen
  \bibfield  {author} {\bibinfo {author} {\bibfnamefont {J.}~\bibnamefont
  {Dunkley}} \emph {et~al.},\ }\bibfield  {title} {\bibinfo {title} {{The
  Atacama Cosmology Telescope: likelihood for small-scale CMB data}},\ }\href
  {https://doi.org/10.1088/1475-7516/2013/07/025} {\bibfield  {journal}
  {\bibinfo  {journal} {JCAP}\ }\textbf {\bibinfo {volume} {07}},\ \bibinfo
  {pages} {025}},\ \Eprint {https://arxiv.org/abs/1301.0776} {arXiv:1301.0776
  [astro-ph.CO]} \BibitemShut {NoStop}%
\bibitem [{\citenamefont {Balkenhol}(2025)}]{Balkenhol:2024obe}%
  \BibitemOpen
  \bibfield  {author} {\bibinfo {author} {\bibfnamefont {L.}~\bibnamefont
  {Balkenhol}},\ }\bibfield  {title} {\bibinfo {title} {Compressed
  '{{CMB-lite}}' {{Likelihoods Using Automatic Differentiation}}},\ }\bibfield
  {journal} {\bibinfo  {journal} {The Open Journal of Astrophysics}\ }\textbf
  {\bibinfo {volume} {8}},\ \href {https://doi.org/10.33232/001c.129886}
  {10.33232/001c.129886} (\bibinfo {year} {2025}),\ \Eprint
  {https://arxiv.org/abs/2412.00826} {arXiv:2412.00826 [astro-ph]} \BibitemShut
  {NoStop}%
\end{thebibliography}%
\clearpage

\appendix

\section{Power spectrum, projection, and range of validity}\label{app:validity}

In this appendix we derive the power spectrum implied by our model for the fluctuations (Appendix~\ref{app:power_spectrum}), give the relations used to restate our constraints for other spectral shapes together with the accuracy of that restatement (Appendix~\ref{app:projection}), and estimate the systematic uncertainties in our model at large coherence lengths (Appendix~\ref{app:range_of_validity}). Appendix~\ref{app:error_budget} collects these estimates into the uncertainty shown in Figure~\ref{fig:power_spectrum_constraints}.

\subsection{Power spectrum for OU electron fluctuations}\label{app:power_spectrum}

As described in Section~\ref{sec:model}, we model the fluctuations as an Ornstein--Uhlenbeck stochastic process along the line of sight, which fixes a time-time correlation function for the fluctuations (Equations~\ref{eqn:gaussian_correlator} and \ref{eqn:LN_corr}). These correlation functions, along with the assumptions of homogeneity and isotropy, fix the power spectrum of the electron fluctuations to be a squared Lorentzian (for the Gaussian case) or a sum of squared Lorentzians (for the log-normal case). We now show this.

We assume a slowly varying OU process, so that $\abs{\gamma - \gamma'} = \abs{\eta - \eta'}/\Delta_c$ in Equations~\ref{eqn:gaussian_correlator} and \ref{eqn:LN_corr}, and further assume that $\delta_{\rm e}(\vb{x}, \eta)$ is statistically homogeneous and isotropic. These equations specify the correlation function along a line of sight, but by isotropy we can make the replacement $\abs{\eta - \eta'} \to r \equiv \abs{\vb{x} - \vb{x}'}$. For a homogeneous and isotropic field, the power spectrum is related to the correlation function by
\begin{equation}\label{eqn:power_spectrum_def}
    P(k) = \frac{4 \pi}{k}\int_0^{\infty} \dd r\, r\, C(r) \sin kr\,,
\end{equation}
where $C(r) \equiv \eavg{\delta_{\rm e}(\vb{x}) \delta_{\rm e}(\vb{x}')}$ is the correlation function that appears in Equations~\ref{eqn:gaussian_correlator} or \ref{eqn:LN_corr}. We will use
\begin{equation}
    \int_0^\infty \dd r \,r\, e^{-ar} \sin kr = \frac{2 a k}{(a^2 + k^2)^2}\,, \qfor a>0\,.
\end{equation}

We begin with the Gaussian case, taking $C(r) = \sigma_{\rm e}^2 e^{-r / \Delta_c}$. Equation~\ref{eqn:power_spectrum_def} gives
\begin{equation}\label{eqn:pk_gaussian}
    P(k) = \frac{8 \pi \sigma_{\rm e}^2 \Delta_c ^3}{(1 + k^2 \Delta_c^2)^2}\,. \qq{(Gaussian)}
\end{equation}
The spectrum is white, $P \to 8\pi\sigma_{\rm e}^2\Delta_c^3$, for $k\Delta_c \ll 1$ and falls as $k^{-4}$ for $k\Delta_c \gg 1$. In terms of the dimensionless power $\Delta_{\rm e}^2 = k^3 P(k) / (2\pi^2)$ and $x \equiv k \Delta_c$, this is
\begin{equation}\label{eqn:dimless_power_gaussian}
    \Delta_{\rm e}^2 (x) = \frac{4 \sigma_{\rm e}^2}{\pi} \frac{x^3}{(1 + x^2)^2}\,.\qq{(Gaussian)}
\end{equation}
The dimensionless power peaks at $k = \sqrt{3}/\Delta_c$.

\begin{figure}[t]
    \centering
    \includegraphics{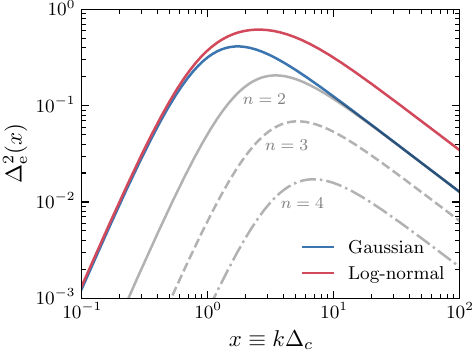}
    \caption{Dimensionless power spectra of Gaussian (blue) and log-normal (red) electron density fluctuations. We also show a few of the individual terms from the harmonic expansion of Equation~\ref{eqn:dimless_power_lognormal}, illustrating that the higher harmonics inject power at smaller scales. The $n=1$ term would coincide with the Gaussian curve. All curves have $\sigma_{\rm e}^2 = 1$.}\label{fig:pk_comparison}
\end{figure}

For the log-normal case, we begin by expanding the exponential:
\begin{align}
    C(r) &= \exp(\sigma_{\rm e}^2 e^{-r / \Delta_c}) - 1 \\
    &= \sum_{n=1}^{\infty} \frac{\sigma_{\rm e}^{2n}}{n!} e^{-nr / \Delta_c}\,.
\end{align}
Each term is a Gaussian-case correlation function with a rescaled correlation length $\Delta_c / n$, so we can use Equation~\ref{eqn:pk_gaussian} term by term. We find
\begin{align}
    P(k) &= \sum_{n=1}^{\infty} \frac{\sigma_{\rm e}^{2n}}{n!} \frac{8 \pi (\Delta_c/n)^3}{(1 + k^2 \Delta_c^2/n^2)^2} \\
    &= 8 \pi \Delta_c^3 \sum_{n=1}^{\infty} \frac{\sigma_{\rm e}^{2n}}{n!\,n^3} \frac{1}{(1 + k^2 \Delta_c^2/n^2)^2}\,, \qq{(LN)} \label{eqn:pk_lognormal}
\end{align}
and, converting to dimensionless power,
\begin{equation}\label{eqn:dimless_power_lognormal}
    \Delta_{\rm e}^2(x) = \frac{4}{\pi}\sum_{n=1}^{\infty}\frac{\sigma_{\rm e}^{2n}}{n!\,n^3} \frac{x^3}{(1+x^2/n^2)^2}\,. \qq{(LN)}
\end{equation}
The higher harmonics in the expansion inject power at smaller correlation lengths $\Delta_c / n$, weighted by $\sigma_{\rm e}^{2n}/(n!\, n^3)$. As anticipated in Section~\ref{sec:motivation}, in the limit $\sigma_{\rm e}^2 \to 0$ the $n=1$ term dominates and we recover the Gaussian result. We illustrate these power spectra in Figure~\ref{fig:pk_comparison}.

\subsection{Projection onto other spectral shapes}\label{app:projection}

Inverting Equation~\ref{eqn:power_spectrum_def} gives the correlation function in terms of the dimensionless power spectrum,
\begin{equation}\label{eqn:corr_from_power}
    C(r) = \int \Delta_{\rm e}^2(k)\, \frac{\sin kr}{kr}\, \dd\ln k\,.
\end{equation}
Setting $r=0$ gives Equation~\ref{eqn:proj_var}. Integrating over $r$ and using $\int_0^\infty (\sin kr / kr)\,\dd r = \pi/2k$ gives Equation~\ref{eqn:proj_corr}, since $\eavg{\delta_{\rm e}^2}\eta_c^{\rm phys} = \int_0^\infty C(r)\,\dd r$.

We write an assumed spectrum as its variance times a normalized shape,
\begin{equation}
    \Delta_{\rm e}^2(k) = \eavg{\delta_{\rm e}^2}\, W(k)\,, \qquad \int W(k)\, \dd\ln k = 1\,,
\end{equation}
so that Equation~\ref{eqn:proj_var} is satisfied identically and Equation~\ref{eqn:proj_corr} becomes
\begin{equation}\label{eqn:eta_from_shape}
    \eta_c^{\rm phys} = \frac{\pi}{2}\int \frac{W(k)}{k}\, \dd\ln k\,.
\end{equation}
The physical coherence length is thus $\pi/2$ times the power-weighted mean of $1/k$, or equivalently one quarter of the power-weighted mean wavelength $2\pi/k$. A given shape has one free scale, and Equation~\ref{eqn:eta_from_shape} fixes it in terms of $\eta_c^{\rm phys}$. For a $\delta$-function spike, $W = \delta_{\rm D}(\ln k - \ln k_s)$, this gives
\begin{equation}
    k_s = \frac{\pi}{2\eta_c^{\rm phys}}\,.
\end{equation}
For a spectrum that is log-normal in $k$ with width $\sigma_s$,
\begin{equation}\label{eqn:lognormal_spectrum}
    W(k) = \frac{1}{\sqrt{2\pi}\sigma_s} \exp\left[-\frac{\ln^2(k/k_s)}{2\sigma_s^2}\right]\,,
\end{equation}
the integral in Equation~\ref{eqn:eta_from_shape} is the mean of $e^{-\ln k}$ over a normal distribution in $\ln k$, equal to $e^{\sigma_s^2/2}/k_s$, so that
\begin{equation}
    k_s = \frac{\pi}{2}\frac{e^{\sigma_s^2/2}}{\eta_c^{\rm phys}}\,.
\end{equation}
In Figure~\ref{fig:power_spectrum_constraints} we show the limit on $\eavg{\delta_{\rm e}^2}$ as a function of $k_s$ for both spectra; the two differ only in where a given $\eta_c^{\rm phys}$ falls in $k_s$, by the factor $e^{\sigma_s^2/2}$.

\subsection{Range of validity}\label{app:range_of_validity}

We now estimate the range of scales over which our model provides a good description of the electron fluctuations. The conditions at large scales concern the transport channel, while at small scales the condition concerns the recombination channel.

\subsubsection{Breakdown at large scales}

In the presence of electron fluctuations, we may write the photon distribution $\Theta = \bar{\Theta} + \var{\Theta}$, where $\var{\Theta}$ is the part of the distribution sourced by scattering with $\delta_{\rm e}$. The observables (i.e., the $C_{\ell}$'s) are quadratic in $\Theta$, so the true quantity of interest when comparing to data is the variance
\begin{equation}\label{eqn:power_decomposition}
    \eavg{\Theta\cdot\Theta} = \bar{\Theta}\cdot\bar{\Theta} + \eavg{\delta\Theta\cdot\delta\Theta}\,.
\end{equation}
As the coherence length $\Delta_c$ grows, the spectra computed by \textsc{classito} depart from Equation~\ref{eqn:power_decomposition} in three ways:

\begin{enumerate}[label=(\roman*)]
    \item \textsc{classito} assumes that quantities such as $\Theta$ are slowly varying over the fast timescale set by $\Delta_c$: this is the separation of scales assumption that enters in the $G(\eta, \eta') \simeq 1$ step of deriving Equation~\ref{eqn:heuristic_correction}. For large $\Delta_c$, this approximation becomes invalid, and the evolution of $\bar{\Theta}$ computed by \textsc{classito} departs from the truth.
    \item \textsc{classito} computes the evolution of the photon distribution averaged over the electron fluctuations, $\bar{\Theta} \equiv \eavg{\Theta}$, and omits the second term in Equation~\ref{eqn:power_decomposition}. CVB note that this term is of the same order as the transport corrections included in the evolution of $\bar{\Theta}$ but defer its computation \citep[Section~3.5 of][]{Chluba:2025jqt}. Below, we argue that the dropped term contributes a shot-noise-like power for long coherence lengths.
    \item For large $\Delta_c$, a single photon mode samples fewer independent patches of the electron field, so its true evolution has some scatter around the ensemble-averaged evolution.
\end{enumerate}

We estimate each of these effects in turn. 

\paragraph{Evolution of the mean ---} Let $\vb{X}$ be the state vector of the photon--baryon system at fixed $\vb{k}$, $\vb{X} = (F_{\gamma \ell}, G_{\gamma \ell}, \delta_{\rm b}, \theta_{\rm b}, \Phi, \Psi)^{\intercal}$, and write it as a mean and fluctuating component $\vb{X} = \bar{\vb{X}} + \var{\vb{X}}$. In this notation, the electron-ensemble-averaged Boltzmann equation (Equation~\ref{eqn:eavg_boltzmann}) is:
\begin{equation}\label{eqn:mean_operator_form}
    \partial_\eta \bar{\vb{X}} = \mathcal{M}\bar{\vb{X}} - \bar{\Gamma}\eavg{\delta_{\rm e}\,\mathcal{C}\,\delta\vb{X}}\,, \qquad \mathcal{M} \equiv \mathcal{K} - \bar{\Gamma}\mathcal{C}\,,
\end{equation}
where $\mathcal{M}$ is the evolution operator of the mean medium, $\mathcal{C}$ is the collision operator of Equation~\ref{eqn:collision_kernel}, now a matrix acting on $\vb{X}$ (we write its action by juxtaposition), and $\mathcal{K}$ contains the remaining free-streaming, gravitational, and baryon terms. To first order in $\delta_{\rm e}$ the fluctuation obeys 
\begin{equation}\label{eqn:fluc_eq}
    \partial_\eta \delta\vb{X} - \mathcal{M}\,\delta\vb{X} = -\bar{\Gamma}\,\delta_{\rm e}\,\mathcal{C}\bar{\vb{X}},
\end{equation}
whose solution is Equation~\ref{eqn:dtheta_solution} with the Green's function now a matrix. That is, $G(\eta,\eta')$ solves $\partial_\eta G = \mathcal{M}G$ with $G(\eta',\eta')=1$ and carries the state of the mean medium from $\eta'$ to $\eta$. Inserting this solution into the last term of Equation~\ref{eqn:mean_operator_form} and using stationarity, $\eavg{\delta_{\rm e}(\eta)\delta_{\rm e}(\eta-r)} = C(r)$,
\begin{equation}\label{eqn:correction_kernel_def}
    -\bar{\Gamma}\eavg{\delta_{\rm e}\,\mathcal{C}\,\delta\vb{X}} = \bar{\Gamma}^2\int_0^\infty \dd r\; C(r)\,\mathcal{J}(r)\,,
\end{equation}
where we have defined the kernel
\begin{equation}\label{eqn:kernel_def}
    \mathcal{J}(r) \equiv \mathcal{C}\,G(r)\,\mathcal{C}\,\bar{\vb{X}}(\eta - r)\,,
\end{equation}
with $G(r) \equiv G(\eta, \eta-r)$. Read from the right, $\mathcal{J}$ describes the perturbation: it is sourced at $\eta - r$ in proportion to the departure of the mean from equilibrium with the scatterers, $\mathcal{C}\bar{\vb{X}}$; it is evolved to $\eta$ by $G(r)$; and the correlated fluctuation at $\eta$ scatters it, the outer $\mathcal{C}$.

The correlation function has width $\Delta_c$, while $\mathcal{J}$ varies on the scales $1/\bar{\Gamma}$ and $1/k$ of the photon evolution, so the kernel is smooth across the support of $C$. We therefore expand $\mathcal{J}(r) = \mathcal{J}(0) + r \mathcal{J}'(0) + \dots$ and insert into the integrand of Equation~\ref{eqn:correction_kernel_def},
\begin{equation}
    \begin{split}
    \int_0^\infty \dd r\; C(r)\mathcal{J}(r) &= \mathcal{J}(0) \int_0^\infty \dd r\; C(r) \\
    &\qquad + \mathcal{J}'(0) \int_0^\infty \dd r\; rC(r) + \dots
    \end{split}
\end{equation}
Defining $\bar{r} \equiv \left(\int_0^\infty r C\,\dd r\right) /  \left(\int_0^\infty C\,\dd r\right)$ we can rewrite Equation~\ref{eqn:correction_kernel_def} as
\begin{equation}\label{eqn:moment_expansion}
    -\bar{\Gamma}\eavg{\delta_{\rm e}\,\mathcal{C}\,\delta\vb{X}} = \bar{\Gamma}^2 \eavg{\delta_{\rm e}^2}\eta_c^{\rm phys} \left(\mathcal{J}(0) + \bar{r}\mathcal{J}'(0) + \dots \right),
\end{equation}
with $\eavg{\delta_{\rm e}^2}\eta_c^{\rm phys} = \int_0^\infty C\,\dd r$ by Equation~\ref{eqn:phys_corr}. The separation of scales assumption comes from approximating $\mathcal{J}(r) \approx \mathcal{J}(0)$, which neglects the evolution of the sourced perturbation between scatters, and results in Equation~\ref{eqn:heuristic_correction} since $\mathcal{J}(0) = \mathcal{C}^2 \bar{\vb{X}}(\eta)$. As $\Delta_c$ grows, this approximation is no longer good, and the correction is thus $\mathcal{J}(r) - \mathcal{J}(0) \approx r \mathcal{J}'(0)$ at leading order in $r$.

To estimate $\mathcal{J}'(0)$, we write the kernel in terms of $\bar{\vb{X}}(\eta)$, so that $\mathcal{J}(r)$ is directly comparable to $\mathcal{J}(0)$. At zeroth order in $\delta_{\rm e}$ the mean is itself evolved by $G$, so $\bar{\vb{X}}(\eta - r) = G^{-1}(r)\,\bar{\vb{X}}(\eta)$ and
\begin{equation}\label{eqn:operator_evolution}
    \mathcal{J}(r) = \mathcal{C}\left(G(r)\,\mathcal{C}\,G^{-1}(r)\right)\bar{\vb{X}}(\eta)\,.
\end{equation}
Comparing to $\mathcal{J}(0) = \mathcal{C}\,\mathcal{C}\,\bar{\vb{X}}(\eta)$, the inner operators $G\mathcal{C}G^{-1}$ and $\mathcal{C}$ differ to the extent that $\comm{G}{\mathcal{C}} \neq 0$. These operators fail to commute precisely when the evolution of $\bar{\vb{X}}$ over an interval $r$ becomes important: it no longer suffices to approximate the perturbation as being sourced at $\eta$, the time of the second scatter. Treating $\mathcal{M}$ as constant over the interval $r$, so that $G(r) \simeq e^{\mathcal{M}r}$, we expand
\begin{equation}
    G(r)\mathcal{C}G^{-1}(r) = \mathcal{C} + r \comm{\mathcal{M}}{\mathcal{C}} + \dots
\end{equation}
and, noting that $\comm{\mathcal{M}}{\mathcal{C}} = \comm{\mathcal{K}}{\mathcal{C}}$ so that $\mathcal{J}'(0) = \mathcal{C}\comm{\mathcal{K}}{\mathcal{C}}\bar{\vb{X}}(\eta)$, we arrive at 
\begin{equation}\label{eqn:next_order}
    -\bar{\Gamma}\eavg{\delta_{\rm e}\,\mathcal{C}\,\delta\vb{X}} = \bar{\Gamma}^2\eavg{\delta_{\rm e}^2}\eta_c^{\rm phys} \left(\mathcal{C}^2 + \bar{r}\mathcal{C}\comm{\mathcal{K}}{\mathcal{C}} + \dots \right)\bar{\vb{X}}(\eta)\,.
\end{equation}
The ratio of the second term to the first is the fractional error of our calculation, so we must evaluate the commutator.

\begin{figure}[t]
    \centering
    \includegraphics{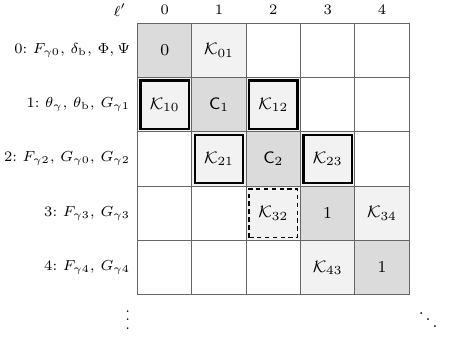}
    \caption{Block structure of the collision operator $\mathcal{C}$ (diagonal blocks) and the streaming operator $\mathcal{K}$ (off-diagonal blocks $\mathcal{K}_{\ell\ell'}$, mapping block $\ell'$ into block $\ell$), with the components of $\vb{X}$ ordered by the multipole at which scattering couples them (rows). We do not show the $\mathcal{K}_{11}$ block. The blocks with a solid outline build the corrections to the dipole and quadrupole, Equations~\ref{eqn:commutator_block1} and \ref{eqn:commutator_block2}; $\mathcal{K}_{32}$ (dashed) builds the correction to the octupole, which we neglect in tight coupling. Blocks linking unit blocks cancel in the sum.}\label{fig:commutator_blocks}
\end{figure}

We proceed by first writing
\begin{equation}\label{eqn:commutator_split}
    \mathcal{C}\comm{\mathcal{K}}{\mathcal{C}}\bar{\vb{X}} = \mathcal{C}\mathcal{K}\left(\mathcal{C}\bar{\vb{X}}\right) - \mathcal{C}^2\left(\mathcal{K}\bar{\vb{X}}\right)\,.
\end{equation}
The first term streams the perturbation, which is created in proportion to the departure from kinetic equilibrium $\var{\vb{X}} \propto \mathcal{C}\bar{\vb{X}}$; the second streams the mean and projects afterwards. To compare them we order the components of $\vb{X}$ by the multipole at which scattering couples them. The collision terms of Equation~\ref{eqn:collision_kernel} give
\begin{equation}\label{eqn:collision_blocks}
    \begin{gathered}
        \mathcal{C} = 0\oplus\mathsf{C}_1\oplus\mathsf{C}_2\oplus\mathbf{1}\oplus\mathbf{1}\oplus\cdots,\\[6pt]
        \mathsf{C}_1 = \begin{pmatrix} 1 & -1 & 0 \\ -R & R & 0 \\ 0 & 0 & 1 \end{pmatrix},\qquad
        \mathsf{C}_2 = \begin{pmatrix} \tfrac{9}{10} & -\tfrac{1}{10} & -\tfrac{1}{10} \\[2pt] -\tfrac{1}{2} & \tfrac{1}{2} & -\tfrac{1}{2} \\[2pt] -\tfrac{1}{10} & -\tfrac{1}{10} & \tfrac{9}{10} \end{pmatrix},
    \end{gathered}
\end{equation}
with $\mathsf{C}_1$ acting on $(\theta_\gamma,\theta_{\rm b}, G_{\gamma 1})$ and $\mathsf{C}_2$ on $(F_{\gamma2}, G_{\gamma0}, G_{\gamma2})$, which scattering couples through $\Pi = F_{\gamma2} + G_{\gamma0} + G_{\gamma2}$. We write $\bar{\vb{X}}_\ell$ for the part of $\bar{\vb{X}}$ in block $\ell$ and $\mathcal{K}_{\ell\ell'}$ for the part of $\mathcal{K}$ mapping block $\ell'$ into block $\ell$ (see Figure~\ref{fig:commutator_blocks}). Streaming couples only neighboring blocks; the one diagonal block of $\mathcal{K}$ is $\mathcal{K}_{11}$, the Hubble drag on the baryons, and the sums below run over $\ell' \in \{\ell-1, \ell, \ell+1\}$. With these definitions, Equation~\ref{eqn:commutator_split} in block $\ell$ reads
\begin{equation}\label{eqn:commutator_blocks}
    \left(\mathcal{C}\comm{\mathcal{K}}{\mathcal{C}}\bar{\vb{X}}\right)_\ell
    = \mathsf{C}_\ell \sum_{\ell' = \ell,\ell\pm1} \left(\mathcal{K}_{\ell\ell'}\,\mathsf{C}_{\ell'} - \mathsf{C}_\ell\,\mathcal{K}_{\ell\ell'}\right)\bar{\vb{X}}_{\ell'}\,.
\end{equation}
The bracket vanishes when both $\mathsf{C}$ blocks are the identity ($\ell,\ell'\geq3$), and for $\mathsf{C}_0 = 0$. The correction is thus confined to the dipole and quadrupole, and to the octupole through $\mathcal{K}_{32}$, which we address below (Figure~\ref{fig:commutator_blocks}). Written out, with $\mathsf{C}_0 = 0$ and $\mathsf{C}_3 = \mathbf{1}$,
\begin{align}
    \begin{split}
    \left(\mathcal{C}\comm{\mathcal{K}}{\mathcal{C}}\bar{\vb{X}}\right)_1 &= -\mathsf{C}_1^2\,\mathcal{K}_{10}\bar{\vb{X}}_0 + \mathsf{C}_1\left(\mathcal{K}_{11}\mathsf{C}_1 - \mathsf{C}_1\mathcal{K}_{11}\right)\bar{\vb{X}}_1 \\
    &\qquad + \mathsf{C}_1\left(\mathcal{K}_{12}\mathsf{C}_2 - \mathsf{C}_1\mathcal{K}_{12}\right)\bar{\vb{X}}_2\,, \label{eqn:commutator_block1}
    \end{split}\\
    \begin{split}
    \left(\mathcal{C}\comm{\mathcal{K}}{\mathcal{C}}\bar{\vb{X}}\right)_2 &= \mathsf{C}_2\left[\mathcal{K}_{21}\mathsf{C}_1\bar{\vb{X}}_1 - \mathsf{C}_2\mathcal{K}_{21}\bar{\vb{X}}_1 \right. \\ 
    &\qquad \left. + (\mathbf{1} - \mathsf{C}_2)\mathcal{K}_{23}\bar{\vb{X}}_3\right]\,. \label{eqn:commutator_block2}
    \end{split}
\end{align}
The first term of Equation~\ref{eqn:commutator_block1} is the streaming of the mean field's monopole, generating a dipole. However, because Thomson scattering conserves photon number, the perturbation has no monopole and therefore its dipole lacks this source. Similarly, in Equation~\ref{eqn:commutator_block2}, the second term represents the streaming of the mean field's dipole into a quadrupole. Since Thomson scattering is internal to the fluid, it cannot generate a bulk velocity, and $\var \vb{X}_1 \propto \mathsf{C}_1 \bar{\vb{X}}_1 \propto (\theta_\gamma - \theta_{\rm b})$. Therefore, while streaming does source a quadrupole in the perturbation, its source is much reduced in magnitude compared to the source of the mean field's quadrupole, which includes contributions from the bulk velocity. This asymmetry between the two terms is the reason for the correction.

To evaluate these terms we need the relative sizes of the different $\bar{\vb{X}}_\ell$'s. At zeroth order in $\delta_{\rm e}$, $\bar{\vb{X}}$ obeys $\partial_\eta\bar{\vb{X}} = \mathcal{M}\bar{\vb{X}}$, so in block $\ell$,
\begin{equation}\label{eqn:mean_block_eq}
    \partial_\eta\bar{\vb{X}}_\ell = \sum_{\ell'=\ell,\ell\pm1}\mathcal{K}_{\ell\ell'}\bar{\vb{X}}_{\ell'} - \bar{\Gamma}\,\mathsf{C}_\ell\bar{\vb{X}}_\ell\,.
\end{equation}
Scattering alone relaxes each eigenvector of $\mathsf{C}_\ell$ with eigenvalue $c$ at the rate $\nu = c\,\bar{\Gamma}$:
\begin{equation}\label{eqn:tc_rates}
    \nu_1 = (1+R)\,\bar{\Gamma}\,, \qquad \nu_2 = \bar{\Gamma}\ \text{or}\ 0.3\,\bar{\Gamma}\,,
\end{equation}
for the eigenvectors of $\mathsf{C}_1$ and $\mathsf{C}_2$. While the photons are tightly coupled, these rates exceed the rate $k$ at which streaming changes, $\nu\gg k$. To see what this implies, apply $\mathsf{C}_\ell$ to Equation~\ref{eqn:mean_block_eq}. Since $\mathsf{C}_\ell$ and $\bar{\Gamma}$ vary slowly compared with $1/\nu$, we find 
\begin{equation}\label{eqn:departure_eq}
    \partial_\eta\left(\mathsf{C}_\ell\bar{\vb{X}}_\ell\right) = \mathsf{C}_\ell\sum_{\ell'=\ell,\ell\pm1}\mathcal{K}_{\ell\ell'}\bar{\vb{X}}_{\ell'} - \bar{\Gamma}\,\mathsf{C}_\ell^2\bar{\vb{X}}_\ell\,.
\end{equation}
On an eigenvector of $\mathsf{C}_\ell$ with rate $\nu$, writing $a = \mathsf{C}_\ell \bar{\vb{X}}_\ell$ for clarity and $S$ for the streaming term, Equation~\ref{eqn:departure_eq} reads $\partial_\eta a = S - \nu a$, with solution
\begin{equation}\label{eqn:departure_solution}
    a(\eta) = \int^\eta \dd\eta'\, e^{-\nu(\eta-\eta')}\,S(\eta') = \frac{S}{\nu} - \frac{\dot{S}}{\nu^2} + \dots
\end{equation}
The source varies at the acoustic rate, $\dot{S}\sim kS$, so $a\simeq S/\nu$ up to corrections of relative order $k/\nu$. To leading order in $k/\nu$ we may therefore treat $a$ as constant and write Equation~\ref{eqn:departure_eq} as
\begin{equation}\label{eqn:tc_balance}
    \mathsf{C}_\ell\Big(\sum_{\ell'=\ell,\ell\pm1}\mathcal{K}_{\ell\ell'}\bar{\vb{X}}_{\ell'}\Big) \simeq \bar{\Gamma}\,\mathsf{C}_\ell^2\bar{\vb{X}}_\ell\,, \qquad \ell \geq 1\,.
\end{equation}
This is the first-order tight-coupling approximation: in block $1$ it gives the slip, $\theta_\gamma-\theta_{\rm b}\simeq k^2\delta_\gamma/4\nu_1$, in block $2$ the shear, $\sigma_\gamma\simeq(16/45)\,\theta_\gamma/\bar{\Gamma}$, and in each higher block a multipole smaller than the last by $k/\bar{\Gamma}$. As expected in tight coupling, the mean field is thus a monopole and a bulk velocity, with any departure from kinetic equilibrium with the scatterers suppressed by $k/\nu$. In Equations~\ref{eqn:commutator_block1} and \ref{eqn:commutator_block2} the terms involving the slip, $\bar{\vb{X}}_2$, or $\bar{\vb{X}}_3$ are therefore subleading. What remains, $-\mathsf{C}_1^2\left(\mathcal{K}_{10}\bar{\vb{X}}_0 + \mathcal{K}_{11}\bar{\vb{X}}_1\right)$ and $-\mathsf{C}_2^2\mathcal{K}_{21}\bar{\vb{X}}_1$, is $-\mathsf{C}_\ell$ times the left side of Equation~\ref{eqn:tc_balance}, so we arrive at
\begin{align*}
    \left(\mathcal{C}\comm{\mathcal{K}}{\mathcal{C}}\bar{\vb{X}}\right)_\ell &\simeq -\mathsf{C}_\ell\big(\bar{\Gamma}\,\mathsf{C}_\ell^2\bar{\vb{X}}_\ell\big) \\
    &= -\bar{\Gamma}\,\mathsf{C}_\ell^3\,\bar{\vb{X}}_\ell \qquad (\ell = 1,2)\numberthis\label{eqn:tc_commutator}
\end{align*}
On an eigenvector of $\mathsf{C}_\ell$ with rate $\nu$, the second term of Equation~\ref{eqn:next_order} is therefore $-\nu\bar{r}$ times the first: the perturbation sourced by one fluctuation decays at the rate $\nu$ before the photons reach the correlated fluctuation, and the leading term neglects this decay. The correction acts on the dipole, with $\nu_1=(1+R)\bar{\Gamma}$, and on the quadrupole, with $\nu_2=\bar{\Gamma}$ or $0.3\,\bar{\Gamma}$. We neglect the corresponding term for the octupole, $\mathcal{K}_{32}(\mathsf{C}_2 - \mathbf{1})\bar{\vb{X}}_2$, which feeds back into the quadrupole only at relative order $(k/\bar{\Gamma})^2$.

To estimate how this affects our limits, we treat spectra that include the correction as the truth and ask how far our model, which omits it, misestimates the transport amplitude. We compute such spectra with \textsc{classito} by dividing each shape function acting on the dipole and quadrupole by $1+\nu\bar{r}$, i.e., $f_S(\tau_\ell,\sigma_{\rm e}^2)\to f_S(\tau_\ell,\sigma_{\rm e}^2)/(1+\tau_\ell\bar{r}/\Delta_c)$, and leaving the corrections to $\ell'\geq3$ unchanged. This form agrees with Equation~\ref{eqn:next_order} to first order in $\bar{r}$, remains positive when $\nu\bar{r}$ is not small, and for an exponential correlation function ($\bar{r}=\Delta_c$) is the suppression CVB find when they retain $\mathcal{K}$ in the moment equations \citep[Section~6.1.1 of][]{Chluba:2025jqt}. At fixed $\Delta_c$, as for our conditional limits, we then fit these spectra with the template of our model, $S_{\rm tr}=\partial C_\ell/\partial\zeta_{\rm e}$ computed with the uncorrected shape functions, marginalizing over \lcdm, $\sigma_{\rm e}^2$, and the calibrations (Equation~\ref{eqn:fisher_bias}). Because the true corrections are weaker, the fit returns a smaller $\zeta_{\rm e}$ than the input, and our limits on the transport channel are too tight by the same fraction, $\epsilon_{\rm mean}$. For the baseline model $\epsilon_{\rm mean}$ is proportional to $\bar{r}$, with a coefficient of $0.11$--$0.18\,{\rm Mpc}^{-1}$ over $0.05\leq\Delta_c\leq2$~Mpc for either dataset. This coefficient is an average of the rates at $z_{\rm ref}$, $\nu_1\approx0.18\,{\rm Mpc}^{-1}$ and $\nu_2\approx0.02$--$0.07\,{\rm Mpc}^{-1}$, weighted by the sensitivity of the data. We adopt
\begin{equation}\label{eqn:eps_mean}
    \epsilon_{\rm mean} = \nu_\star\,\bar{r}\,,\qquad \nu_\star = 0.2\,{\rm Mpc}^{-1}\,,
\end{equation}
with $\bar{r}$ that of the log-normal OU process at the $\sigma_{\rm e}^2$ of the limit curve. For the power-law models of Section~\ref{sec:timedep}, which have larger $\tau_c$ before last scattering, we use the computed $\epsilon_{\rm mean}$ directly.

This estimate uses the tight-coupling form of the kernel throughout. Once the photons free stream ($\nu\lesssim k$), the departure from equilibrium in blocks $1$ and $2$ is no longer suppressed, and the correction becomes of order $k\bar{r}$ times the leading term, with no definite sign. The multipoles $\ell\approx2000$--$4000$ that constrain the transport channel have $k\approx0.15$--$0.3\,{\rm Mpc}^{-1}$, comparable to $\nu_\star$, so near last scattering this omitted term could change $\nu_\star$ by order unity in either direction. We therefore regard $\nu_\star$ as uncertain by about a factor of two and treat $\epsilon_{\rm mean}$ as a symmetric error (Appendix~\ref{app:error_budget}).

\paragraph{Fluctuating power ---} The second term of Equation~\ref{eqn:power_decomposition} is the power spectrum of $\var{\Theta}$, averaged over both the electron field and the primordial perturbations. We compute it to first order in $\delta_{\rm e}$, and treat it as an unmodeled template in the likelihood in order to estimate the bias from its omission.

Consider a shell of thickness $\Delta \eta$ at comoving distance $D(\eta)\simeq D_A^\star$ from the observer. Solving Equation~\ref{eqn:fluc_eq} to first order along a line of sight $\vu{n}$, the shell changes the observed temperature by
\begin{equation}\label{eqn:shell_kick}
    \begin{gathered}
    \var{\Theta}_\eta(\vu{n}) = -e^{-\bar{\tau}(\eta)}\;\delta\tau(\vu{n})\;c(\vu{n})\,, \\ 
    \delta\tau(\vu{n})\equiv\int_{\rm shell}\dd\eta\;\bar{\Gamma}\,\delta_{\rm e}\,,\qquad
    c(\vu{n})\equiv\mathcal{C}[\bar{\Theta}]\big(D(\eta)\,\vu{n},\,\vu{n},\,\eta\big)\,,
    \end{gathered}
\end{equation}
where $\delta\tau$ is the excess optical depth of the shell and $c$ the departure of the mean from equilibrium with the scatterers there. Here we have adopted the free-streaming propagator $G=e^{-\bar{\tau}}$, which neglects the scattering of the perturbation's own low multipoles back into the line of sight. This is exact for $\ell' \geq3$, on which $\mathcal{C}$ acts as the identity (Figure~\ref{fig:commutator_blocks}), and is a good approximation for the dipole and quadrupole once $\bar{\Gamma}\lesssim k$, which is the epoch that contributes below. We see that $\var{\Theta}_{\eta}$ is the product of two random fields on the sky, $\delta\tau$ and $c$, which are independent because $\delta_{\rm e}$ is assumed to be uncorrelated with the primordial perturbations that source $c$.

Since the template matters only at $\ell\gtrsim10^3$, we work in the flat-sky limit. The power spectrum of the product of the two fields is the convolution
\begin{equation}\label{eqn:product_spectrum}
    C_\ell = \int\frac{\dd^2L}{(2\pi)^2}\;C^\tau_{\abs{\vb*{\ell}-\vb{L}}}\;C^c_{L}\,,
\end{equation}
where $\vb{L}$ runs over the modes of $c$ and $\vb*{\ell}-\vb{L}$ is the mode of $\delta\tau$ that pairs with it. We need the spectrum of $\delta\tau$, and the total variance and rms wavenumber of $c$.

The angular power spectrum of $\delta\tau$ on the shell follows from Equation~\ref{eqn:shell_kick} and the Limber approximation. For $\Delta\eta\gg\Delta_c$ the excess optical depth is a projection of the electron field through the shell, which retains only modes transverse to the line of sight, with $L=k_\perp D_A^\star$:
\begin{equation}\label{eqn:limber_shell}
    C^\tau_L = \frac{\bar{\Gamma}^2\,\Delta\eta}{D_A^{\star2}}\,P_{\rm e}\!\left(L/D_A^\star;\,\eta\right)\,,
\end{equation}
with $P_{\rm e}$ the spectrum of Appendix~\ref{app:power_spectrum} evaluated with the coherence length $\Delta_c(\eta)$ of the shell. Using the Gaussian spectrum (Equation~\ref{eqn:pk_gaussian}, which is also the leading term for log-normal fluctuations) and expanding for small $k\Delta_c$,
\begin{equation}\label{eqn:ctau_white}
    C^\tau_L = \left(4 \pi \theta_c^2 \right)\left(2 \zeta_{\rm e} \bar{\Gamma} \Delta \eta \right)\left(1 - 2 L^2 \theta_c^2 + \dots \right)\,,
\end{equation}
where $\theta_c(\eta) \equiv \Delta_c(\eta) / D_A^{\star}$ is the angular size of a patch on the shell. The spectrum of the excess optical depth is white above the patch resolution, with an amplitude set by the solid angle of a patch (up to a numerical factor) times the variance of the excess optical depth.

The variance of the collision term for spectrum type $X = T, P$ is
\begin{align*}
    V^X_\mathcal{C} &\equiv \int \frac{\dd[2]L}{(2\pi)^2}\, C_L^{c,X}\\
    &= \int\dd\ln k\;\mathcal{P}_{\mathcal{R}}(k)\sum_{\ell'}(2\ell'+1)\abs{c^X_{\ell'}(k,\eta)}^2\,,\numberthis \label{eqn:VC_def}
\end{align*}
where the $c^X_{\ell'}$ are the multipoles of $\mathcal{C}[\bar{\Theta}]$ at fixed $\vb{k}$, with all quantities in units of $\Theta$ ($\Theta_{\ell'} = F_{\gamma\ell'}/4$, $\Theta_{{\rm P}\ell'} = G_{\gamma\ell'}/4$, and $\Pi$ rescaled accordingly). In the notation of Section~\ref{sec:modified_BH}, for temperature
\begin{equation}
    \begin{gathered}
        c^T_0=0\,,\quad c^T_1=\Theta_1-\frac{v_{\rm b}}{3}\,,\\ c^T_2=\Theta_2-\frac{\Pi}{10}\,,\quad c^T_{\ell'\geq3}=\Theta_{\ell'}\,,
    \end{gathered}
\end{equation}
and for polarization
\begin{equation}
    \begin{gathered}
        c^P_0=\Theta_{{\rm P}0}-\frac{\Pi}{2}\,,\quad c^P_2=\Theta_{{\rm P}2}-\frac{\Pi}{10}\,,\quad c^P_{\ell'=1,\,\ell'\geq3}=\Theta_{{\rm P}\ell'}\,.
    \end{gathered}
\end{equation}
The rms wavenumber $k_{\mathcal{C}}$ is defined by the same integral with an additional factor of $k^2$, divided by $V^X_{\mathcal{C}}$; at $z=1000$ it is $k_{\mathcal{C}} = 0.11\,{\rm Mpc}^{-1}$ for both $T$ and $P$.

Equation~\ref{eqn:product_spectrum} simplifies when $C^\tau$ varies slowly across the band of $\vb{L}$ that carries the power of $c$. Expanding $C^\tau_{\abs{\vb*{\ell}-\vb{L}}}$ about $\vb*{\ell}$ and averaging over the orientation of $\vb{L}$, the term linear in $\vb{L}$ vanishes and
\begin{equation}\label{eqn:collapse}
    C_\ell = V^X_{\mathcal{C}}\left[C^\tau_\ell + \frac{\langle L^2\rangle}{4}\,\nabla_{\ell}^2\, C^\tau_\ell + \dots\right]\,,
\end{equation}
where $\nabla^2_\ell$ is the two-dimensional Laplacian in $\vb*{\ell}$ and $\langle L^2\rangle$ is the mean of $L^2$ weighted by $C^c_L$. A mode $\vb{k}$ of $c$ appears on the shell at $L = k_\perp D_A^\star$, so $\langle L^2\rangle \leq (k_{\mathcal{C}} D_A^\star)^2$. For the spectrum of Equation~\ref{eqn:pk_gaussian} the second term is then at most $2(k_{\mathcal{C}}\Delta_c)^2$ of the first at any $\ell$, and we retain only the leading term. This bound is $2.4\%$ at $\Delta_c = 1$~Mpc, but in the baseline model $\Delta_c$ grows through recombination, to $3.6\,\Delta_c(z_{\rm ref})$ at $z = 1000$. Therefore, for $\Delta_c(z_{\rm ref}) \gtrsim 1$~Mpc the expansion parameter is no longer small toward the end of recombination, an epoch that still contributes to the integral of Equation~\ref{eqn:incoherent_power} below. Because the convolution conserves the total power of the product, the neglected terms redistribute the template in $\ell$ near its turnover rather than change its amplitude. We do not include this in the band, and return to it in Appendix~\ref{app:error_budget}.

\begin{figure}[t]
    \centering
    \includegraphics{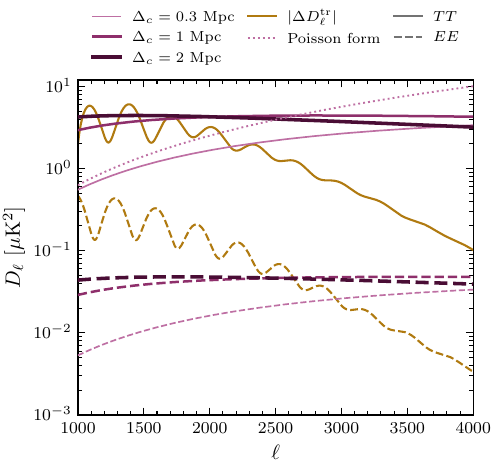}
    \caption{The omitted fluctuating power against the transport corrections, both for the baseline model at $\zeta_{\rm e}(z_{\rm ref})=0.01$ and $\sigma_{\rm e}^2=0.3$. Purple: $D_\ell^{\rm inc}$ (Equation~\ref{eqn:incoherent_power}, $z\geq900$) in $TT$ (solid) and $EE$ (dashed) for $\Delta_c=0.3$, $1$, and $2$~Mpc. Gold: $\abs{\Delta D_\ell^{\rm tr}}$, the change in the $TT$ and $EE$ spectra due to the transport corrections. Dotted: the Poisson form, Equation~\ref{eqn:poisson_form}, for $\Delta_c=0.3$~Mpc, with the effective coherence length $\Delta_{c, \rm eff} \equiv 7.3\,\Delta_c(z_{\rm ref})$, the value that matches the amplitude of the integrated template. The incoherent power term exceeds the transport corrections at $\ell\gtrsim2000$ in both $TT$ and $EE$. It is Poisson-like ($\propto\ell^2$) where the patches are unresolved and turns over at $\ell\sim D_A^\star/\Delta_{c,\rm eff}$, which falls within the observed range for $\Delta_c\gtrsim1$~Mpc.}\label{fig:incoherent}
\end{figure}

In total, the angular power spectrum of a single shell is
\begin{equation}
     C_\ell^{\eta} = e^{-2\bar{\tau}}\;C^\tau_\ell\;V^X_\mathcal{C}(\eta)\,.
\end{equation}
Shells thicker than $\Delta_c$ have independent $\delta \tau$ contributions, so we can add the contributions from each shell. Finally, we find that the incoherent power from the unresolved patches has the spectrum:
\begin{subequations}
\begin{align}\label{eqn:incoherent_power}
    C_\ell^{\rm inc,TT} &= \frac{1}{D_A^{\star2}}\int\dd\eta\;\bar{\Gamma}^2e^{-2\bar{\tau}}\;V_\mathcal{C}^T(\eta)\;P_{\rm e}\!\left(\ell/D_A^\star;\,\eta\right)\,,
    \\
    C_\ell^{\rm inc,EE} &= \frac{1}{2}\,C_\ell^{\rm inc,TT}\big|_{V_\mathcal{C}^T \to V_\mathcal{C}^P}\,.
\end{align}
\end{subequations}
The factor of $1/2$ multiplying $C_\ell^{\rm inc, EE}$ arises because the power is divided equally between $E$ and $B$ and there is no correlation with temperature, so $C_\ell^{\rm inc, TE} = 0$.

\begin{table*}[t]
\caption{The omitted fluctuating power. $\Delta{\rm Var}^{\rm inc}\equiv\sum_\ell(2\ell+1)C_\ell^{\rm inc}/4\pi$ is the total variance of the omitted term in temperature, from Equation~\ref{eqn:incoherent_power} restricted to $z\geq900$, at $\zeta_{\rm e}(z_{\rm ref})=0.01$ and $\sigma_{\rm e}^2=0.3$, and $\Delta{\rm Var}^{\rm tr}\equiv\sum_\ell(2\ell+1)\abs{\Delta C_\ell^{\rm tr}}/4\pi$ the same integral of the transport corrections at the same $\zeta_{\rm e}$; the ratio is given for $TT$ and for $EE$. The last six columns give the bias $b$ on $\zeta_{\rm e}$ (Equation~\ref{eqn:fisher_bias}) at $\Delta_c=0.3$, $1$, and $2$~Mpc, marginalized over $\Lambda$CDM, $\sigma_{\rm e}^2$, and the calibration parameters, with the templates evaluated at the $\sigma_{\rm e}^2$ of each dataset's limit curve.}\label{tab:incoherent}
\begin{ruledtabular}
\begin{tabular}{l c c c ccc ccc}
 & $\Delta{\rm Var}^{\rm inc}_{TT}$ & \multicolumn{2}{c}{$\Delta{\rm Var}^{\rm inc}/\Delta{\rm Var}^{\rm tr}$} & \multicolumn{3}{c}{$b$ (PR3)} & \multicolumn{3}{c}{$b$ (SPA)} \\
Model & $[\mu{\rm K}^2]$ & $TT$ & $EE$ & $0.3$ & $1$ & $2$ & $0.3$ & $1$ & $2$ \\
\hline
constant $\tau_c$ & $12.5$ & $0.81$ & $0.29$ & $-0.02$ & $-0.10$ & $-0.12$ & $+0.28$ & $+0.02$ & $-0.11$ \\
$n=-3$ & $4.7$ & $0.10$ & $0.08$ & $<\!0.001$ & $-0.003$ & $-0.02$ & $+0.007$ & $+0.06$ & $+0.09$ \\
$n=-4$ & $5.1$ & $0.13$ & $0.09$ & $<\!0.001$ & $-0.004$ & $-0.02$ & $+0.01$ & $+0.08$ & $+0.10$ \\
$n=-5$ & $5.5$ & $0.18$ & $0.11$ & $<\!0.001$ & $-0.007$ & $-0.04$ & $+0.02$ & $+0.12$ & $+0.11$ \\
\end{tabular}
\end{ruledtabular}
\end{table*}

We evaluate Equation~\ref{eqn:incoherent_power} with the multipoles of the mean hierarchy, truncated at $\ell'=50$ on 60 logarithmically spaced wavenumbers in $10^{-3}\le k\le1\,{\rm Mpc}^{-1}$, with $D_A^\star=13869$~Mpc and the log-normal $P_{\rm e}$ (Equation~\ref{eqn:pk_lognormal}). Because the shape functions and $P_{\rm e}$ are non-linear in $\sigma_{\rm e}^2$ at fixed $\zeta_{\rm e}$, all templates are evaluated at the $\sigma_{\rm e}^2$ of the point on each dataset's limit curve, $\sigma_{\rm e}^2=\ln\big(1+\eavg{\delta_{\rm e}^2}_{\rm lim}(\Delta_c)\big)$. The integral is restricted to $z\geq900$. In the baseline model $\Delta_c(\eta)=\tau_c/\bar{\Gamma}$ grows without bound as recombination completes, reaching $17\,\Delta_c(z_{\rm ref})$ by $z=900$, and since $P_{\rm e}\propto\Delta_c^3$ in the white regime the integrand would otherwise be dominated by an epoch at which the separation of scales assumed throughout has failed. The cut retains $95\%$ of the weight $\int\bar{\Gamma}^2e^{-2\bar{\tau}}V_{\mathcal{C}}\,\dd\eta$ and marks the edge of the model's domain, and we treat the result for this model as an estimate (Appendix~\ref{app:error_budget}). The power-law models of Section~\ref{sec:timedep} are insensitive to the cut. 

To translate the omitted term into an error on the constraints, let $S_{\rm tr}$ be the transport corrections and $S_{\rm inc}$ be Equation~\ref{eqn:incoherent_power}, both per unit $\zeta_{\rm e}$. Data containing $\zeta_{\rm e}(S_{\rm tr}+S_{\rm inc})$ but fit with $S_{\rm tr}$ alone return $\zeta_{\rm e}(1+b)$, with $b$ given by Equation~\ref{eqn:fisher_bias} for each likelihood, marginalizing over \lcdm, $\sigma_{\rm e}^2$, and the calibration parameters with the priors of the analysis. We take 
\begin{equation}\label{eqn:eps_inc}
    \epsilon_{\rm inc}\equiv\abs{1/(1+b)-1}
\end{equation}
as the fractional error on the transport channel. Table~\ref{tab:incoherent} gives $b$ at $\Delta_c=0.3$, $1$, and $2$~Mpc. For PR3 $b<0$ throughout, reaching $-0.12$ at $\Delta_c\approx1.6$~Mpc. For SPA $b$ is positive below $\Delta_c\approx1$~Mpc, where the limit is conservative, and negative above.

The bias is modest although $D_\ell^{\rm inc}$ exceeds the transport corrections at $\ell\gtrsim2000$ for the coherence lengths in the table. We show $D_\ell^{\rm inc}$ for $TT$ and $EE$ in Figure~\ref{fig:incoherent}, for three values of $\Delta_c$. To leading order in $\ell\theta_c$, $P_{\rm e}\to8\pi\sigma_{\rm e}^2\Delta_c^3$ and Equation~\ref{eqn:incoherent_power} is white in $C_\ell$, so that
\begin{equation}\label{eqn:poisson_form}
    D_\ell^{\rm inc}\simeq2\,\Delta{\rm Var}\,(\ell\theta_{c,\rm eff})^2\,,
\end{equation}
where $\theta_{c,\rm eff} = \Delta_{c,\rm eff}/D_A^\star$ is defined by this relation. This shares the typical Poisson form of unresolved point sources, with the coherent fluctuating patches as the sources. We show this as a dotted line in Figure~\ref{fig:incoherent}. The foreground models for the CMB temperature bandpowers from the three experiments used here all include a Poisson-like foreground template \citep{Planck:2019nip,AtacamaCosmologyTelescope:2025blo,SPT-3G:2025bzu}, and when the foreground-marginalized likelihoods are constructed the covariance in this direction is inflated \citep{Dunkley:2013vu,Balkenhol:2024obe}. The omitted term also has no $TE$ component and a smaller $EE$ component. For PR3 at $\Delta_c=1$~Mpc, with only $\zeta_{\rm e}$ free, $TT$ alone gives $b=-0.33$ and $EE$ alone $-0.07$; all spectra together give $-0.14$, and marginalizing over the other parameters $-0.10$. With a diagonal covariance the single-parameter value would be $-0.54$. The remaining uncertainty on $S_{\rm inc}$ is the free-streaming propagator for the $\ell'\leq2$ components of $c$, which contribute $80\%$ of $V_{\mathcal{C}}$ and are corrected at relative order $\bar{\Gamma}/k\approx0.1$--$0.2$ over the contributing epoch; we take the template to be uncertain by $15\%$.

\paragraph{Scatter ---} The fluctuating part also makes the spectrum of a single realization scatter about its average. A single $a_{\ell m}$ averages over $N\approx(D_A^\star/\ell)^2\Delta\eta_\star/\Delta_c^3$ independent patches, with $\Delta\eta_\star\approx20$~Mpc the thickness of the last-scattering surface, and even if each patch perturbed the photon distribution at order unity the scatter $[N(2\ell+1)]^{-1/2}$ at $\ell=4000$ would be $0.07\%$ and $0.2\%$ for $\Delta_c=1$ and $2$~Mpc. We neglect this.

\subsubsection{Breakdown at small scales}

At small scales, our model breaks down due to the approximations made in the ``separate universe'' treatment of the modified recombination history detailed in Section~\ref{sec:modified_rec}. At sufficiently small scales, two of these approximations fail: static perturbations and the locality of the recombination process. The latter fails once the typical fluctuation size is smaller than the mean free path of Lyman-$\alpha$ photons, in which case neighboring regions mix and the problem can no longer be factorized into separate recombination calculations. This occurs at $k \gtrsim 10^3\,\text{Mpc}^{-1}$ \citep{Lee:2021bmn,Schiff:2025vjf} and is not the limiting condition in our model, so we do not discuss it further.

The limiting scale is instead set by the breakdown of the static perturbation assumption, which happens below the baryon Jeans scale. This can be seen by considering the terms that we drop in Section~\ref{sec:modified_rec}, in particular the velocity-divergence modification to the local expansion rate, $H \to H + \theta_{\rm b}/3a$. The size of the neglected term relative to the (global) expansion rate is
\begin{align}
   \frac{\abs{\theta_{\rm b}}/3a}{H} &= \frac{\abs{\dot{\delta}_{\rm b}}}{3aH}
   \sim \frac{k c_s}{aH}\, \delta_{\rm b}\,,
\end{align}
where we used $\theta_{\rm b} = -\dot{\delta}_{\rm b}$ and $\dot{\delta}_{\rm b} \sim k c_s \delta_{\rm b}$ for an oscillating mode, with $c_s$ the baryon sound speed. 
The neglected dynamical effects become important at $k_J \sim aH/c_s$, the scale at which the sound-crossing time is comparable to the Hubble time and perturbations are no longer static on the relevant timescales, as our approach assumes. At $z > 150$, the baryon Jeans scale is $k_J \approx 10^2\,\text{Mpc}^{-1}$ \citep{Lee:2021bmn}.

The modification to the recombination dynamics is sensitive to the overall variance of the fluctuations, so an appropriate comparison is between $k_J$ and the peak of the power spectra in Appendix~\ref{app:power_spectrum}. The Gaussian spectrum peaks at $k_{\rm peak} = \sqrt{3} / \Delta_c$ (Appendix~\ref{app:power_spectrum}), a slightly smaller wavenumber than the peak of the log-normal spectrum. We therefore take $k_{\rm peak} \lesssim 100\,\text{Mpc}^{-1}$ as an upper limit, corresponding to $\Delta_c \approx 10\,\text{kpc}$. This lower bound on $\Delta_c$ could be extended to $\sim 1\,\text{kpc}$ if the treatment of \citet{Lee:2021bmn} were used, or beyond using the non-local treatment of \citet{Schiff:2025vjf}. We leave such extensions for future work. Since the existing studies indicate that the neglected effects accelerate recombination further (Section~\ref{sec:modified_rec}), we expect our limits to be conservative at these scales, but we do not attempt to quantify this.

\subsection{Uncertainty shown in Figure~\ref{fig:power_spectrum_constraints}}\label{app:error_budget}

Both biases are fractional errors on the transport corrections, and we combine them as
\begin{equation}\label{eqn:eps_tot}
    \epsilon_{\rm tot} = \sqrt{\epsilon_{\rm mean}^2 + \epsilon_{\rm inc}^2}\,,
\end{equation}
with $\epsilon_{\rm mean}$ from Equation~\ref{eqn:eps_mean} and $\epsilon_{\rm inc}$ from Equation~\ref{eqn:eps_inc}, for the baseline constant-$\tau_c$ model. We evaluate $\epsilon_{\rm mean}$ directly along each dataset's upper-limit curve, and $\epsilon_{\rm inc}$ at twelve coherence lengths between $0.05$ and $3$~Mpc, interpolated in $\ln\Delta_c$. Neither affects the recombination channel, which is local and responds to the variance of the fluctuations on all scales.

To propagate $\epsilon_{\rm tot}$ to the limits we use the fact that, at fixed $\eavg{\delta_{\rm e}^2}$, the transport corrections depend on the coherence length only through $\eavg{\delta_{\rm e}^2}\eta_c^{\rm phys}\propto\Delta_c$. A fractional error on their strength is therefore equivalent to the same fractional error on $\Delta_c$. Writing $L(\Delta_c) \equiv \eavg{\delta_{\rm e}^2}_{\rm lim}(\Delta_c)$ for the limit curve, we bound the systematic error by the curve displaced horizontally by that amount in either direction,
\begin{equation}\label{eqn:band}
    L^{\pm}(\Delta_c) = L\big(\Delta_c\left[1\mp\epsilon_{\rm tot}(\Delta_c)\right]\big)\,.
\end{equation}
This displacement vanishes where the limit is set by the recombination channel and the curve is flat. We take it to be symmetric: while the photons are tightly coupled $\epsilon_{\rm mean}$ weakens the limits, but its sign is not established near last scattering, and the sign of $\epsilon_{\rm inc}$ differs between the datasets.

The band shown in Figure~\ref{fig:power_spectrum_constraints} combines this systematic estimate with the statistical uncertainty of the limit curve, the $68\%$ interval $[L_{16}, L_{84}]$ from 100 block-bootstrap refits of the quantile regression of Section~\ref{sec:results}. The systematic displaces the curve along $\Delta_c$, while the bootstrap scatters it in $L$ at fixed $\Delta_c$, so the two are combined at fixed $\Delta_c$ as independent errors in $\ln L$,
\begin{equation}\label{eqn:band_total}
    L_{\pm} = L\,\exp\!\left[\pm\sqrt{\delta_{\rm sys,\pm}^2 + \delta_{\rm stat,\pm}^2}\right]\,,
\end{equation}
with $\delta_{\rm sys,\pm} = \big|\ln\big(L^{\pm}/L\big)\big|$ from Equation~\ref{eqn:band}, $\delta_{\rm stat,+} = \ln(L_{84}/L)$, and $\delta_{\rm stat,-} = \ln(L/L_{16})$. Where the recombination channel sets the limit, the band is the statistical interval alone.

In the right panel we construct the band in the same way, as a function of $\Delta_c$, and map the curve and both edges to $k_s$ through the relations of Appendix~\ref{app:projection}. For the systematic error this is, to good approximation, $k_s\to k_s/(1\mp\epsilon_{\rm tot})$, and for the statistical part we neglect the weak dependence of $k_s$ on $L$ through $\eta_c^{\rm phys}(\sigma_{\rm e}^2)$. For the restated spectra $\epsilon_{\rm mean}$ uses the $\bar{r}$ of the spectrum in question, which follows from Equation~\ref{eqn:corr_from_power},
\begin{equation}\label{eqn:rbar_spectral}
    \bar{r} = \frac{2}{\pi}\,\frac{\int W k^{-2}\,\dd\ln k}{\int W k^{-1}\,\dd\ln k}\,,
\end{equation}
giving $\bar{r}/\eta_c^{\rm phys} = 1$ for the Gaussian OU process, $\simeq 1 + \sigma_{\rm e}^2/8$ for the log-normal OU process (at most $1.06$ on the limit curves), $4/\pi^2 \simeq 0.41$ for the $\delta$-function spike, and $(4/\pi^2)\,e^{\sigma_s^2} \simeq 1.10$ for the log-normal spectrum with $\sigma_s = 1$. For $\epsilon_{\rm inc}$ we use the value computed for the OU process at the same $\Delta_c$. This is conservative for the $\delta$-function spike, which has no power at $k\lesssim k_s$ and hence no fluctuating power in the observed range.

For the limits on the OU process, $\epsilon_{\rm mean}$ is $6\%$, $20\%$, and $40\%$ at $\Delta_c=0.3$, $1$, and $2$~Mpc for either dataset, and $\epsilon_{\rm inc}$ is $2\%$, $12\%$, and $14\%$ for PR3 and $22\%$, $2\%$, and $13\%$ for SPA, so the mean evolution dominates for PR3 at all $\Delta_c$ shown and for SPA above $\Delta_c \sim 1$~Mpc, below which the fluctuating power dominates. In the right panel $\epsilon_{\rm mean}\approx0.13\,(k_s/{\rm Mpc}^{-1})^{-1}$ for the $\delta$-function spike and $0.57\,(k_s/{\rm Mpc}^{-1})^{-1}$ for the log-normal spectrum with $\sigma_s=1$. We truncate both panels at $\Delta_c = 2$~Mpc, where $\nu_\star\bar{r}\approx0.4$ for the OU process and the log-normal spectrum and a first-order estimate ceases to be meaningful. For the $\delta$-function spike, whose $\bar{r}$ is smaller, this is conservative. In the right panel the truncation corresponds to $k_s\approx0.8$ and $1.3\,{\rm Mpc}^{-1}$ for the two spectra.

These estimates rest on the following assumptions. Both errors are of second order in $\delta_{\rm e}$, and $\epsilon_{\rm mean}$ is of first order in $\bar{r}$, with a rate $\nu_\star$ calibrated using the tight-coupling form of the kernel and uncertain by about a factor of two near last scattering (Appendix~\ref{app:range_of_validity}).

The fluctuating power is computed from $\var{\Theta}$ at first order in $\delta_{\rm e}$, in the Limber approximation and with the free-streaming propagator, which we estimate makes the template uncertain by roughly $15\%$. It also retains only the leading term of the expansion in Equation~\ref{eqn:collapse}, which is no longer valid at $\Delta_c(z_{\rm ref}) \gtrsim 1$~Mpc. For the baseline model $\epsilon_{\rm inc}$ further depends on the restriction to $z\geq900$, without which it would change by a factor of two to five. For this model $\epsilon_{\rm inc}$ is therefore an estimate whose reliability decreases above $\Delta_c\approx1$~Mpc. It is subdominant to $\epsilon_{\rm mean}$ there, so that a factor-of-two error in $\epsilon_{\rm inc}$ would increase $\epsilon_{\rm tot}$ by at most about a third (PR3 at $1$~Mpc), and by $15\%$ at $2$~Mpc.

Finally, the two errors are added in quadrature as independent uncertainties; were they fully correlated the total would be larger by at most $\sqrt{2}$. The small-scale breakdown is not a quantifiable uncertainty of this kind, and we restrict the figure to $\Delta_c\geq10$~kpc. Reducing these uncertainties through more detailed modeling of the large-$\Delta_c$ regime is left for future work.

\section{Code details}\label{app:code}

\textsc{classito} is a fork of \textsc{class} v3.3 \citep{Blas:2011rf}. Its companion package \textsc{classito-cobaya}\footnote{\url{https://github.com/gplynch619/classito-cobaya}} interfaces it with user-defined models (Appendix~\ref{app:cobaya}).

\subsection{Inputs and required settings}\label{app:code_inputs}

The corrections are computed by a new \textsc{class} module named \code{ito} which tabulates $\sigma_{\rm e}^2(z)$, $\tau_c(z)$, and the shape functions of Equations~\ref{eqn:f_gaussian}--\ref{eqn:f_pol} on the same redshift grid as the thermodynamics module. The \code{ito} module is initialized in two stages, with the quantities needed for the $\Delta X_{\rm e}/X_{\rm e}$ correction of Section~\ref{sec:modified_rec} initialized before the \code{thermodynamics} module, and the rest of the \code{ito} quantities (e.g., $\tilde{\Gamma}(z)$) initialized after. The modified hierarchy and line-of-sight sources of Section~\ref{sec:modified_BH} are implemented in the \code{perturbations} module.

The following settings are recommended when using the module:

\begin{lstlisting}[style=settings]
    has_ito = yes
    gauge = newtonian
    radiation_streaming_approximation = 3
    tight_coupling_trigger_tau_c_over_tau_h = 0.0005
    tight_coupling_trigger_tau_c_over_tau_k = 4.0e-05
    start_small_k_at_tau_c_over_tau_h = 0.0004
    start_sources_at_tau_c_over_tau_h = 0.006
\end{lstlisting}
The modified hierarchy is only implemented in the Newtonian gauge, and the radiation streaming approximation \citep{Blas:2011rf} must be turned off (\code{radiation_streaming_approximation = 3}). Tensor perturbations are not implemented. We have also not implemented the tight-coupling approximation (TCA) for the modified hierarchy, as we found the standard \code{ndf15} stiff integrator used by \textsc{class} to be adequate even when the full hierarchy is integrated from early times. However, the TCA cannot be disabled entirely because the scalar initial conditions assume it holds when the integration of each mode begins. By default, \textsc{class} begins integration when $aH/\bar{\Gamma} \leq$ \code{start_small_k_at_tau_c_over_tau_h} and $k/aH \leq$ \code{start_large_k_at_tau_h_over_tau_k} \citep{Blas:2011rf}. The first trigger, which bounds $aH/\bar{\Gamma}$ and is independent of $k$, should therefore be set modestly above the former, and the second, which bounds $k/\bar{\Gamma}$, modestly above the product of the two. Since the first sets the latest time at which any mode can remain tightly coupled, lowering it together with its starting threshold is the most direct way to end the approximation earlier. The above settings were used in conjunction with the high-precision settings needed by ACT and SPT-3G.

Three parameterizations of the fluctuations are supported:
\begin{enumerate}
    \item {\it Constant}: \code{var_e} $= \sigma_{\rm e}^2$ and \code{tau_c} $= \tau_c$, both constant in time. This is the baseline model of our analysis (Section~\ref{sec:new_params}).
    \item {\it Power law}: setting \code{tau_c_type = time_dependent} with index
    \code{tau_c_power_law_n} $= n$ gives the coherence length of Equation~\ref{eqn:time_dep_corr}, normalized so that $\tau_c(z_{\rm ref}=1100)$ equals the input \code{tau_c}.
    \item {\it Tabulated}: with \code{ito_input_mode = table}, the user supplies
    $\sigma_{\rm e}^2(z)$ and the comoving coherence length $\Delta_c(z)$ (in Mpc) at a set of redshift nodes, either as inline lists (\code{ito_z_table},
    \code{ito_var_e_table}, \code{ito_Delta_c_table}) or as a three-column file
    (\code{ito_table_file}). The nodes are splined onto the internal grid and held at their endpoint values outside the tabulated range; the code forms
    $\tau_c(z) = \Delta_c(z)\, \bar{\Gamma}(z)$ internally once the thermodynamics history is available. This is the mode intended for user-supplied clumping models.
\end{enumerate}
In the tabulated mode, $\Delta X_{\rm e}/X_{\rm e}$ is obtained by evaluating the precomputed separate-universe table of Section~\ref{sec:modified_rec} along the trajectory $(z, \sigma_{\rm e}^2(z))$. Both Gaussian and log-normal statistics for the fluctuations are implemented, selected by \code{electron_dist = 'gaussian'} or \code{electron_dist = 'lognormal'}.

\subsection{Interface for user-defined clumping models}\label{app:cobaya}

To make the code directly usable for constraining specific generation mechanisms, we provide a companion package, \textsc{classito-cobaya}, which connects the tabulated mode to the \textsc{cobaya} sampling framework \citep{Torrado:2020dgo}. The package supplies a subclass of \textsc{cobaya}'s \textsc{class} \code{Theory} interface that declares a requirement, \code{ito_clumping}, together with a template theory class that provides it. A user implements their model as a single method returning
$\bigl(\sigma_{\rm e}^2(z), \Delta_c(z)\bigr)$ on a redshift grid chosen by the wrapper, and declares the model's sampled parameters. For example, a minimal model with a constant variance and coherence length is
\begin{lstlisting}[style=settings, language=Python]
import numpy as np
from classito_cobaya import ClumpingModel

class MyClumpingModel(ClumpingModel):

    def get_can_support_params(self):
        return ["A_e", "Delta_0"]

    def clumping_functions(self, z, **params):
        sigma_e2 = params["A_e"] * np.ones_like(z)
        Delta_c  = params["Delta_0"] * np.ones_like(z)   # comoving Mpc
        return sigma_e2, Delta_c
\end{lstlisting}
and its priors are set in the \code{params} block of the input file like any
other sampled parameter:
\begin{lstlisting}[style=settings]
params:
  A_e:     {prior: {min: 0., max: 1.}, latex: '\sigma_{\rm e}^2'}
  Delta_0: {prior: {min: 0., max: 5.}, latex: '\Delta_c\,[{\rm Mpc}]'}
\end{lstlisting}

The dependency chain is then likelihood $\rightarrow C_\ell \rightarrow$ \code{ito_clumping} $\rightarrow$ model, structured in the input file as
\begin{lstlisting}[style=settings, language=Python]
theory:
  classito_cobaya.classito:
    extra_args: {has_ito: 'yes',
                 ito_input_mode: 'table', ...}
  my_module.MyClumpingModel:
    provides: [ito_clumping]
\end{lstlisting}
so that the sampled parameters reach the Boltzmann code through the model.
Tables are passed to \textsc{classito} as inline parameter lists, or through temporary files when they exceed \textsc{class}'s limit on parameter length.

Constraining a generation mechanism with the pipeline of this paper thus requires only the mapping from that mechanism's parameters to the variance and coherence length of the electron fluctuations as functions of redshift. Of course, many generating mechanisms may impact other aspects of the cosmological model, and a self-consistent evolution will require directly modifying the code. Nonetheless, the interface is a useful first step toward a full treatment in particular cases.
\end{document}